\pdfoutput=1
\documentclass[reprint,twocolumn,superscriptaddress,numerical,prx,floats]{revtex4-2}

\usepackage{bbold}
\usepackage{amsmath}
\usepackage{bm}
\usepackage{physics}
\usepackage{calrsfs}
\usepackage{graphicx}
\usepackage{bbm}
\usepackage{xcolor}
\usepackage{hyperref}
\usepackage{booktabs}
\usepackage{tensor}
\usepackage{empheq}
\usepackage{amsmath} 
\usepackage[protrusion=true, expansion=true, tracking=true, kerning=true, spacing=true, final]{microtype}
\usepackage{hyphenat}
\usepackage{tikz}
\usepackage{pgfplots} 

\newcommand{\lpartial}{\overleftarrow{\partial}}
\newcommand{\rpartial}{\overrightarrow{\partial}}

\DeclareMathAlphabet{\pazocal}{OMS}{zplm}{m}{n}

\usetikzlibrary{calc,arrows.meta}
\pgfplotsset{compat=1.17}
\usetikzlibrary{decorations.markings}
\tikzset{
    set arrow inside/.code={\pgfqkeys{/tikz/arrow inside}{#1}},
    set arrow inside={end/.initial=stealth, opt/.initial=},
    /pgf/decoration/Mark/.style={
        mark/.expanded=at position #1 with
        {
            \noexpand\arrow[\pgfkeysvalueof{/tikz/arrow inside/opt}]{\pgfkeysvalueof{/tikz/arrow inside/end}}
        }
    },
    arrow inside/.style 2 args={
        set arrow inside={#1},
        postaction={
            decorate,decoration={
                markings,Mark/.list={#2}
            }
        }
    },
}

\begin{document}

\title{Systematic Truncated Wigner Approximation for Open Spin Systems Reveals Long-Range Order Far from Equilibrium}

\author{Igor Timofeev}
\affiliation{Department of Physics and Astronomy, University College London,
Gower Street, London, WC1E 6BT, United Kingdom}
\author{Dawid A. Hryniuk}
\affiliation{Department of Physics and Astronomy, University College London,
Gower Street, London, WC1E 6BT, United Kingdom}
\author{Marzena H. Szymańska}
\email{m.szymanska@ucl.ac.uk}
\affiliation{Department of Physics and Astronomy, University College London,
Gower Street, London, WC1E 6BT, United Kingdom}

\date{\today}

\begin{abstract}

Long-range interactions can protect dissipative order against fluctuations, but determining when they do so in two dimensions remains out of reach for existing many-body methods.
We develop a continuous SU(2)-covariant truncated-Wigner framework that maps the problem onto stochastic dynamics with one Bloch sphere per site.
Derived systematically from the exact Stratonovich–Weyl evolution of the Lindblad equation, it generates both drift and reservoir noise for arbitrary spin $S$.
Benchmarks against exact and tensor-network calculations validate its description of non-equilibrium dynamics and steady states.
Applied to 400-site two-dimensional dissipative anisotropic XY lattices, our method reveals physics missed by mean-field and perturbative approaches: vortex–antivortex proliferation destroys the predicted staggered-XY phase for nearest-neighbour coupling, whereas algebraically decaying interactions stabilize an extended $\mathbbm{Z}_2$-ordered phase.
We obtain the first beyond-mean-field steady-state phase diagram of the long-range interacting model, revealing reentrant loss of order at strong coupling and closure of the ordered region beyond $\alpha\approx 3.5$.
Our new method provides a scalable route to strongly correlated driven-dissipative systems with long-range interactions, higher-dimensional geometries, spin–boson couplings, and spins beyond one half, relevant to modern experiments with magnetic atoms and polar molecules in optical lattices and tweezer arrays.

\end{abstract}

\maketitle

\section*{INTRODUCTION}

\begin{figure*}[t]
    \centering
    \includegraphics[width=\textwidth]{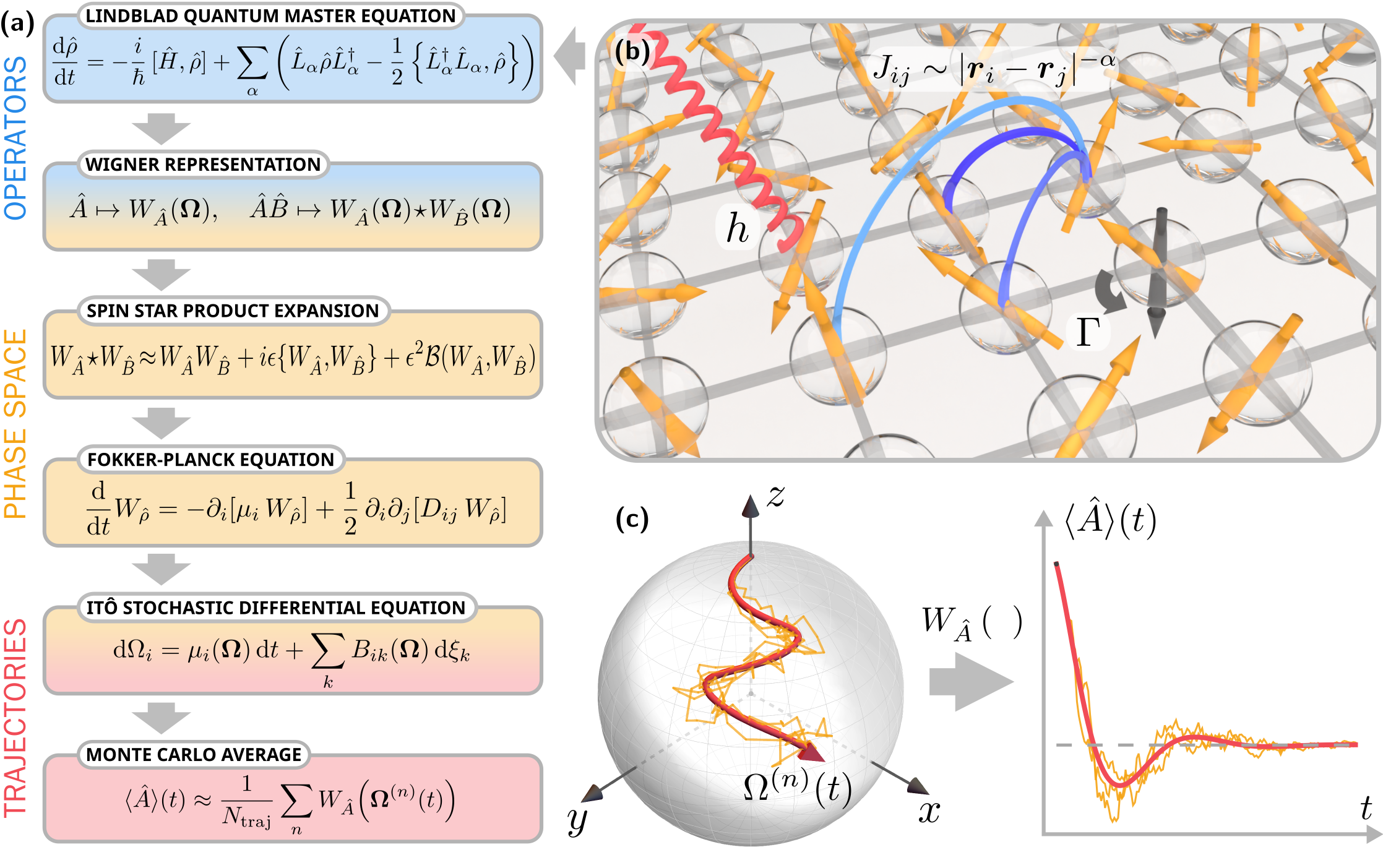}
    \caption{
        Overview of the TWA method for open many-spin systems.
        (a) Mapping the Lindblad master equation onto an ensemble of stochastic trajectories.
        Colours indicate the level of description: operators (blue), phase-space distributions (orange), and trajectories (pink).
        Operators are replaced by Wigner symbols and products by star products.
        Expanding the spin star product to second order in $\epsilon = (2S+1)^{-1}$, Eq.~\eqref{eq:starprod}, yields a Fokker--Planck equation for $W_{\hat{\rho}}$, Eq.~\eqref{eq:FP}, equivalent to an It\^{o} stochastic differential equation for $\bm{\Omega}$, Eq.~\eqref{eq:Ito}.
        Expectation values are obtained as Monte Carlo averages of $W_{\hat{A}}$ over the trajectories.
        (b) Physical setting: a two-dimensional spin lattice (orange Bloch vectors) in an array of trapping sites, with algebraically decaying interactions $J_{ij} \sim |\bm{r}_i - \bm{r}_j|^{-\alpha}$; example nearest-, next-nearest-, and next-next-nearest-neighbour couplings are shown by arcs of increasing length.
        Each spin experiences a local field $h$ (red coil) and spontaneous decay at rate $\gamma$ (black).
        (c) Left: stochastic trajectories $\bm{\Omega}^{(n)}(t)$ for one site on the Bloch sphere, indexed by $n$ and generated by Eq.~\eqref{eq:Ito} (yellow), with their average (red).
        Right: $W_{\hat{A}}$ evaluated along each trajectory gives noisy single-trajectory estimates (thin yellow curves), whose ensemble average (thick red curve) relaxes to the steady-state value (grey dashed line).}
    \label{fig:method_flowchart}
\end{figure*}

The competition between coherent many-body dynamics and environment-induced processes such as dissipation, noise, or external driving gives rise to some of the richest phenomena in modern physics~\cite{Carusotto2013,Sieberer2016,Fazio2025,Sieberer2025}.
Far from equilibrium, this competition is known to produce dissipative symmetry breaking and phase transitions~\cite{Baumann2010,Kirton2017,Diehl2008}, metastable states~\cite{Carr2013}, fluctuation-induced switching~\cite{Fitzpatrick2017,Macieszczak2016}, or spatially patterned order~\cite{Schauss2012,Lee2011,PhysRevLett.110.257204}, with important aspects of the resulting dynamics sharing close counterparts in classical stochastic and complex systems~\cite{Fontaine2022,Marcuzzi2016,Sieberer2016,Sieberer2025}.
Long-ranged interactions enrich this behaviour further by coupling distant degrees of freedom, reshaping the structure of correlations and propagation of critical fluctuations~\cite{Richerme2014,FossFeig2015}, allowing large-scale ordered phases to emerge~\cite{Defenu2023,Defenu2024}.
At the same time, modern experimental platforms provide increasingly precise control over coherent long-range couplings~\cite{Bohnet2016,BrowaeysLahaye2020} as well as engineered environmental channels~\cite{Barreiro2011,Diehl2008}, making them directly pertinent to quantum simulation~\cite{BlattRoos2012}, metrology~\cite{Pezze2018}, sensing~\cite{Degen2017}, and information processing~\cite{Verstraete2009}.
Their theoretical description nevertheless remains exceptionally demanding.
The many-body state space grows exponentially, while non-local connectivity in long-range interacting higher-dimensional lattices severely restricts many of the numerical methods that are successful in one and two dimensions~\cite{Weimer2021,Kshetrimayum2017,McKeever2021}.
Formulating scalable approaches that treat coherent evolution, incoherent dissipation, and complex long-range interactions on an equal footing therefore remains an outstanding problem of non-equilibrium many-body science.

Spin systems with long-range algebraically decaying interactions, $J_{ij}\sim|\bm{r}_i-\bm{r}_j|^{-\alpha}$, in which the exponent $\alpha$ interpolates between collective and effectively local dynamics~\cite{Defenu2023,Defenu2024}, provide a particularly important setting.
Couplings of this form arise naturally across atomic, molecular, or optical platforms.
Phonon-mediated Ising interactions with tunable $0<\alpha<3$ are engineered in trapped-ion crystals~\cite{Britton2012,Richerme2014,Bohnet2016,Zhang2017}, resonant dipole--dipole ($\alpha=3$) and van der Waals ($\alpha=6$) interactions in programmable Rydberg tweezer arrays~\cite{Labuhn2016,Bernien2017,BrowaeysLahaye2020}, while dipolar exchange occurs between polar molecules or magnetic atoms in optical lattices~\cite{Yan2013,Lepoutre2019,Patscheider2020}, the latter with potentially large spin quantum numbers $S$.
All are subject to noise acting on timescales comparable to the interactions, including spontaneous emission, finite Rydberg lifetimes, photon loss, or magnetic-field fluctuations.
Several applications rest on this competition, including metrologically useful spin squeezing in Rydberg arrays, optical clocks, and ion crystals~\cite{Bornet2023,Eckner2023,Franke2023}, native multi-qubit gates mediated by the interactions themselves~\cite{Levine2019}, dissipative preparation of correlated steady states~\cite{Barreiro2011,Diehl2008,Verstraete2009}, and analogue quantum simulation of many-body dynamics~\cite{Bernien2017,Zhang2017}.

Despite this experimental progress, computational tools capable of adequately treating correlations of such open long-range spin systems remain scarce.
Existing many-body approaches have largely treated algebraically decaying interactions in one dimension~\cite{Sulz2024,Hryniuk2024}, while methods developed for genuinely two-dimensional open systems have predominantly been applied to local or nearest-neighbour models~\cite{Kshetrimayum2017,McKeever2021}.
Only very recently have variational and tensor-network approaches begun to address the simultaneous combination of dissipation, two-dimensional geometry, and long-range interactions~\cite{Hryniuk2026,dunham2025efficienttimeevolution2d}, though both face significant limitations in the attainable interaction profiles or system sizes.
As a consequence, much of what is known about two-dimensional, long-range interacting, driven-dissipative lattices still rests on mean-field arguments, whose reliability is least assured in precisely the regime where spatial correlations and strong fluctuations matter most.

Phase-space methods have become a standard set of techniques for the simulation of large many-body systems, since they trade the exponentially large Hilbert space for an ensemble of classical trajectories whose cost grows only linearly with system size.
For bosonic fields, the truncated Wigner approximation (TWA), which is obtained by truncating the exact evolution equation of the Wigner function~\cite{Wigner1932,Hillery1984} at second order in derivatives and sampling the resulting Fokker--Planck equation, has been applied with great success to interacting ultracold gases and to nonlinear quantum optics~\cite{Steel1998,Sinatra2002,Blakie2008,Polkovnikov2003,Polkovnikov2010,Tosca2025}.
Together with the generalized $P$ representation~\cite{Glauber1963,Sudarshan1963,DrummondGardiner1980,Deuar2021,Ferrier2025}, it also extends naturally to open systems, where the coupling to a Markovian reservoir simply adds drift and diffusion terms to the phase-space equations of motion~\cite{GardinerZoller2004}.
Driven-dissipative exciton--polariton condensates have become a prominent application of this open-system formulation~\cite{WoutersSavona2009,Carusotto2013}.
TWA-based stochastic field theories have captured coherence and critical fluctuations across the condensation threshold in direct comparison with experiment~\cite{Luders2021,Alnatah2024}, as well as fluctuation-dominated phenomena such as vortex-mediated Berezinskii--Kosterlitz--Thouless (BKT) physics~\cite{Dagvadorj2015,Comaron2021} and Kardar--Parisi--Zhang (KPZ) scaling~\cite{Fontaine2022}.

For spin systems, however, most practical progress has instead been made in a \emph{discretized} phase space~\cite{Wootters1987}.
In the discrete TWA (DTWA), initial quantum fluctuations are sampled from a Wootters phase-point distribution and each trajectory is propagated with the classical mean-field equations~\cite{Schachenmayer2015PRX,Schachenmayer2015NJP}.
DTWA has proven remarkably accurate for closed spin-$1/2$ lattices with long-range couplings, reproducing experimental results with Rydberg and dipolar atoms~\cite{Orioli2018}.
It has been generalized to arbitrary spin length and to SU($N$) degrees of freedom~\cite{DavidsonPolkovnikov2015,Zhu2019,Nagao2021} as well as to clusters of exactly treated spins~\cite{Wurtz2018}.
Its accuracy is nevertheless difficult to establish \emph{a priori}, as its dynamics are postulated to be mean-field rather than obtained as the lowest order of a controlled expansion in a small parameter.
Moreover, its validity is typically restricted to short and intermediate times~\cite{Czischek2019,Sundar2019,Khasseh2020,Kunimi2021}, with the long-time steady state potentially falling outside this range.
Dissipative dynamics are also difficult to incorporate into the method, with most prescriptions being heuristic~\cite{Huber2022}.

Consequently, continuous phase-space formulations are potentially more systematic.
An SU(2)-covariant Wigner formalism for spin-$1/2$ operators expressed in terms of Pauli matrices has previously been derived, from which TWA is recovered by qualitatively motivated discarding of specific terms~\cite{Mink2022}; approximate collective rules better suited to superradiant emission were introduced subsequently~\cite{MinkFleischhauer2023}; related phase-space treatments of local and collective dissipators have been formulated~\cite{Merkel2021}; and the correspondence with the Keldysh path-integral formulation has recently been explored~\cite{Hosseinabadi2025,Noel2026,Hartmann2026}.
A complementary variational route has also been developed, in which a mixture of spin-coherent states with signed weights is propagated in the Husimi-$Q$ representation under the Dirac--Frenkel principle~\cite{Tosca2026}.

These continuous constructions are, however, formulated for spin-$1/2$ throughout~\cite{Merkel2021,Mink2022,MinkFleischhauer2023,Noel2026,Hartmann2026}.
Moreover, while they yield exact dynamics for single spins, the truncations required to obtain a Fokker--Planck equation incorporating lattice interactions are based on ad-hoc ``truncated correspondence rules'' and not on a systematic expansion of the non-commutative operator product in a small parameter.
Larger spin quantum numbers have so far been reached only through bosonic embeddings supplemented by approximate sampling distributions, having been demonstrated only for collective or one-dimensional geometries~\cite{Huber2020,Huber2021,Hosseinabadi2025}.
Consequently, a continuous phase-space formulation that is simultaneously systematically derivable from an exact equation of motion with controlled truncation in a small quantization parameter, valid across a broad range of relaxation timescales, and applicable to experimentally relevant two-dimensional lattices of arbitrary spin $S$ with short- and long-range interactions together with non-unitary dissipation, remains lacking.
Constructing such a framework is the purpose of this work.

Here, we derive a continuous, $\mathrm{SU}(2)$-covariant truncated-Wigner approximation for open spin lattices directly from the exact Stratonovich--Weyl evolution of the Lindblad master equation.
A systematic expansion in small $\epsilon=(2S+1)^{-1}$ yields Fokker--Planck and associated stochastic equations whose drift and reservoir noise follow from the microscopic Hamiltonian and dissipators.
With one Bloch sphere per site, the method applies to arbitrary spin $S$ and captures transient and steady-state dynamics in two-dimensional systems with short- or long-range interactions.
It reproduces exact benchmarks and, at substantially lower computational cost, the steady-state magnetization in the dissipative anisotropic XY model on a square lattice obtained with an iPEPO tensor network algorithm.
Unlike in the translationally invariant iPEPO approach used for comparison, individual TWA trajectories directly expose vortex--antivortex proliferation, confirming the Keldysh field-theory prediction for the suppression of the mean-field staggered-XY phase~\cite{PhysRevB.93.014307}.
In contrast, for algebraically decaying interactions, the method reveals a novel extended $\mathbbm{Z}_2$-ordered phase, its reentrant loss at strong coupling, and its closure near $\alpha\approx3.5$.

The remainder of the paper is organized as follows.
Section~\ref{sec:method} develops the continuous spin phase-space construction and derives the corresponding Fokker--Planck and stochastic equations.
Section~\ref{sec:results} first benchmarks the method against exact solutions for a single spin governed by the Lipkin--Meshkov--Glick model and then applies it to the two-dimensional dissipative anisotropic XY model.
For nearest-neighbour interactions, we compare with mean-field and iPEPO results and analyse the proliferation of vortex--antivortex pairs.
We then map the steady-state phase diagram for algebraically decaying interactions and identify the extent and reentrant boundary of the $\mathbbm{Z}_2$-ordered phase.
Conclusions and perspectives are presented in Sec.~\ref{sec:discussion}.

\section{METHOD}
\label{sec:method}

\subsection{Phase-space approach to open many-spin systems -- overview}

We are tasked with simulating the dynamics of an open Markovian quantum many-body spin system, described by the reduced density matrix $\hat{\rho}$ evolving under a Lindblad quantum master equation (QME)~\cite{Lindblad1976,GKS1976},
\begin{equation}
    \frac{\dd}{\dd t}\hat{\rho} = -\frac{i}{\hbar} [\hat{H}, \hat{\rho}] + \sum_\alpha \pazocal{D}_\alpha [\hat{\rho}].
    \label{eq:Lindblad}
\end{equation}
In the above, $\hat{H}$ is the Hamiltonian of the reduced system describing its coherent dynamics, while the dissipators $\pazocal{D}_\alpha$ encode the environment-induced non-unitary dynamics of the system, and are given in terms of Lindblad jump operators $\hat{L}_\alpha$ as
\begin{equation}
    \pazocal{D}_\alpha [\hat{\rho}] = \hat{L}_\alpha \hat{\rho} \hat{L}_\alpha^\dagger - \frac{1}{2}\{\hat{L}_\alpha^\dagger \hat{L}_\alpha, \hat{\rho}\}.
\end{equation}
We will treat this problem by mapping the QME onto a Fokker--Planck equation for the Wigner distribution of $\hat{\rho}$ on the $N$-body spin phase space $(\mathbbm{S}^{2})^{\times N}$, identifying a corresponding It\^{o} stochastic differential equation (SDE), and then simulating the resulting stochastic trajectories.

Before developing our new phase-space approach for dissipative many-spin systems, we find it instructive to begin by briefly recapitulating the basic notions of the phase-space formulation of quantum mechanics.

\subsection{Stratonovich–Weyl correspondence}

A phase-space formulation of quantum mechanics involves the mapping of quantum-mechanical operators, including the state $\hat{\rho}$ of the system, to phase-space quasi-distributions (the prefix quasi- accounts for the fact that the distribution may be non-positive).
Given a linear operator $\hat{A}$ acting on a Hilbert space $\pazocal{H}$, one seeks a corresponding quasi-distribution function $W_{\hat{A}}(\Omega)$ on a suitable phase space $\pazocal{M}$, with the argument $\Omega$ denoting coordinates in $\pazocal{M}$.
Formally, the mapping is accomplished via the Stratonovich--Weyl correspondence, a prescription ensuring the resultant functions obey properties expected of phase-space distributions, known as the Stratonovich postulates~\cite{stratonovich1957distributions} (for a detailed exposition, see Appendix~\ref{app:SW}).
Because such phase-space mappings are in essence a basis change to a representation of the underlying canonical group (here, SU(2)) in terms of phase-space functions and because such representations can be constructed in myriad ways, the flavour of mapping is essentially determined by the covariance postulate.
The Stratonovich--Weyl mapping preserves the geometric structure of the canonical group, which we believe to be appropriate given our choice of a continuous manifold as underlying phase space.
A general expression for the Wigner symbol $W_{\hat{A}}$ of the operator $\hat{A}$ satisfying the Stratonovich postulates is
\begin{equation}
    W_{\hat{A}}(\Omega) = \Tr\{ {\hat{A}} \Delta(\Omega) \},
    \label{eq:Wigner_def}
\end{equation}
where $\Delta(\Omega)$ is an appropriately defined Stratonovich--Weyl operator kernel, with the inverse
\begin{equation}
    {\hat{A}} = \int_{\pazocal{M}} \dd\Omega \, W_{\hat{A}}(\Omega) \Delta(\Omega).
    \label{eq:Wigner_inv}
\end{equation}
The Wigner symbol of the density matrix $\hat{\rho}$ is known as the Wigner distribution.
Choosing the same $\Delta(\Omega)$ to appear in both the forward and inverse mappings is what makes the resulting distributions `Wigner': variants with asymmetry in the forwards and backwards operators can be constructed~\cite{VARILLY1989107}, giving the analogues of the bosonic Sudarshan-P and Husimi-Q distributions, but are unsuitable for our present aims and will not be discussed further.

The product of operators becomes the ``star product'' of the respective Wigner symbols, i.e.\ for two operators $\hat{A}$ and $\hat{B}$ acting on $\pazocal{H}$
\begin{equation}
    {\hat{A}}{\hat{B}} \mapsto W_{{\hat{A}}{\hat{B}}}(\Omega) = W_{\hat{A}}(\Omega) \star W_{\hat{B}}(\Omega).
\end{equation}
The star product serves to encode the non-commutative nature of operator multiplication.
Its formal definition can be found in, e.g., Ref.~\cite{Klimov_2002}; here, we will only state it for the specific settings considered.
We will also need the expectation value of an operator $\hat{A}$ in the state ${\hat{\rho}}$ at time $t$, which in the Schr\"odinger picture is calculated as
\begin{align}
    \langle {\hat{A}}\rangle_{\hat{\rho}}(t) = \Tr\{ {\hat{A}} {\hat{\rho}} (t) \}
    = \int_{\pazocal{M}} \dd\Omega W_{\hat{A}}(\Omega) W_{\hat{\rho}}(\Omega; t).
	\label{eq:exp_val}
\end{align}

\subsection{Bosons in phase space}

We first illustrate the general Stratonovich--Weyl construction on the familiar case of $N$ canonical (bosonic) degrees of freedom to make explicit the relationship between the widely used Wigner--Weyl calculus and the kernel language employed throughout this work.
For a single bosonic mode with Hilbert space $\pazocal{H} = L^2(\mathbbm{R})$, the phase space is the flat plane $\pazocal{M} = \mathbbm{R}^2$, parametrized by the position $x$ and momentum $p$.
We label the coordinates by $\Omega \equiv (x,p)$ and use the conventionally normalized flat phase-space measure $\dd\Omega = \dd x \dd p / (2 \pi \hbar)$.
The relevant canonical group is the three-dimensional Heisenberg group $H_3(\mathbbm{R})$, unitarily represented by
$
    \Gamma(\alpha, \gamma) = \exp\!\left(\alpha\hat{a}^\dagger - \alpha^*\hat{a} - i \gamma\right) .
$
Since the contribution of $\gamma$ is a trivial phase factor (and thus does not contribute to conjugation in the covariance postulate), we may focus on the components of these operators known as the displacement operators
    $\hat{D}(x,p) = \exp\!\left[\tfrac{i}{\hbar}\left(p\hat{x} - x\hat{p}\right)\right]
    = \exp\!\left(\alpha\hat{a}^\dagger - \alpha^*\hat{a}\right)$, 
 which furnish a projective representation of the two-dimensional translation group.

The Stratonovich--Weyl kernel in this case is the displaced-parity operator~\cite{Grossmann1976,Royer1977,klimov2009group}, which up to a choice of constants is
\begin{subequations}
\label{eq:GR_kernel}
\renewcommand{\theequation}{\theparentequation.\arabic{equation}}
\begin{equation}
    \Delta(\Omega) = 2\,\hat{D}(\Omega)\,\hat{\Pi}\,\hat{D}^\dagger(\Omega),
\end{equation}
\begin{equation}
    \hat{\Pi} = (-1)^{\hat{a}^\dagger\hat{a}} = \int \dd y \,\ket{-y}\!\bra{y},
\end{equation}
\end{subequations}
i.e.\ the parity operator about the phase-space point $(x,p)$.
This general form is an immediate consequence of the covariance postulate once the operator's value at $(0,0)$ has been fixed: writing $\Delta_0 \equiv \Delta(0,0) = 2\hat{\Pi}$, by the postulate one has $\Delta(\Omega) = \hat{D}(\Omega)\,\Delta_0\,\hat{D}^\dagger(\Omega)$.
This is the group-covariant realization of the abstract kernel of the previous section.
Upon inserting position eigenstates, the resultant Wigner function $W_{\hat{A}}(x,p)$ defined in Eq.~\eqref{eq:Wigner_def} reduces to the celebrated form~\cite{Wigner1932,Hillery1984}
\begin{equation}
    W_{\hat{A}}(x,p) = \int_{\mathbbm{R}} \dd x'\, e^{ipx'/\hbar}\,
    \bra{x - x'/2}\hat{A}\ket{x + x'/2},
    \label{eq:boson_W_explicit}
\end{equation}
with the inverse (Weyl transform) given by the same kernel,
\begin{equation}
    \hat{A} = \int_{\mathbbm{R}^2} \dd\Omega\, W_{\hat{A}}(x,p)\,\Delta(x,p).
    \label{eq:boson_inverse}
\end{equation}
One verifies directly that $\Delta(x,p)$ satisfies the Stratonovich postulates: in particular the standardization $\int\dd\Omega\,W_{\hat{A}} = \Tr\hat{A}$ and the traciality $\int\dd\Omega\,W_{\hat{A}}W_{\hat{B}} = \Tr(\hat{A}\hat{B})$, so that $\int\dd\Omega\,W_{\hat{\rho}} = 1$ for any normalized state.
The single-mode operators map to the phase-space coordinates themselves, i.e.\ $W_{\hat{x}} = x$, $W_{\hat{p}} = p$, $W_{\hat{a}} = \alpha$, and $W_{\hat{a}^\dagger} = \alpha^*$, while quadratic operators acquire the symmetric-ordering correction characteristic of the Weyl map, e.g.\ $W_{\hat{a}^\dagger\hat{a}} = |\alpha|^2 - \tfrac{1}{2}$.

For $N$ modes the phase space is $\mathbbm{R}^{2N}$ with coordinates $\boldsymbol{\Omega} \equiv (\Omega_1,\dots,\Omega_N)$ and product measure $\dd^N\boldsymbol{\Omega} = \prod_i \dd\Omega_i$.
Since the modes act on distinct Hilbert-space factors, the kernel factorizes as a tensor product,
\begin{equation}
    \Delta^{(N)}(\boldsymbol{\Omega}) = \bigotimes_{i=1}^{N} \Delta_i(\Omega_i),
    \label{eq:Nkernel}
\end{equation}
and the $N$-mode Wigner function of a state $\hat{\rho}$ acting on $\pazocal{H}^{\otimes N}$ is $W_{\hat{\rho}}(\boldsymbol{\Omega}) = \Tr\{\hat{\rho}\,\Delta^{(N)}(\boldsymbol{\Omega})\}$, normalized as $\int\dd^N\boldsymbol{\Omega}\,W_{\hat{\rho}} = 1$.
Finally, the star product takes the Groenewold--Moyal closed form~\cite{Groenewold1946, Moyal1949}
\begin{equation}
    W_{\hat{A}} \star W_{\hat{B}}
    = W_{\hat{A}}\,
    \exp\!\left[\frac{i\hbar}{2}\left(
    \overleftarrow{\partial}_x \overrightarrow{\partial}_p
    - \overleftarrow{\partial}_p \overrightarrow{\partial}_x \right)\right]
    W_{\hat{B}},
    \label{eq:moyal_boson}
\end{equation}
the exponential of a single first-order bidifferential operator built from the symplectic form, whose leading antisymmetric term reproduces the Poisson bracket, $W_{\hat{A}} \star W_{\hat{B}} - W_{\hat{B}} \star W_{\hat{A}} = i\hbar\{W_{\hat{A}}, W_{\hat{B}}\} + O(\hbar^3)$.

\subsection{Spins in phase space}

Having established the construction on flat phase space, we now turn to the case of central interest in this work, ensembles of distinguishable spins.
For a single spin-$S$ particle with Hilbert space $\pazocal{H} = \mathbbm{C}^{2S+1}$, the covariance postulate requires our chosen phase space to be a homogeneous space for the associated canonical group, which is now SU(2).
A natural choice is SU(2)/U(1) = $\mathbbm{S}^2$, which may intuitively be thought of as the space one obtains by quotienting out the unphysical global phase freedom of the quantum theory from the largest possible phase space.
Parametrizing this space by the azimuthal angle $\phi \in [0,2\pi)$ and polar angle $\theta \in [0,\pi]$, we label the coordinates by $\Omega \equiv (\theta,\phi)$ and use the SU(2)-invariant measure normalized to the state count, $\dd\Omega = \frac{2S+1}{4\pi}\,\sin\theta\,\dd\theta\,\dd\phi$.
This is the natural analogue of the flat measure $\dd x \dd p/(2\pi\hbar)$ used above.

One possible self-dual kernel $\Delta^{(S)}(\Omega)$ satisfying the Stratonovich--Weyl axioms~\cite{VARILLY1989107} reads in the $\hat{S}^z$ eigenstate basis $\ket{S,m}$ as
\begin{equation}
    \Delta^{(S)}(\Omega) = \sum_{r, s=-S}^{S} \tensor[]{Z}{^{(S)}^r_{s}}(\Omega) \ket{S, r}\bra{S, s},
    \label{eq:spin_kernel}
\end{equation}
where the matrix elements are
\begin{equation}
    \tensor[]{Z}{^{(S)}^r_{s}} (\Omega) = \sum_{l=0}^{2S} \sum_{m=-l}^{l} \tfrac{2l+1}{2S+1}
    \tensor[]{C}{^{(SlS)}^r^m_s} \, \tensor[]{Y}{^{(l)}_m}(\Omega),
    \label{eq:spin_kernel_matrix}
\end{equation}
with $\tensor[]{C}{^{(jkl)}^m^n_p} = \langle j, m; k, n | l, p \rangle$ denoting the Clebsch--Gordan coefficients and $\tensor[]{Y}{^{(l)}_m}(\Omega)$ the Racah-normalized spherical harmonics.
The forward and inverse maps $W_{\hat{A}}(\Omega) = \Tr\{\hat{A}\,\Delta^{(S)}(\Omega)\}$ and $\hat{A} = \int\dd\Omega\,W_{\hat{A}}\,\Delta^{(S)}$ take the same form as Eqs.~\eqref{eq:Wigner_def} and \eqref{eq:Wigner_inv}, and the single-spin operators map to the polar representation of the Cartesian coordinates on a sphere of radius $\sqrt{S(S+1)}$:
    $W_{\hat{S}^x} = \sqrt{S(S+1)}\sin\theta\cos\phi,
    W_{\hat{S}^y} = \sqrt{S(S+1)}\sin\theta\sin\phi,
    W_{\hat{S}^z} = \sqrt{S(S+1)}\cos\theta$.

Finally, we state the formal expression for the star product of SU(2) spin operators due to Klimov and Espinoza~\cite{Klimov_2002},
\begin{equation}
\label{eq:moyal_spin}
\begin{split}
    W_A \star W_B = &N_S \sum_{j=0}^\infty a_j \int \frac{\dd\psi}{2\pi}\,
    \tilde{F}^{-1}(\pazocal{J}^2)
    \\ &\times [(S^+)^j \tilde{F}(\pazocal{J}^2)W_A][(S^-)^j \tilde{F}(\pazocal{J}^2)W_B],
\end{split}
\end{equation}
where $N_S = \sqrt{2S+1}$, $a_j = \frac{(-1)^j}{j!(2S+j+1)!}$, $\tilde{F}$ is a complicated function the exact form of which need not be written down, and $\pazocal{J}^2$, $S^\pm$ furnish a differential representation of the SU(2) Lie algebra on the space spanned by columns of Wigner D-matrices:
$
    \pazocal{J}^2 \,D^l_{m m'}(\phi, \theta,\psi) = l(l+1) \times D^l_{m m'}(\phi, \theta,\psi),
$ and
$
    S^\pm D^l_{m m'}(\phi, \theta,\psi) = \pm\sqrt{l(l+1) - m'(m'\pm1)} \times D^l_{m {m'\pm1}}(\phi, \theta,\psi).
$
Notably, unlike the Groenewold--Moyal star product for flat phase space in Eq.~\eqref{eq:moyal_boson}, the SU(2) star product does not admit a simple form purely as a function of its Poisson bracket.
We summarize and compare the phase-space constructions for bosons and spins in Table~\ref{tab:phase_space_comparison}.

\begin{table}
\raggedright
\renewcommand{\arraystretch}{1.3}
\begin{tabular}{ l c c }
\toprule
 & Bosons & Spins \\
\midrule
Hilbert space & $L^2(\mathbbm{R})$, $\dim=\infty$ & $\mathbbm{C}^{2S+1}$, $\dim=2S{+}1$ \\
Phase space & $\mathbbm{R}^{2}:(x,p)$ & $\mathbbm{S}^{2}:(\theta,\phi)$ \\
Measure $\dd\Omega$ & $\dfrac{\dd x\,\dd p}{2\pi\hbar}$ & $\dfrac{2S+1}{4\pi}\sin\theta\,\dd\theta\,\dd\phi$ \\
Canonical grp. & $H_3(\mathbbm{R})$ & SU(2) \\
Lie Algebra & $[\hat{x},\hat{p}]=i\hbar$ & $[\hat{S}_i,\hat{S}_j]=i\epsilon_{ijk}\hat{S}_k$ \\
Covariance op. & Displacement  $\hat{D}(x,p)$ & Rotation $\hat{R}(\theta,\phi, \psi)$ \\
Kernel $\Delta$ &
  \begin{tabular}{@{}c@{}}Eq.~\eqref{eq:GR_kernel}\\ (Grossmann--Royer)\end{tabular} &
  \begin{tabular}{@{}c@{}}Eq.~\eqref{eq:spin_kernel}\\ (V\'arilly--Gracia-Bond\'ia)\end{tabular} \\
Star product &
  \begin{tabular}{@{}c@{}}Eq.~\eqref{eq:moyal_boson}\\ (Groenewold--Moyal)\end{tabular} &
  \begin{tabular}{@{}c@{}}Eq.~\eqref{eq:moyal_spin}\\ (Klimov--Espinoza)\end{tabular} \\
Coord. symbols & $W_{\hat{x}}=x,\ W_{\hat{p}}=p$ &
  $W_{\hat{\mathbf{S}}}=\sqrt{S(S{+}1)}\,\mathbf{\hat{n}}$ \\
Semicl. param. & $\hbar$ & $1/(2S+1)$ \\
\bottomrule
\end{tabular}
\caption{
    Comparison of the phase-space formulation for single-particle bosons and spins.
    The $N$-body case is built factor-wise, with Hilbert-space dimension scaling as $\infty$ (bosons) and $(2S+1)^N$ (spins).}
\label{tab:phase_space_comparison}
\end{table}

\subsection{Spin star product expansion}

While the star product in Eq.~\eqref{eq:moyal_spin} is an exact expression, it is not practically usable and requires approximation.
In Ref.~\cite{Klimov_2002}, an expansion to first order in the semiclassical parameter $\epsilon = (2S+1)^{-1} \sim \hbar$ was found.
This may be used to obtain valid mean-field equations of motion for unitary systems, but faces obstacles in the driven-dissipative setting: when mapping to phase-space symbols, applying the first-order product to the dissipator component of the Lindblad master equation generates $O(\epsilon^2)$ terms (see Appendix~\ref{app:LMG} for an example derivation).
For the truncation to be consistent, to produce a valid Fokker--Planck equation, and thus to be amenable to stochastic simulation,  $O(\epsilon^2)$ terms must then be retained in the product.
In this work, we derive this second-order expansion, which we find yields a valid Fokker--Planck equation, providing an efficient semiclassical description of the quantum many-body problem.

The resultant expression for the expanded star product can be compactly stated as
\begin{equation}
    \begin{split}
    W_{\hat{A}} \star W_{\hat{B}} = {}& W_{\hat{A}} W_{\hat{B}} + i\epsilon\{W_{\hat{A}}, W_{\hat{B}}\} \\
    &+ \epsilon^2\, \pazocal{B}(W_{\hat{A}}, W_{\hat{B}}) + O(\epsilon^3),
    \end{split}
    \label{eq:starprod}
\end{equation}
where, in terms of the canonical coordinates $(\theta, \phi)$, $\{f, g\}$ is the Poisson bracket on the sphere,
\begin{equation}
\{f, g\} = \partial_\phi f \,\partial_{\cos\theta} g - \partial_{\cos\theta} f \partial_\phi g, 
\end{equation} 
and $\pazocal{B}$ is a symmetric bilinear differential operator, of second order in each of its arguments.
Denoting by $\partial^a$ the derivative monomials of first and second order, $\partial^a \in \{\partial_\theta, \partial_\phi, \partial_\theta^2, \partial_\theta\partial_\phi, \partial_\phi^2\}$, it can be written as
\begin{equation}
    \pazocal{B}(W_{\hat{A}}, W_{\hat{B}}) = \sum_{a, b} \pazocal{B}_{ab}(\theta)\, \partial^a W_{\hat{A}}\, \partial^b W_{\hat{B}},
    \label{eq:B_table_def}
\end{equation}
with the coefficients $\pazocal{B}_{ab}(\theta)$ collected in Table~\ref{tab:bilinear} and derived in Appendix~\ref{app:starproduct}.

For composite systems (e.g.\ spins on a lattice), the overall star product is the tensor product of the single-site star products, leading to cross terms between different sites at second order in $\epsilon$ that encode inter-site correlations.
We state the result at the level of the operator rather than of its action: $\leftarrow$ and $\rightarrow$ mark the slots occupied by the left and right symbol, so that each derivative is labelled by the symbol it differentiates and $W_{\hat{A}} \star W_{\hat{B}}$ is recovered by inserting $W_{\hat{A}}$ and $W_{\hat{B}}$ into the respective slots.
Operators carrying different site labels act on disjoint coordinates and hence commute, so their juxtaposition denotes composition without ambiguity.
With $\{\leftarrow,\rightarrow\}_j$ and $\pazocal{B}_j(\leftarrow,\rightarrow)$ applying to the coordinates $(\theta_j, \phi_j)$, and $J$ the set of one-body subsystems,
\begin{equation}
\label{eq:star-expansion}
\begin{split}
    \star = \bigotimes_{j\in J}\star_j
    &= 1 + i\epsilon \sum_{j\in J}\{\leftarrow,\rightarrow\}_j \\
    &\quad - \epsilon^2
    \sum_{\substack{j,k\in J\\ j < k}}
    \{\leftarrow,\rightarrow\}_j\,\{\leftarrow,\rightarrow\}_k \\
    &\quad + \epsilon^2 \sum_{j\in J}\pazocal{B}_j(\leftarrow,\rightarrow)
    + O(\epsilon^3).
\end{split}
\end{equation}
Note that, at this order, all terms with an even power of $\epsilon$ are symmetric, while those with an odd power are antisymmetric.
While we do not make use of it in the simulations of this paper, in Appendix~\ref{tensorstarproduct} we also show how this tensorial form may easily be extended to incorporate bosonic modes into a description of a system with both types of degree of freedom.

\begin{table}[t]
\centering
\begin{tabular}{ |c|c|c|c|c|c| }
\hline
\scalebox{0.9}{$\pazocal{B}_{ab}$} & \scalebox{0.9}{$\partial_\theta$} & \scalebox{0.9}{$\partial_\phi$} & \scalebox{0.9}{$\partial_\theta^2$} & \scalebox{0.9}{$\partial_\theta\partial_\phi$} & \scalebox{0.9}{$\partial_\phi^2$} \\
\hline
\scalebox{0.9}{$\partial_\theta$} & \scalebox{0.9}{$\frac{1}{2}$} & \scalebox{0.9}{$0$} & \scalebox{0.9}{$-\frac12 \cot\theta$} & \scalebox{0.9}{$0$} & \scalebox{0.9}{0} \\
\scalebox{0.9}{$\partial_\phi$} & \scalebox{0.9}{$\ast$} & \scalebox{0.9}{$\frac12 (1 + \cos^2\theta)\csc^4\theta$} & \scalebox{0.9}{$0$} & \scalebox{0.9}{$-\cot\theta \csc^2\theta$} & \scalebox{0.9}{$0$} \\
\scalebox{0.9}{$\partial_\theta^2$} & \scalebox{0.9}{$\ast$} & \scalebox{0.9}{$\ast$} & \scalebox{0.9}{$0$} & \scalebox{0.9}{$0$} & \scalebox{0.9}{$-\frac{1}{2}\csc^2\theta$} \\
\scalebox{0.9}{$\partial_\theta\partial_\phi$} & \scalebox{0.9}{$\ast$} & \scalebox{0.9}{$\ast$} & \scalebox{0.9}{$\ast$} & \scalebox{0.9}{$\csc^2\theta$} & \scalebox{0.9}{$0$} \\
\scalebox{0.9}{$\partial_\phi^2$} & \scalebox{0.9}{$\ast$} & \scalebox{0.9}{$\ast$} & \scalebox{0.9}{$\ast$} & \scalebox{0.9}{$\ast$} & \scalebox{0.9}{$0$} \\
\hline
\end{tabular}
\caption{
    Coefficients $\pazocal{B}_{ab}(\theta)$ of the second-order bilinear operator $\pazocal{B}$ of Eqs.~\eqref{eq:starprod} and \eqref{eq:B_table_def}, in the coordinates $\theta$ and $\phi$.
    Rows label the derivative monomial $\partial^a$ acting on the left symbol $W_{\hat{A}}$, columns the derivative monomial $\partial^b$ acting on the right symbol $W_{\hat{B}}$, ordered by differential order (first, second).
    The coefficients depend on $\theta$ only, and the table is symmetric under $a \leftrightarrow b$.
    Zeroth-order entries are absent, i.e.\ $\pazocal{B}(1, W_{\hat{A}}) = \pazocal{B}(W_{\hat{A}}, 1) = 0$.
    See Appendix~\ref{app:starproduct} for the derivation.}
\label{tab:bilinear}
\end{table}

\subsection{Truncated Wigner Approximation for open many-spin dynamics}

By replacing the operators with their respective Wigner functions, the Lindblad QME \eqref{eq:Lindblad} can be expressed in phase space as
\begin{equation}
\label{eq:wigner-lindblad}
\begin{split}
    \frac{\dd}{\dd t}W_{\hat{\rho}}
    &= -\frac{i}{\hbar}[W_{\hat{H}}, W_{\hat{\rho}}]_{\star} 
    + \sum_\alpha \Big(
    W_{\hat{L}_\alpha} \star W_{\hat{\rho}} \star W_{\hat{L}_\alpha}^* \\&\qquad
    - \frac{1}{2}
    \{W_{\hat{L}_\alpha}^* \star W_{\hat{L}_\alpha}, W_{\hat{\rho}}\}_{\star}
    \Big),
\end{split}
\end{equation}
where we denoted $W_{\hat{A}}^* = W_{\hat{A}^\dagger}$ and the star-commutator and star-anticommutator as $[W_{\hat{A}}, W_{\hat{B}}]_{\star} = W_{\hat{A}} \star W_{\hat{B}} - W_{\hat{B}} \star W_{\hat{A}}$ and $\{W_{\hat{A}}, W_{\hat{B}}\}_{\star} = W_{\hat{A}} \star W_{\hat{B}} + W_{\hat{B}} \star W_{\hat{A}}$, respectively.
Making use of our expansion in Eq.~\eqref{eq:star-expansion}, each star product can then be expanded in powers of $\epsilon$ up to $\epsilon^2$.
In practice this is most conveniently done in the Bopp representation~\cite{Bopp1956}, in which star multiplication by the symbol of a spin operator, from the left or from the right, is realized as a single differential operator acting on $W_{\hat{\rho}}$; the corresponding operators for $\hat{S}^\pm$ and $\hat{S}^z$ are tabulated to $O(\epsilon^2)$ in Table~\ref{tab:bopp} of Appendix~\ref{app:starproduct}.
To this order, left and right star multiplication differ only by $\epsilon \to -\epsilon$, so that only odd powers of $\epsilon$ survive in the star-commutator and only even powers in the star-anticommutator; the coherent part of Eq.~\eqref{eq:wigner-lindblad} therefore contributes to the drift alone, and diffusion arises at order $\epsilon^2$ from the dissipators.
To map the resultant expression onto an easily soluble SDE, we must also discard all terms of $O(\epsilon^3)$, as in the usual TWA~\cite{Carusotto2013}.
Upon performing this truncation, one obtains a Fokker--Planck equation for the measure-weighted Wigner function $\overline{W}_{\hat{\rho}} = W_{\hat{\rho}} \sin\theta$,
\begin{equation}
    \frac{\dd }{\dd t} \overline{W}_{\hat{\rho}} = -\partial_i\!\left[\mu_i\, \overline{W}_{\hat{\rho}}\right] + \tfrac{1}{2}\partial_i\partial_j\!\left[D_{ij}\, \overline{W}_{\hat{\rho}}\right],
    \label{eq:FP}
\end{equation}
where the drift vector $\mu$ and diffusion matrix $D$ are real functions of $(\bm{\theta},\bm{\phi})$ determined by the Hamiltonian and dissipators of the master equation considered; explicit expressions for $\mu$ and $D$ will be derived for the models studied in this work in Sec.~\ref{sec:results}.

We collect the phase-space coordinates into $\bm{\Omega} = (\theta_1,\phi_1,\dots,\theta_N,\phi_N)^\top$.
When the diffusion matrix is positive semi-definite, it factorizes as $D = B B^\top$.
Then, Eq.~\eqref{eq:FP} is equivalent to the $2N$-dimensional It\^{o} SDE~\cite{Gardiner2009}
\begin{equation}
    \dd \Omega_i = \mu_i(\bm{\Omega})\,\dd t + \sum_k B_{ik}(\bm{\Omega})\,\dd\xi_k,
    \label{eq:Ito}
\end{equation}
with $\bm{\xi}$ a vector of independent Wiener processes, $\langle \xi_i\,\xi_j\rangle = \delta_{ij}\, t$, in the sense that the density of the process \eqref{eq:Ito} evolves precisely as the density of trajectories obeying \eqref{eq:FP}.
Expectation values \eqref{eq:exp_val} then become ensemble averages over trajectories $\{\bm{\Omega}^{(n)}(t)\}_{n=1}^{N_{\text{traj}}}$ sampled from the initial distribution $W_{\hat{\rho}}(\bm{\Omega}; 0)$ and propagated independently under \eqref{eq:Ito},
$\langle \hat{A}\rangle(t) \approx N_{\mathrm{traj}}^{-1}\sum_n
W_{\hat{A}}(\bm{\Omega}^{(n)}(t))$, converging to the exact result as
$N_{\mathrm{traj}}\to\infty$ (up to the $\epsilon^2$ truncation of the star product).
Note that the $\sin\theta$ component of the measure required for the integral
\begin{equation}
    \langle \hat{A}\rangle(t) = \int W_{\hat{A}}(\bm{\Omega}) W_{\hat{\rho}}(\bm{\Omega};t)\sin\theta\, \dd\theta\, \dd\phi
\end{equation}
has here already been absorbed into the measure-weighted Wigner function, and thus does not appear explicitly in the trajectory average.
When the initial Wigner function $W_{\hat{\rho}}(\bm{\Omega}; 0)$ is not everywhere non-negative (which is always the case for pure spin states), one samples from its absolute value $|W_{\hat{\rho}}|$.
One may then either weight each trajectory by the sign of $W_{\hat{\rho}}$ at the initial point, or simply allow the trajectories to reach a driven-dissipative steady state and normalize the resulting average of unweighted trajectories by the total mass $\int |\overline{W}_{\hat{\rho}}|\, \dd^N\bm{\theta}\, \dd^N\bm{\phi}$.
The former approach becomes computationally infeasible for large numbers of spins as the probability of any given trajectory being positively weighted approaches $\frac12$ for many simple initial states as the number of spins increases.
The latter works by relying on the positive and negative components of the distribution to independently attain the same steady state with opposite signs, so that $|\overline{W}_{\hat{\rho}}|$ differs from the true distribution by only a factor.
It is available so long as the system possesses a steady state, which many driven-dissipative systems do.
We summarize and illustrate the method in Fig.~\ref{fig:method_flowchart}.

\section{NUMERICAL RESULTS}
\label{sec:results}

In this section, we demonstrate the capabilities of our TWA algorithm for interacting spin systems, starting with the dissipative Lipkin--Meshkov--Glick model before moving on to the many-body two-dimensional dissipative anisotropic XY model with short- and long-range interactions, where we report the onset of a novel symmetry-broken phase in the non-equilibrium steady state.
We set $\hbar=1$ throughout.

\subsection{Single spin: Lipkin--Meshkov--Glick model} \label{sec:LMGNumerical}

As a first test of the method, we consider a dissipative Lipkin--Meshkov--Glick (LMG) model~\cite{LMG1965}, describing a single spin $S$ evolving under the Hamiltonian
\begin{equation}
    \hat{H} = -J (\hat{S}^x)^2 + \Delta \hat{S}^z,
\end{equation}
and subject to the Lindblad jump operator $\hat{L}  = \sqrt{\Gamma} \hat{S}^-$.
Defining the dimensionless quantities $\tau = t\Gamma$, $\tilde{J} = \frac{J}{\Gamma}$, and $\tilde{\Delta} = \frac{\Delta}{\Gamma}$, the Lindblad QME takes the form
\begin{equation}
    \begin{split}
    \frac{\dd}{\dd \tau}\hat{\rho} = {}& -i[-\tilde{J}(\hat{S}^x)^2 + \tilde{\Delta}\hat{S}^z, \hat{\rho}] \\
    &\quad + \hat{S}^- \hat{\rho} \hat{S}^+ - \frac{1}{2}\{\hat{S}^+ \hat{S}^-, \hat{\rho}\}.
    \end{split}
\end{equation}
In phase space this reads
\begin{equation}
\begin{split}
    \frac{\dd}{\dd \tau}W_{\hat{\rho}} &= -i[-\tilde{J}W_{\hat{S}^x} \star W_{\hat{S}^x} + \tilde{\Delta}W_{\hat{S}^z}, W_{\hat{\rho}}]_{\star} \\
    &\quad + W_{\hat{S}^-} \star W_{\hat{\rho}} \star W_{\hat{S}^+} \\
    &\quad - \frac{1}{2}\{W_{\hat{S}^+} \star W_{\hat{S}^-}, W_{\hat{\rho}}\}_{\star}.
\end{split} \label{eq:LMG_QME_FULL}
\end{equation}
Expanding the star products to second order in $\epsilon$ and truncating at this order gives a Fokker--Planck equation for the measure-weighted Wigner function
\begin{equation}
    \frac{\dd}{\dd \tau}\overline{W}_{\hat{\rho}} = -\partial_i\!\left[\mu_i\, \overline{W}_{\hat{\rho}}\right] + \tfrac{1}{2}\partial_i\partial_j\!\left[D_{ij}\, \overline{W}_{\hat{\rho}}\right],
\end{equation}
with drift vector and diffusion matrix (identifying the first coordinate as $\theta$ and the second as $\phi$)
\begin{subequations}
\renewcommand{\theequation}{\theparentequation.\arabic{equation}}
\label{eq:LMG_FP_coeffs}
\begin{align}
    \mu &= \lambda\begin{pmatrix}
        \frac12 \lambda \cot(\theta) + \pazocal{S}\sin(\theta)\left[1 + \tilde{J}\sin(2\phi)\right]\\[2pt]
        \tilde{\Delta} + 2\tilde{J}\pazocal{S}\,\cos(\theta)\cos^2(\phi)
    \end{pmatrix},
    \label{eq:LMG_drift} \\[4pt]
    D &= \lambda^2 \begin{pmatrix}
        1 & 0\\[2pt]
        0 & \cot^2(\theta)
    \end{pmatrix},
    \label{eq:LMG_diff}
\end{align}
\end{subequations}
where $\lambda = 2\pazocal{S}\epsilon$ and $\pazocal{S}=\sqrt{S(S+1)}$.
For an explicit derivation of the above, see Appendix~\ref{app:LMG}.
Since $D$ is diagonal, the corresponding It\^{o} SDE follows immediately,
\begin{equation}
	\begin{pmatrix}
		\dd \theta \\
		\dd\phi
	\end{pmatrix}
	=
	\mu \, \dd\tau
	+ \lambda
	\begin{pmatrix}
		1 & 0 \\[2pt]
		0 & \cot(\theta)
	\end{pmatrix}
	\begin{pmatrix}
		\dd\xi_1\\
		\dd\xi_2
	\end{pmatrix}.
    \label{eq:LMG_SDE}
\end{equation}
The $\theta$-component of Eq.~\eqref{eq:LMG_SDE} is regular on $[0,\pi]$, except for the term
$
    \frac12 \lambda^2 \cot(\theta),
$
a strongly repulsive term driving the stochastic process away from the poles $\theta \in \{0, \pi\}$.
Rescaling $\lambda^2 \dd\tau = \dd\tau'$, near the points we see that
$
    \dd\theta \sim \frac12 \cot(\theta) \dd\tau',
$
so that there the process is asymptotically Bessel-$\frac12$~\cite{lawler2018notes}.
It will thus almost surely not reach the poles in finite time if started away from them.
The azimuthal diffusion $D_{\phi\phi}$ is also irregular at the poles, but this is the usual coordinate degeneracy of $\phi$ at these points and is immaterial given that they are never reached.
In our implementation $\theta \in \{0,\pi\}$ are treated explicitly as reflecting boundaries to absorb the finite-step-size overshoot of the Stability-Optimized Stochastic Runge--Kutta method~\cite{rossler2010runge}.

\begin{figure}
    \includegraphics[width=\columnwidth]{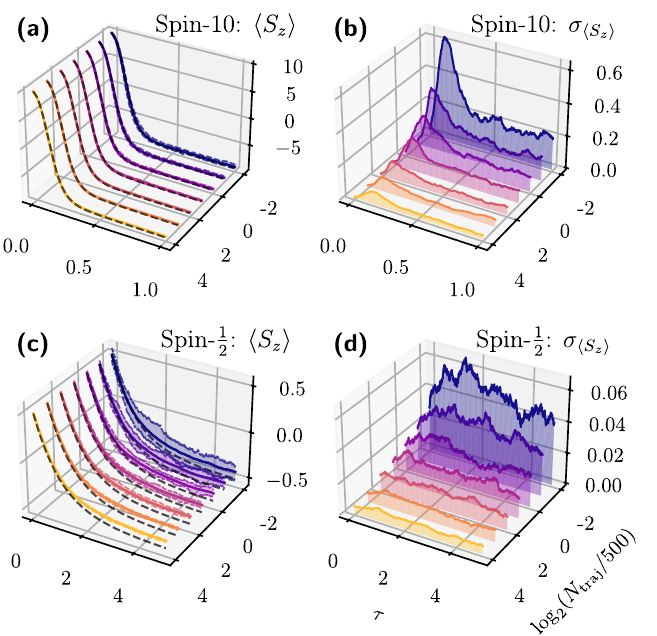}
    \caption{
        (a) Relaxation dynamics of $\langle \hat{S}^z \rangle$ for the dissipative LMG model for $S=10$ ($\tilde{J} = 1$, $\tilde{\Delta} = 1$) obtained by simulating Eq.~\eqref{eq:LMG_SDE}, averaging over varying numbers of trajectories, from $N_{\text{traj}} = 125$ to $N_{\text{traj}} = 8000$, compared to exact numerical results from integrating the Lindblad QME (grey dashed lines).
        For each $N_{\text{traj}}$, 30 batches of $N_{\text{traj}}$ trajectories are computed and $\langle \hat{S}^z \rangle$ calculated for each.
        The shaded envelope indicates the maximum and minimum values attained by $\langle \hat{S}^z \rangle$ across these batches.
        (b) Corresponding standard deviation of $\langle \hat{S}^z \rangle$ across these batches over time.
        (c--d) As in (a--b), but for $S=\frac{1}{2}$.}
    \label{fig:LMG}
\end{figure}

In Fig.~\ref{fig:LMG} we study the non-equilibrium relaxation dynamics of the $\langle \hat{S}^z \rangle$ magnetization, starting from the pure state $\ket{S,S}\bra{S,S}$ for $S=10, \frac{1}{2}$.
Its measure-weighted Wigner function follows from the diagonal matrix elements of the spin kernel, Eq.~\eqref{eq:spin_kernel_matrix} with $r = s = S$; since $\tensor[]{Y}{^{(l)}_0}(\theta,\phi) = P_l(\cos(\theta))$, this is a polynomial in $\cos(\theta)$ alone:
\begin{equation}
    \overline{W}_{\hat{\rho}}(\theta, \phi; 0) = \sin(\theta)\sum_{l=0}^{2S} \tfrac{2l+1}{2S+1} \tensor[]{C}{^{(SlS)}^S^0_S}\, P_l(\cos(\theta)),
    \label{eq:LMG_initial}
\end{equation}
with $P_l$ the Legendre polynomials.
Being independent of $\phi$, Eq.~\eqref{eq:LMG_initial} is sampled by drawing $\phi$ uniformly on $[0,2\pi)$ and $\theta$ from the one-dimensional density $\propto |\overline{W}_{\hat{\rho}}(\theta;0)|$ on $(0,\pi)$, with each trajectory weighted by $\mathrm{sgn}\,\overline{W}_{\hat{\rho}}$ at its initial point.
While for $S=10$ the distribution is positive on most of $[0,\pi]$, for $S=\frac12$ Eq.~\eqref{eq:LMG_initial} reduces to $\tfrac12\sin(\theta)(1 + \sqrt{3}\,\cos(\theta))$, which is negative for $\theta > \pi - \tan^{-1}(\sqrt{2})$.
We then evolve the phase-space coordinates under the SDE in Eq.~\eqref{eq:LMG_SDE} and compute $\langle \hat{S}^z \rangle$ by Monte Carlo averaging over the resultant trajectories.
We find good agreement in both cases with exact numerical results obtained from integrating the Lindblad QME, with a systematic reduction of the statistical noise as the number of trajectories increases.
We also observe that the accuracy of the method improves for larger spins; the relative errors in the steady-state value of $\langle \hat{S}^z \rangle$ for $N_{\text{traj}} = 8000$ are of the order of 2\% ($S=10$) and 15\% ($S=\frac{1}{2}$), respectively.
This is expected given the nature of our small expansion parameter $\epsilon = 1/(2S + 1)$.

\subsection{Lattice: dissipative anisotropic XY model}

As the main focus of this work, we apply our TWA method to a dissipative anisotropic XY model~\cite{PhysRevLett.110.257204} with both short- and long-range interactions on a square lattice of $N=n^2$ spins with $S=\frac{1}{2}$ and cyclic boundary conditions, described by the Hamiltonian
\begin{equation}
    \hat{H} = \frac{ J}{\pazocal{N}[V]}\sum_{j< k} V(j,k) \left(\hat{S}_j^+ \hat{S}_k^+ + \hat{S}_j^- \hat{S}_k^-\right),
    \label{eq:H_XY}
\end{equation}
with the local Lindblad jump operators modelling spontaneous independent decay of the spins with rate $\Gamma$, $\hat{L}_j = \sqrt{\Gamma} \hat{S}_j^-$, for each site $j$ on the lattice, labelled in row-major order.
Here, $J$ is the interaction strength, $V(i,j)$ is a function encoding the form of the interactions (always with no self-interaction), and $\pazocal{N}[V] = \frac{1}{2}\sum_{j\neq 1} V(1,j)$ is a Kac normalization factor~\cite{kac1963van}.
The latter permits the mean-field phase transition values of $J$ to be independent of the form of $V(i,j)$ in the thermodynamic limit, and keeps this limit well-defined when we consider algebraically decaying interactions.
Defining the dimensionless quantities $\tau = t\Gamma$ and $\tilde{J} = \frac{J}{\Gamma}$, the Lindblad QME takes the form
\begin{equation}
\begin{split}
    \frac{\dd}{\dd \tau}\hat{\rho} &= -i\frac{\tilde{J}}{\pazocal{N}[V]}\sum_{j<k} V(j,k) \left[ \hat{S}_j^+ \hat{S}_k^+ + \hat{S}_j^- \hat{S}_k^-, \hat{\rho}\right] \\ &\qquad + \sum_j \left(\hat{S}_j^- \hat{\rho} \hat{S}_j^+ - \frac{1}{2}\{\hat{S}_j^+ \hat{S}_j^-, \hat{\rho}\}\right).
\end{split}
\label{eq:XY_QME}
\end{equation}

We assign the canonical phase-space coordinates $(\theta_j, \phi_j)$ to each site $j$, and collect them into $\bm{\Omega} = (\theta_1, \phi_1, \theta_2, \phi_2, \ldots, \theta_N, \phi_N)$.
Since the two spin operators in each interaction term of Eq.~\eqref{eq:H_XY} act on distinct sites, the tensor product kernel \eqref{eq:star-expansion} gives their symbol exactly at order $\epsilon^0$:
\begin{equation}
    W_{\hat{H}} \! = \! \frac{2 \tilde{J}\pazocal{S}^2}{\pazocal{N}[V]}
    \sum_{j < k} V(j,k)
    \sin(\theta_j)\sin(\theta_k)\cos(\phi_j \! + \! \phi_k).
    \label{eq:XY_WH}
\end{equation}
Furthermore, the only antisymmetric components (and thus those that contribute to the star commutator) of that kernel are the single-variable Poisson brackets at order $O(\epsilon)$.
The coherent part of the Lindbladian thus contributes the following set of Liouvillian terms to the drift:
\begin{equation}
    -\partial_{\theta_j}\left[\mu_{\hat{H}, \theta_j}\overline{W}_{\hat{\rho}}\right] -\partial_{\phi_j}\left[\mu_{\hat{H}, \phi_j}\overline{W}_{\hat{\rho}}\right] = \epsilon \{W_{\hat{H}}, W_{\hat{\rho}}\}_j.
\end{equation}
Explicitly, defining $\pazocal{S} = \sqrt{S(S+1)}$ and $\lambda = 2 \pazocal{S} \epsilon$ as before,
\begin{equation}
\begin{split}
    \mu_{\hat{H},j} = -\frac{2\tilde{J}\pazocal{S}\lambda}{\pazocal{N}[V]} \sum_{k\neq j}& V(j,k)\sin(\theta_k) \\
    &\quad \times \begin{pmatrix}
        \sin(\phi_j+\phi_k)\\[2pt]
        \cot(\theta_j) \cos(\phi_j + \phi_k)
    \end{pmatrix}.
\end{split}
\end{equation}
As the dissipator terms are strictly local to each lattice site, they produce a drift contribution and diffusion matrix that, for each site index, have the same structure as those of the LMG model considered above:
\begin{align}
    \mu_{\pazocal{D},j} &= \begin{pmatrix}
        \frac12 \lambda^2 \cot(\theta_j) + \lambda\pazocal{S}\sin(\theta_j)\\[2pt]
        0
    \end{pmatrix},\\[4pt]
    D_j &= \lambda^2 \begin{pmatrix}
        1 & 0\\[2pt]
        0 & \cot^2(\theta_j)
    \end{pmatrix}.
\end{align}
As in the single-spin case, the resulting stochastic process governing the variables of each lattice site $j$ is asymptotically Bessel-$\frac12$ near the poles $\theta_j \in \{0, \pi\}$, so that all trajectories will almost surely not reach any site's poles in finite time provided they are not initialized at them.
The poles are thus unattainable site by site, independently of the interaction, and no boundary condition is required beyond reflecting the finite-step-size overshoot of the integrator.

\subsection{Short-range interactions: mean-field analysis}

We will begin by treating the problem at a mean-field level.
The Bloch equations for this model with nearest-neighbour interactions were reported in Ref.~\cite{PhysRevLett.110.257204}, and may be easily generalized to our general interaction function $V(i,j)$:
\begin{subequations}
\renewcommand{\theequation}{\theparentequation.\arabic{equation}}
\label{eq:XY_bloch}
\begin{align}
	\frac{\dd\langle \hat{S}^x_i \rangle}{\dd \tau} &=  - \frac{2\tilde{J}}{\pazocal{N}[V]} \sum_{j \neq i} V(i,j) \langle \hat{S}^z_i \rangle \langle \hat{S}^y_j \rangle - \frac12 \langle \hat{S}^x_i \rangle,\\
	\frac{\dd\langle \hat{S}^y_i \rangle}{\dd \tau} &=  - \frac{2\tilde{J}}{\pazocal{N}[V]} \sum_{j \neq i} V(i,j) \langle \hat{S}^z_i \rangle \langle \hat{S}^x_j \rangle - \frac12 \langle \hat{S}^y_i \rangle,\\
	\frac{\dd\langle \hat{S}^z_i \rangle}{\dd \tau} &= \begin{aligned}[t]
	    &\frac{2\tilde{J}}{\pazocal{N}[V]} \sum_{j \neq i} V(i,j) \left( \langle \hat{S}^x_i \rangle \langle \hat{S}^y_j \rangle + \langle \hat{S}^y_i \rangle \langle \hat{S}^x_j \rangle \right) \\
	    &\qquad - \left(\langle \hat{S}^z_i \rangle + \frac12\right).
	\end{aligned}
	\refstepcounter{equation}\tag*{\raisebox{-1.1\baselineskip}{(\theequation)}}\label{MFAXY}
\end{align}
\end{subequations}
A trivial fixed point in the mean-field dynamics is given by $\langle \hat{S}^x_i \rangle = \langle \hat{S}^y_i \rangle = 0$ and $\langle \hat{S}^z_i \rangle = -\frac{1}{2}$ for all $i$, corresponding to a pure state with all spins polarized downwards; this is referred to in Ref.~\cite{PhysRevLett.110.257204} as the paramagnetic phase.
For nearest-neighbour~\cite{PhysRevLett.110.257204} and infinite-range $V(i,j) = 1$~\cite{lee2014dissipative} interactions (the latter not possessing a well-defined thermodynamic limit), linear stability analysis around this point shows that the paramagnetic phase is stable for $\tilde{J} < \tilde{J}_0$ and unstable otherwise.
Such analysis is unavailable for general $V(i,j)$ (and would be more involved): we will return to this in the section on long-range interactions.

In the nearest-neighbour case, for $\tilde{J}>\tilde{J}_0$ the model enters the so-called staggered-XY (sXY) phase, characterized by a checkerboard pattern of alternating sublattices $A$ and $B$ with $S_A^x = S_B^y$ and $S_A^y = S_B^x$, and with the spins oriented at angles $\vartheta$ and $-\vartheta$ relative to the $x=y$ line~\cite{PhysRevLett.110.257204}.
This corresponds to the breaking of a global continuous symmetry
\begin{gather}
    \hat{S}_A^+ \to e^{i\vartheta} \hat{S}_A^+, \quad \hat{S}_A^- \to e^{-i\vartheta} \hat{S}_A^-,\\
    \hat{S}_B^+ \to e^{-i\vartheta} \hat{S}_B^+, \quad \hat{S}_B^- \to e^{i\vartheta} \hat{S}_B^-,
    \label{globalnnsymm}
\end{gather}
so that mean-field theory returns not a single steady state but a continuous family parametrized by $\vartheta$.
In this phase, the per-site longitudinal magnetization is $\langle \hat{S}^z \rangle = -\frac{1}{8\tilde{J}}$, while the per-site magnitude of the transverse magnetization on either sublattice is
\begin{equation}
    m = \sqrt{\langle \hat{S}^{x}_{A/B}\rangle^2+\langle \hat{S}^{y}_{A/B}\rangle^2}
    = \sqrt{\frac{1}{8\tilde{J}} - \frac{1}{32\tilde{J}^2}}. 
\end{equation}
The orientation of the two sublattices is parametrized by $\vartheta$ as
\begin{subequations}
\renewcommand{\theequation}{\theparentequation.\arabic{equation}}
\label{eq:XY_sublattice}
\begin{gather}
	\begin{pmatrix} \langle \hat{S}^x_A \rangle & \langle \hat{S}^y_A \rangle \end{pmatrix} = \begin{pmatrix} m \cos\left( \frac{\pi}{4} \! + \!  \vartheta\right) & m \sin\left( \frac{\pi}{4} \! + \!  \vartheta \right)  \end{pmatrix},\\
	\begin{pmatrix} \langle \hat{S}^x_B \rangle & \langle \hat{S}^y_B \rangle \end{pmatrix} = \begin{pmatrix}  m \sin\left( \frac{\pi}{4} \! + \!  \vartheta \right) & m \cos\left( \frac{\pi}{4} \! + \! \vartheta\right)  \end{pmatrix}.
\end{gather}
\end{subequations}
At $\vartheta=0$ the two Cartesian components are equal and the sublattices are aligned.

\begin{figure}
    \includegraphics[width=\columnwidth]{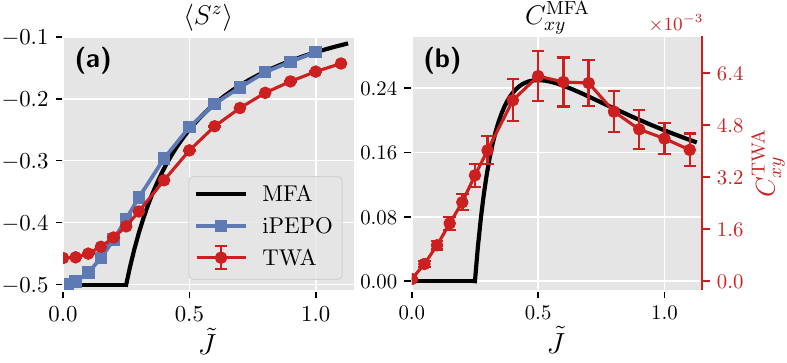}
    \caption{
        Steady state of the dissipative anisotropic XY model with nearest-neighbour interactions on a $20\times 20$ square lattice with periodic boundaries, shown as a function of the dimensionless interaction strength $\tilde{J}$.
        TWA results are averaged over 8000 independent trajectories and evaluated at $\tau=30$.
        (a) Site-averaged longitudinal magnetization $\langle \hat{S}^z \rangle$, compared to iPEPO tensor network results reproduced from Ref.~\cite{mckeever2022tensor} and to mean-field values obtained by numerically integrating Eqs.~\eqref{eq:XY_bloch}; the error bars are within the marker size.
        (b) Normalized correlator $C_{xy}$ of Eq.~\eqref{eq:Cxy_definition}, compared to the mean-field prediction of Eq.~\eqref{Eq: SxSy MFA}, which vanishes below $\tilde{J}_0=1/4$.
        Separate vertical axes are used for the mean-field (left) and TWA (right) results, with scales differing by a factor of approximately $40$.}
    \label{fig: Short-range XY}
\end{figure}

In Fig.~\ref{fig: Short-range XY}(a), we study the site-averaged steady-state magnetization of this model as a function of the interaction strength $\tilde{J}$.
Our numerical TWA results for a finite lattice of $20\times 20$ spins demonstrate the same qualitative behaviour as those obtained by iPEPO (which we take to be exact): the magnetization $\langle \hat{S}^z \rangle$ changes smoothly as a function of $\tilde{J}$, contrary to the sharp transition predicted by mean-field theory.
Taking values between $-0.45$ and $-0.4$ for $\tilde{J}<\tilde{J}_0$ and trending towards $0$ with increasing $\tilde{J}$, our results exhibit a maximum error of approximately $10\%$ relative to iPEPO.
Given the comparatively lower computational cost and ability to probe processes such as vortex formation (see Sec.~\ref{sec:vortexprolif}) that are inaccessible to iPEPO, we believe this to be an acceptable trade-off.
For small $\tilde{J}$ the error is caused by the same mechanism as in Sec.~\ref{sec:LMGNumerical}: a Fokker--Planck equation will always yield an everywhere-positive steady state, which cannot exactly represent the Wigner distribution of the spin-down state for low spin values.
While the origin of the error at larger $\tilde{J}$ is not certain, we believe the mechanism to be analogous.
Importantly, the apparent similarity between iPEPO and mean-field theory at large $\tilde{J}$ should not be taken as an indication that this regime is described by mean-field theory and that our method thus gives inaccurate results in the mean-field limit.
As we will see shortly, mean-field theory incorrectly characterizes the phase of the system in this regime, so the cause of the agreement between mean-field and iPEPO analysis at large $\tilde{J}$ for this specific observable is unclear.

Because our TWA simulations are done on finite lattices, in contrast to iPEPO we cannot directly observe transverse magnetization: $\langle \hat{S}^x \rangle$ and $\langle \hat{S}^y \rangle$ are observed to vanish up to statistical noise.
This is because spontaneous symmetry breaking cannot occur in a finite lattice~\cite{Strocchi2019}, so we are restricted to observing the superposition of all steady states connected by the symmetry Eq.~\eqref{globalnnsymm}, with a resulting average zero transverse magnetization.
To overcome this limitation, we turn to the correlated properties of the model: by considering a higher-order moment involving $\hat{S}^x$ and $\hat{S}^y$, the average of which over $\vartheta$ is non-zero in mean-field theory, we may evaluate our method's ability to probe these components of the spin.

To this end we define the real, normalized correlator
\begin{equation}
    C_{xy} \equiv \frac{4}{N^2}
    \left\langle \left\{\hat{M}^x,\hat{M}^y\right\}\right\rangle, \quad \hat{M}^a = \sum_j^N \hat{S}^a_j.
    \label{eq:Cxy_definition}
\end{equation}
For spin-$1/2$ the same-site anticommutator vanishes, while the Wigner symbols of the cross-site anticommutators are simply products.
The chosen normalization gives the bound $|C_{xy}|\leq 1$, corresponding to a state fully polarized along $x=\pm y$.
By Eq.~\eqref{eq:XY_sublattice}, the total mean-field approximation (MFA) spin is:
\begin{equation}
    M^a_{\mathrm{MFA}}(\vartheta) = \frac{N}{2}\left(\langle \hat{S}^a_A\rangle + \langle\hat{S}^a_B \rangle \right),
\end{equation}
and in the MFA the anticommutator is simply double the product of its terms.
Averaging over $\vartheta$ gives a finite value for the correlator in the non-symmetry-broken superposition state:
\begin{align}
    C_{xy}^{\mathrm{MFA}}
    &= \frac{1}{2\pi}\int_0^{2\pi} \dd\vartheta\;
    2m^2\left(\cos(\tfrac{\pi}{4} + \vartheta) + \sin(\tfrac{\pi}{4} + \vartheta)\right)^2
    \notag \\
    & =2m^2
    = \frac{4\tilde{J}-1}{16\tilde{J}^2} .
    \label{Eq: SxSy MFA}
\end{align}
In Fig.~\ref{fig: Short-range XY}(b), we compare the steady-state values of $C_{xy}$ found from TWA for the $20\times 20$ square lattice to the mean-field value of Eq.~\eqref{Eq: SxSy MFA}.
The TWA correlator is smaller by a factor of approximately $40$, while retaining the same functional dependence on $\tilde{J}$.
We take this as evidence that the sXY phase is an artifact of the MFA, corroborating earlier field-theoretic~\cite{PhysRevB.93.014307} and tensor-network findings~\cite{mckeever2022tensor}.

\subsection{Short-range interactions: vortex proliferation} \label{sec:vortexprolif}

The universality class of the nearest-neighbour anisotropic XY model was identified in Ref.~\cite{PhysRevB.93.014307} as that of the two-dimensional classical XY model.
Its BKT transition~\cite{Berezinskii1971,KosterlitzThouless1973} separates a low-temperature phase with quasi-long-range order and algebraically decaying correlations from a high-temperature disordered phase with exponentially decaying correlations, the latter produced by the unbinding and proliferation of vortices.
The same analysis found the effective temperature of the model to lie above the associated BKT critical temperature throughout, so that the sXY order is expected to be destroyed by vortex formation for any interaction strength.

Our approach gives direct access to the azimuthal variable $\phi_j$ and produces individual trajectories, so these defects can be observed directly.
For every nearest-neighbour bond connecting the $A$ and $B$ sublattices, we calculate the relative azimuthal angle $\chi=(\phi_B-\phi_A)\bmod 2\pi$, which specifies the local orientation of the staggered-XY order parameter, and associate it with the bond midpoint.
A vortex is identified when this angle winds by $2\pi$ around a plaquette of the resulting midpoint lattice, which is rotated by $\pi/4$ relative to the spin lattice.
The winding number $w_p$ is the sum of the successive differences of $\chi$ around the plaquette, each wrapped into $(-\pi,\pi]$, divided by $2\pi$.
In the simulations considered here the observed defects have elementary winding $w_p=\pm1$, corresponding to vortices and antivortices, respectively.

\begin{figure}
    \includegraphics[width=\columnwidth]{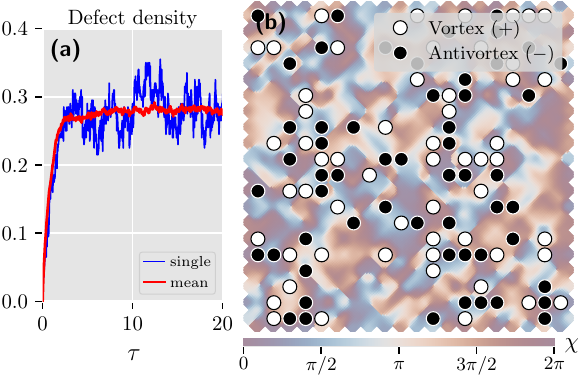}
    \caption{
        (a) Density of topological defects, $n_{\mathrm{def}} = N_{\mathrm{def}}/N$, in time over $0\leq\tau\leq20$ at $\tilde{J}=0.5$, shown for the single trajectory of panel (b) and averaged over $40$ trajectories, each initialized without defects.
        The mean approaches $n_{\mathrm{def}}\approx0.28$ at late times.
        (b) Snapshot of the local orientation $\chi$ of the staggered-XY order parameter, inferred from the relative azimuthal angles of neighbouring spins on the $A$ and $B$ sublattices and displayed on the rotated midpoint lattice, taken from that trajectory at $\tau=30$.
        The colour scale spans $\chi\in[0,2\pi)$.
        Vortices and antivortices, identified as midpoint-lattice plaquettes carrying winding number $+1$ and $-1$ respectively, are marked at the plaquette centres as indicated in the legend.
        The snapshot contains $56$ vortices and $56$ antivortices.}
    \label{fig: Vortices}
\end{figure}

Figure~\ref{fig: Vortices}(a) follows the density of elementary defects, $n_{\mathrm{def}}=N_{\mathrm{def}}/N$ with $N_{\mathrm{def}}=\sum_p|w_p|$, starting from a configuration containing none.
The mean defect density grows rapidly to approximately $0.26$ by $\tau=2$, followed by slower relaxation and late-time fluctuations around $n_{\mathrm{def}}\approx0.28$.
Figure~\ref{fig: Vortices}(b) shows the local orientation of the staggered-XY order parameter for the same trajectory at $\tau=30$, in which extended domains of aligned orientation are separated by vortices and antivortices.
The resulting increase in defect density is consistent with the vortex-proliferation mechanism proposed in Ref.~\cite{PhysRevB.93.014307} for the suppression of the sXY phase.

\subsection{Long-range interactions: steady-state phase diagram}

Having demonstrated the capabilities of our prescription and numerical implementation for open spin systems with nearest-neighbour interactions, we now consider the general case of the dissipative anisotropic XY model with long-ranged algebraically decaying interactions
$
    V(i,j) = r^{-\alpha}_{i,j}
$
as defined by Eq.~\eqref{eq:H_XY}, together with the local jump operators $\hat{L}_j = \sqrt{\Gamma}\hat{S}_j^-$, again concentrating on two-dimensional square lattices.
Here $r_{i,j}$ is the Euclidean distance on the lattice between sites $i$ and $j$, and $\alpha$ the power-law exponent.
Characterizing the non-equilibrium steady-state properties of higher-dimensional spin lattices with long-ranged interactions represents a significant and ongoing challenge, even for advanced tensor network~\cite{dunham2025efficienttimeevolution2d} and quantum Monte Carlo techniques~\cite{Hryniuk2026}.
In this section, we demonstrate the applicability of the TWA method introduced in this work to this general class of problems and showcase its potential for uncovering new physics.

\begin{figure}[t]
    \centering
    \includegraphics[width=\columnwidth]{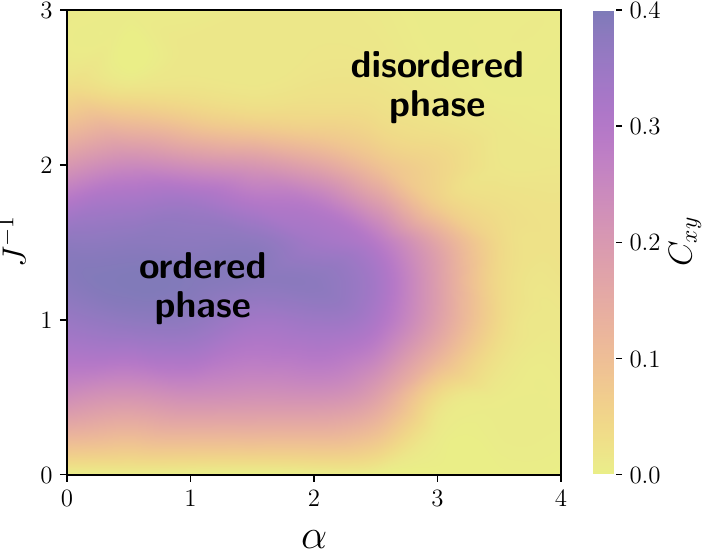}
    \caption{
        Steady-state correlator $C_{xy}$ of Eq.~\eqref{eq:Cxy_definition} for the dissipative anisotropic XY model with long-range algebraically decaying interactions on a $20\times 20$ square lattice, evaluated on a grid of $(\alpha,\tilde{J}^{-1})$ points, each averaged over 30 trajectories at $\tau=40$, and shown as a colour map.
        The correlator becomes finite near $\tilde{J}\approx0.4$ and is largest, $C_{xy}\approx0.4$, near $\tilde{J}\approx1$.
        For $\alpha\lesssim2$ it remains finite up to the largest $\tilde{J}$ shown, decaying towards zero as $\tilde{J}\to\infty$.
        For larger $\alpha$ the ordered region instead closes at a finite upper value of $\tilde{J}$, forming a dome that spans $0.5\lesssim\tilde{J}\lesssim2.0$ at $\alpha=3$ and narrows with increasing $\alpha$ until it is no longer resolved above $\alpha\approx3.5$.}
    \label{fig:phase_diagram}
\end{figure}

Given the known behaviour of systems with nearest-neighbour~\cite{PhysRevLett.110.257204} and infinite-range~\cite{lee2014dissipative} interactions, we may by analogy expect a line of phase transitions from a paramagnetic to a ferromagnetic phase at around $\tilde{J}=\frac{1}{4}$ as $\alpha$ is varied between the two limits.
Unlike for nearest-neighbour interactions, the long-ranged model no longer possesses a global U(1) symmetry associated with the sublattices, but it retains the smaller $\mathbbm{Z}_2$ symmetry associated with simultaneously inverting the signs of $S_i^x$ and $S_i^y$: it is this symmetry that is spontaneously broken in the ferromagnetic phase.
In Fig.~\ref{fig:phase_diagram}, we study the steady-state correlator $C_{xy}$, which is invariant under the remaining $\mathbbm{Z}_2$ symmetry and is expected to be non-zero for a uniform probabilistic mixture of ferromagnetic states.
Its values are evaluated over a grid of $(\alpha,\tilde{J}^{-1})$ points, averaged over 30 trajectories each, and plotted as a heatmap.
The correlator is large over a single connected region of the diagram, marking a $\mathbbm{Z}_2$ symmetry-broken phase.
At fixed $\alpha$, increasing $\tilde{J}$ takes the system into this phase near $\tilde{J}\approx0.4$, in qualitative accord with the line of phase transitions at $\tilde{J}_0=1/4$ suggested by the mean-field results.

\begin{figure}[t]
    \centering
    \includegraphics[width=\columnwidth]{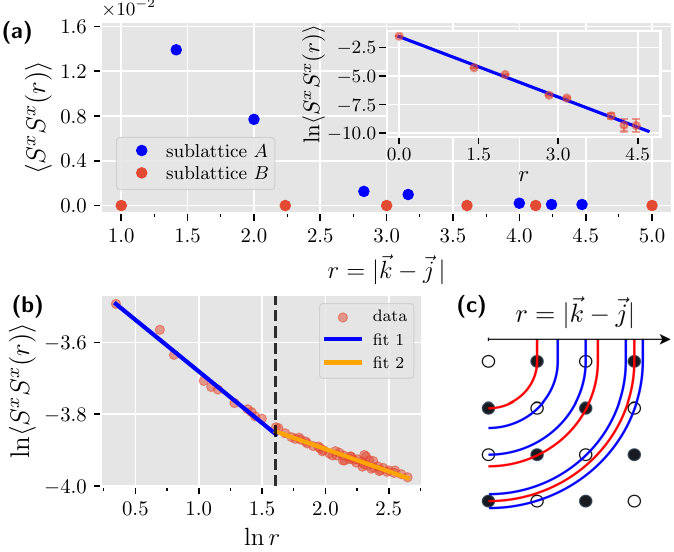}
    \caption{
        Steady-state correlation function $\langle \hat{S}^x_j \hat{S}^x_k \rangle$ of the dissipative anisotropic XY model on a $20\times 20$ square lattice, averaged over lattice sites as a function of the Euclidean separation $r_{jk} = |\bm{k}-\bm{j}|$.
        (a) Correlations resolved by sublattice for nearest-neighbour interactions at $\tilde{J} = 0.3$, obtained from $50560$ trajectories.
        Pairs lying on the same sublattice ($A$, blue) remain correlated, whereas pairs on opposite sublattices ($B$, red) are uncorrelated to within the sampling error, exposing the checkerboard structure of the sXY phase.
        Inset: the same intra-sublattice data on a semi-logarithmic scale, where they are well fitted by the exponential decay $\langle \hat{S}^x_j \hat{S}^x_k \rangle \approx 0.21\,e^{-r_{jk}/0.57}$.
        (b) For long-range algebraically decaying interactions with $\alpha=3$ at $\tilde{J} = 0.847$, from $16000$ trajectories, for $j$ and $k$ on the same sublattice.
        On a log-log scale the data are well fitted by two power laws, $\sim r_{jk}^{-0.288}$ for $r_{jk}<5$ and $\sim r_{jk}^{-0.125}$ beyond.
        (c) Construction of the separations $r_{jk}$ on the lattice, with arcs of constant $r_{jk}$ coloured by the sublattice of the connected sites as in (a).
        Adapted from Ref.~\cite{mckeever2022tensor}.}
    \label{fig:fits}
\end{figure}

The behaviour at large $\tilde{J}$ separates two regimes.
For $\alpha\lesssim2$ the correlator remains finite throughout the range studied, decaying towards zero only as $\tilde{J}\to\infty$, where the dissipation that stabilizes the ordered state is suppressed.
For larger $\alpha$, by contrast, the ordered region closes at a finite upper value of $\tilde{J}$, so that increasing the coupling beyond $\tilde{J}\approx2$ returns the system to a disordered state.
Such reentrance lies outside the reach of mean-field theory, which predicts order for every $\tilde{J}>\tilde{J}_0$ at any $\alpha$.
The resulting dome narrows with increasing $\alpha$ until, beyond $\alpha\approx3.5$, no ordered region is resolved at the accessible system size.
The emergence of the order in this model therefore requires interactions of sufficiently long range, with increasing $\alpha$ progressively enhancing the fluctuations neglected by the mean-field treatment until, in the nearest-neighbour limit considered above, they suppress the ordered phase altogether.
The steady-state phase diagram of the two-dimensional dissipative anisotropic XY model with long-range interactions has not been characterized previously  beyond mean-field theory in the literature.
We note, however, that the reemergence of the symmetry-broken phase discussed here is consistent with earlier unpublished DTWA observations~\cite{rabl2023stochastic}.

The nature of the ordering in each regime is reflected in the two-point correlation function $\langle \hat{S}^x_j \hat{S}^x_k \rangle$, shown in Fig.~\ref{fig:fits}.
For nearest-neighbour interactions at $\tilde{J}=0.3$, Fig.~\ref{fig:fits}(a), the correlations decay exponentially with separation, $\langle \hat{S}^x_j \hat{S}^x_k \rangle \approx 0.21\,e^{-r_{jk}/\xi}$ with a decay length $\xi=0.57$.
The same exponential decay law is found at these parameters with iPEPO tensor networks, with $\xi=0.93$~\cite{mckeever2022tensor}.
Both approaches therefore place the nearest-neighbour steady state in a regime of short-ranged correlations, consistent with the suppression of $C_{xy}$ seen in Fig.~\ref{fig: Short-range XY}(b).
With long-range interactions the decay is instead algebraic.
At $\alpha=3$ and $\tilde{J}=0.847$, Fig.~\ref{fig:fits}(b), the data follow $\sim r_{jk}^{-0.288}$ out to $r_{jk}\approx5$ and $\sim r_{jk}^{-0.125}$ beyond, the crossover to slower decay at larger separations plausibly being a finite-size effect.
Correlations of this form are indicative of the long-range order underlying the symmetry-broken region of Fig.~\ref{fig:phase_diagram}.

\section{DISCUSSION}
\label{sec:discussion}

In this work, we have introduced a continuous SU(2) phase-space framework for the dynamics of open interacting spin systems with arbitrary local spin quantum number $S$.
Starting directly from the Lindblad master equation and the Stratonovich--Weyl correspondence, the exact phase-space evolution is expressed through spin star products and organized systematically in powers of the semiclassical parameter $\epsilon=(2S+1)^{-1}$.
Truncation at second order produces a Fokker--Planck equation whose drift and diffusion follow directly from the Hamiltonian and Lindblad jump operators.
The resulting It\^{o} equations involve only two stochastic coordinates per spin and admit straightforward parallelization over independent trajectories, while placing no intrinsic restriction on spatial dimension, interaction range, or lattice geometry.
The same stochastic evolution describes both transient relaxation and stationary behaviour, providing direct access to non-equilibrium steady states of large many-body systems through long-time propagation.

We have benchmarked the method on two models.
For the Lipkin--Meshkov--Glick model, a single spin with nonlinear coherent interactions subject to dissipation, the stochastic trajectories reproduce observables for both large and small spin quantum numbers.
For the two-dimensional dissipative anisotropic XY model with nearest-neighbour interactions on lattices of $400$ spins, the longitudinal magnetization agrees qualitatively with iPEPO tensor-network results~\cite{mckeever2022tensor}, while the mean-field staggered-XY order is found to be strongly suppressed, with exponentially decaying correlations.
Access to individual trajectories further allows the underlying mechanism to be resolved directly, and the observed proliferation of vortex--antivortex pairs is consistent with the vortex-driven destruction of the staggered-XY phase anticipated on field-theoretic grounds~\cite{PhysRevB.93.014307}.
We then studied the same model with long-range algebraically decaying interactions, a regime that has remained inaccessible to other methods.
Mapping the steady state as a function of interaction strength and decay exponent, we located an extended $\mathbbm{Z}_2$ symmetry-broken phase, bounded in both interaction strength and interaction range, whose reentrant boundary at strong coupling and closure beyond $\alpha\approx3.5$ lie outside the reach of mean-field theory.

Several directions follow immediately.
The favourable scaling with spin quantum number makes the method particularly well suited to two- and three-dimensional higher-spin platforms with algebraic dipolar tails, building on experimental realizations of long-range spin-1 trapped-ion chains~\cite{Senko2015}, spin-3 dipolar dynamics in three-dimensional optical lattices~\cite{Lepoutre2019}, and coherent spin-1 encodings in multilevel polar-molecule tweezer arrays~\cite{Hepworth2025}.
Within quantum technologies, future work can include quantifying the decoherence-limited generation of spin squeezing in extended arrays~\cite{Bornet2023,Eckner2023,Franke2023}, characterizing collective emission and subradiance in higher-dimensional emitter ensembles~\cite{Rui2020,RubiesBigorda2023,MinkFleischhauer2023}, or designing dissipatively prepared steady states~\cite{Barreiro2011,Diehl2008,Verstraete2009}, where the interplay between engineered and intrinsic decay channels sets the fidelity that can be reached.
On the methodological side, the truncated Wigner approximation is long established for bosonic fields~\cite{Steel1998,Sinatra2002,Blakie2008,Polkovnikov2010} and rests on the same Stratonovich--Weyl footing as the systematic spin construction introduced here: in Appendix~\ref{tensorstarproduct} we show how the two combine into a single star product.
A unified treatment of hybrid spin--boson systems therefore appears within reach, which would give direct access to cavity~\cite{Ritsch2013} and waveguide~\cite{Sheremet2023} QED, optomechanical arrays~\cite{Aspelmeyer2014}, and trapped-ion crystals in which the mediating phonon modes are retained rather than adiabatically eliminated~\cite{Britton2012,Bohnet2016}.

\section*{CODE and DATA AVAILABILITY}

The method introduced in this manuscript (implemented in \textit{Julia}~\cite{doi:10.1137/141000671}) and the simulation data presented are available upon request.\\\\

\section*{ACKNOWLEDGEMENTS}

We thank Jack Dunham, who kindly reproduced iPEPO data from Ref.~\cite{mckeever2022tensor} and offered helpful comments on the capabilities of iPEPO versus our method.
The authors acknowledge support from the Engineering and Physical Sciences Research Council (EPSRC) Grants No. EP/N509577/1, EP/T517793/1, EP/V026496/1, EP/S019669/1, EP/Z000572/1, and EP/R513143/1.
We further acknowledge support from the EU QuantERA Network PROTEQT and INTERPOL consortia, and the EUCENTRAL project (this project is funded by the European Union under Horizon Europe (project No. 101186579)).
The authors acknowledge the use of the UCL HPC facilities and associated support services in the completion of this work.

\section*{Appendices}

\appendix

\section{Stratonovich postulates}
\label{app:SW}

A Stratonovich--Weyl correspondence is a map $\hat{A}\mapsto W_{\hat{A}}$ from operators on $\pazocal{H}$ to functions on the phase space $\pazocal{M}$, carrying a symmetry group $G$ with unitary irreducible representation $\pi_g$ and $G$-invariant measure $\dd\Omega$.
It is required to satisfy the Stratonovich postulates~\cite{stratonovich1957distributions, VARILLY1989107}:
\begin{enumerate}
    \item \emph{Linearity.} $\hat{A}\mapsto W_{\hat{A}}$ is linear and one-to-one.
    \item \emph{Reality.} $W_{\hat{A}^\dagger}(\Omega)=\overline{W_{\hat{A}}(\Omega)}$, so $W_{\hat{A}}$ is real for Hermitian $\hat{A}$.
    \item \emph{Standardization.} $\int_{\pazocal{M}}\dd\Omega\,W_{\hat{A}}(\Omega)=\Tr\hat{A}$.
    \item \emph{Traciality.} $\int_{\pazocal{M}}\dd\Omega\,W_{\hat{A}}(\Omega)W_{\hat{B}}(\Omega)=\Tr(\hat{A}\hat{B})$.
    \item \emph{Covariance.} $W_{\pi_g\hat{A}\pi_g^\dagger}(\Omega)=W_{\hat{A}}(g^{-1}\!\circ\Omega)$.
\end{enumerate}
Standardization with $\hat{A}=\hat{\rho}$ gives the normalization $\int\dd\Omega\,W_{\hat{\rho}}=1$, and traciality with $(\hat{A},\hat{B})=(\hat{\rho},\hat{O})$ gives the expectation-value formula of the main text.

Writing the map in trace form $W_{\hat{A}}(\Omega)=\Tr\{\hat{A}\,\Delta(\Omega)\}$, the postulates become algebraic conditions on the operator-valued kernel $\Delta(\Omega)$:
\begin{align}
    \text{Reality} &\;\Longleftrightarrow\;
    \Delta(\Omega)=\Delta^\dagger(\Omega), \\
    \text{Standardization} &\;\Longleftrightarrow\;
    \int_{\pazocal{M}}\dd\Omega\,\Delta(\Omega)=\mathbbm{1}, \\
    \text{Traciality} &\;\Longleftrightarrow\;
    \Tr\{\Delta(\Omega)\,\Delta(\Omega')\}=\delta(\Omega,\Omega'), \\
    \text{Covariance} &\;\Longleftrightarrow\;
    \Delta(g\!\circ\Omega)=\pi_g\,\Delta(\Omega)\,\pi_g^\dagger,
\end{align}
with $\delta(\Omega,\Omega')$ the Dirac distribution for the measure $\dd\Omega$.
The standardization condition is a resolution of the identity and fixes the measure normalization.
The traciality condition makes $\Delta$ self-dual, which is what allows the same kernel to implement both the forward map $W_{\hat{A}}=\Tr\{\hat{A}\Delta\}$ and its inverse $\hat{A}=\int\dd\Omega\,W_{\hat{A}}\Delta$.

\begin{widetext}

\section{Spin star product expansion}
\label{app:starproduct}

The exact SU(2) star product of Eq.~\eqref{eq:moyal_spin} is not directly usable.
Here we expand it to second order in the semiclassical parameter $\epsilon = (2S+1)^{-1}$.
Throughout, $W_A\equiv W_{\hat{A}}$ denotes the Wigner symbol of $\hat{A}$, and we use $S$ for the spin, reserving $j$ for product indices.
A more verbose form of the derivation may be found in the first author's thesis~\cite{timofeev2025superfluidity}.

\subsection{Definitions}

We start from the self-dual ($s=s_1=s_2=0$) form of the star product derived in Ref.~\cite{Klimov_2002},
\begin{equation}
    W_A \star W_B = N_S \sum_{j=0}^\infty a_j \int \frac{\dd\psi}{2\pi}\,
    \tilde{F}^{-1}(\pazocal{J}^2)
    \left[\big((S^+)^j \tilde{F}(\pazocal{J}^2)W_A\big)
    \big((S^-)^j \tilde{F}(\pazocal{J}^2)W_B\big) \right],
    \label{eq:klimov39}
\end{equation}
with
\begin{gather}
    \pazocal{J}^2 = -\left[ \partial_\theta^2 + \cot\theta\,\partial_\theta
    + \frac{1}{\sin^2\theta}\left( \partial_\phi^2
    - 2\cos\theta\,\partial_\phi\partial_\psi + \partial_\psi^2 \right) \right],\\
    S^\pm = i e^{\mp i\psi}\left( \pm\cot\theta\,\partial_\psi
    + i\,\partial_\theta \mp \frac{1}{\sin\theta}\,\partial_\phi \right),\\
    a_j = \frac{(-1)^j}{j!\,(2S+j+1)!},\qquad N_S = \sqrt{2S+1},\\
    F(l) = \sqrt{(2S+l+1)!\,(2S-l)!},\qquad
    \tilde{F}(\pazocal{J}^2)\,D^{l}_{mm'} = F(l)\,D^{l}_{mm'},
\end{gather}
where $D^{l}_{mm'}$ are the Wigner $D$-functions.
The operators satisfy (on $\psi$-independent functions) $[\pazocal{J}^2, S^\pm]f(\theta,\phi)=0$ and the product rule
\begin{equation}
    \pazocal{J}^2(fg) = g\,\pazocal{J}^2 f + f\,\pazocal{J}^2 g
    - (S^+ f)(S^- g) - (S^+ g)(S^- f).
    \label{eq:L2product}
\end{equation}
We also define the $\psi$-independent operators
\begin{gather}
    \pazocal{L}^2 = -\left[ \partial_\theta^2 + \cot\theta\,\partial_\theta
    + \frac{1}{\sin^2\theta}\,\partial_\phi^2 \right],\\
    S^{\pm(j)} = \prod_{k=0}^{j-1}\left( k\cot\theta - \partial_\theta
    \mp \frac{i}{\sin\theta}\,\partial_\phi \right),
\end{gather}
the factors of the latter ordered with $k$ increasing to the left, so that, acting on $\psi$-independent functions, $\pazocal{J}^2 f(\theta,\phi) = \pazocal{L}^2 f(\theta,\phi)$ and $(S^\pm)^j f(\theta,\phi) = e^{\mp i j\psi} S^{\pm(j)} f(\theta,\phi)$.
Since symbols are $\psi$-independent, the two factors in Eq.~\eqref{eq:klimov39} contribute $e^{-ij\psi}$ and $e^{+ij\psi}$, whose product is unity: the integrand carries no $\psi$-dependence and $\int\!\dd\psi/2\pi = 1$.
We retain the integral here in the form given in Ref.~\cite{Klimov_2002}, but it is redundant for symbols.

\subsection{Expansion of the deformation operators}

Writing $\epsilon = (2S+1)^{-1}$, the coefficient $F(l)$ factorizes as
\begin{equation}
    F(l) = \frac{(2S+1)!}{\sqrt{2S+1}}
    \exp\!\left[ \tfrac12 \sum_{k=0}^{l} \ln\frac{1+\epsilon k}{1-\epsilon k} \right].
    \label{eq:FL}
\end{equation}
Since $\ln\frac{1+\epsilon k}{1-\epsilon k} = 2\epsilon k + o(\epsilon^2)$,
\begin{equation}
    F(l) = \frac{(2S+1)!}{\sqrt{2S+1}}
    \left[ 1 + \tfrac12\epsilon\, l(l+1)
    + \tfrac18\epsilon^2\, l^2(l+1)^2 + o(\epsilon^2) \right],
\end{equation}
from which
\begin{gather}
    \tilde{F}(\pazocal{J}^2) = \frac{(2S+1)!}{\sqrt{2S+1}}
    \underbrace{\left[ 1 + \tfrac12\epsilon\,\pazocal{J}^2
    + \tfrac18\epsilon^2\,\pazocal{J}^4 + o(\epsilon^2) \right]}_{\tilde{G}(\pazocal{J}^2)},\\
    \tilde{F}^{-1}(\pazocal{J}^2) = \frac{\sqrt{2S+1}}{(2S+1)!}
    \underbrace{\left[ 1 - \tfrac12\epsilon\,\pazocal{J}^2
    + \tfrac18\epsilon^2\,\pazocal{J}^4 + o(\epsilon^2) \right]}_{\tilde{G}^{-1}(\pazocal{J}^2)}.
\end{gather}
The prefactors combine with $N_S$ in \eqref{eq:klimov39} to give $(2S+1)!$, so that
\begin{equation}
    W_A \star W_B = (2S+1)! \sum_{j=0}^\infty a_j \int \frac{\dd\psi}{2\pi}\,
    \tilde{G}^{-1}(\pazocal{J}^2)
    \left[\big((S^+)^j \tilde{G}(\pazocal{J}^2)W_A\big)
    \big((S^-)^j \tilde{G}(\pazocal{J}^2)W_B\big) \right].
    \label{eq:klimov2}
\end{equation}
Expanding the coefficients,
\begin{equation}
    (2S+1)!\,a_j = \frac{(-1)^j}{j!} \prod_{k=1}^{j}
    \left(\epsilon - \epsilon^2 k + o(\epsilon^2)\right),
\end{equation}
so to second order only $j=0,1,2$ contribute, with
\begin{equation}
    c_0 = 1,\qquad c_1 = \epsilon^2 - \epsilon + o(\epsilon^2),\qquad
    c_2 = \tfrac12\epsilon^2 + o(\epsilon^2),
\end{equation}
and
\begin{equation}
    W_A \star W_B = \sum_{j=0}^2 c_j \int \frac{\dd\psi}{2\pi}\,
    \tilde{G}^{-1}(\pazocal{J}^2)
    \left[\big((S^+)^j \tilde{G}(\pazocal{J}^2)W_A\big)
    \big((S^-)^j \tilde{G}(\pazocal{J}^2)W_B\big) \right] + o(\epsilon^2).
    \label{eq:klimov3}
\end{equation}

\subsection{Order-by-order evaluation}

\subsubsection{\texorpdfstring{$j=0$}{j=0}}

Expanding
\begin{equation}
	\left(1 - \tfrac12\epsilon \pazocal{J}^2 + \tfrac18\epsilon^2\pazocal{J}^4\right)\left[\Big(\big(1+\tfrac12\epsilon\pazocal{J}^2+\tfrac18\epsilon^2\pazocal{J}^4\big)W_{\hat A}\Big)\Big(\big(1+\tfrac12\epsilon\pazocal{J}^2+\tfrac18\epsilon^2\pazocal{J}^4\big)W_{\hat B}\Big)\right]
\end{equation}
order by order in $\epsilon$ gives
\begin{equation}
	O(\epsilon^0): \qquad \boxed{W_{\hat A} W_{\hat B}}
\end{equation}
\begin{align}
	O(\epsilon^1): \qquad
	&\tfrac12 W_{\hat B}\, \pazocal{J}^2 W_{\hat A} + \tfrac12 W_{\hat A}\, \pazocal{J}^2 W_{\hat B} - \tfrac12\, \pazocal{J}^2(W_{\hat A} W_{\hat B}) \notag\\
	&\stackrel{\eqref{eq:L2product}}{=} \tfrac12 \left[(S^+ W_{\hat A})(S^- W_{\hat B}) + (S^- W_{\hat A})(S^+ W_{\hat B})\right] \notag\\
	&= \boxed{\tfrac12 \left[(S^{+(1)} W_{\hat A})(S^{-(1)} W_{\hat B}) + (S^{-(1)} W_{\hat A})(S^{+(1)} W_{\hat B})\right]}
\end{align}
At order $\epsilon^2$, repeated use of Eq.~\eqref{eq:L2product} to rewrite $\pazocal{J}^2$ acting on products of symbols in terms of $S^{\pm(1)}$, followed by cancellation amongst the resulting terms, leaves
\begin{equation}
	O(\epsilon^2): \qquad \boxed{
	\begin{aligned}
		\tfrac18 \Big[&(S^{+(1)} S^{+(1)} W_{\hat A})(S^{-(1)} S^{-(1)} W_{\hat B}) + (S^{-(1)} S^{+(1)} W_{\hat A})(S^{+(1)}S^{-(1)} W_{\hat B}) \\
		&+  (S^{+(1)}S^{-(1)} W_{\hat A})(S^{-(1)}S^{+(1)} W_{\hat B}) + (S^{-(1)}S^{-(1)} W_{\hat A})(S^{+(1)}S^{+(1)} W_{\hat B})\\
		&-(\pazocal{Q} W_{\hat A})(S^{-(1)} W_{\hat B}) -  (S^{+(1)} W_{\hat A})(\pazocal{R} W_{\hat B})\\
		&- (\pazocal{R} W_{\hat A})(S^{+(1)} W_{\hat B}) - (S^{-(1)} W_{\hat A})(\pazocal{Q} W_{\hat B})\Big]
	\end{aligned}}
\end{equation}
where we have defined the commutators
\begin{equation}
	\pazocal{Q} \equiv [\pazocal{L}^2, S^{+(1)}], \qquad \pazocal{R} \equiv [\pazocal{L}^2, S^{-(1)}].
\end{equation}

\subsubsection{\texorpdfstring{$j=1$}{j=1}}

Expanding
\begin{equation}
	(\epsilon^2 - \epsilon) \left(1 - \tfrac12 \epsilon \pazocal{J}^2 + \tfrac18 \epsilon^2 \pazocal{J}^4 \right)
	\left[\Big(S^+\big(1 + \tfrac12 \epsilon \pazocal{J}^2 + \tfrac18 \epsilon^2 \pazocal{J}^4 \big) W_{\hat A}\Big) \Big( S^- \big(1 + \tfrac12 \epsilon \pazocal{J}^2 + \tfrac18 \epsilon^2 \pazocal{J}^4 \big) W_{\hat B} \Big) \right]
\end{equation}
order by order gives
\begin{equation}
	O(\epsilon^0): \qquad \boxed{0}
\end{equation}
\begin{equation}
	O(\epsilon^1): \qquad \boxed{- (S^{+(1)} W_{\hat A}) (S^{-(1)} W_{\hat B}) }
\end{equation}
\begin{equation}
	O(\epsilon^2): \qquad \boxed{
	\begin{aligned}
		&(S^{+(1)} W_{\hat A}) (S^{-(1)} W_{\hat B}) +  \tfrac12 (\pazocal{Q} W_{\hat A}) (S^{-(1)} W_{\hat B}) +  \tfrac12 (S^{+(1)} W_{\hat A}) ( \pazocal{R} W_{\hat B}) \\
		&- \tfrac12 (S^{+(1)}S^{+(1)} W_{\hat A}) (S^{-(1)} S^{-(1)} W_{\hat B}) - \tfrac12 (S^{-(1)} S^{+(1)} W_{\hat A}) (S^{+(1)} S^{-(1)}  W_{\hat B})
	\end{aligned}}
\end{equation}

\subsubsection{\texorpdfstring{$j=2$}{j=2}}

Expanding
\begin{equation}
	\tfrac12 \epsilon^2  \left(1 - \tfrac12 \epsilon \pazocal{J}^2 + \tfrac18 \epsilon^2 \pazocal{J}^4 \right)
	\left[\Big((S^+)^2\big(1 + \tfrac12 \epsilon \pazocal{J}^2 + \tfrac18 \epsilon^2 \pazocal{J}^4 \big) W_{\hat A}\Big) \Big( (S^-)^2 \big(1 + \tfrac12 \epsilon \pazocal{J}^2 + \tfrac18 \epsilon^2 \pazocal{J}^4 \big) W_{\hat B} \Big) \right]
\end{equation}
order by order gives
\begin{equation}
	O(\epsilon^0): \ \boxed{0}, \qquad
	O(\epsilon^1): \ \boxed{0}, \qquad
	O(\epsilon^2): \ \boxed{\tfrac12 (S^{+(2)} W_{\hat A})(S^{-(2)} W_{\hat B})}.
\end{equation}

\subsubsection{Sum over \texorpdfstring{$j$}{j}}

Adding the $j=0,1,2$ contributions in Eq.~\eqref{eq:klimov3} order by order gives, at zeroth and first order,
\begin{equation}
	O(\epsilon^0): \qquad \boxed{W_{\hat A} W_{\hat B}},
\end{equation}
\begin{equation}
	O(\epsilon^1): \qquad \boxed{\tfrac12 \left[ (S^{-(1)} W_{\hat A}) (S^{+(1)} W_{\hat B}) - (S^{+(1)} W_{\hat A}) (S^{-(1)} W_{\hat B}) \right] },
\end{equation}
and, at second order, after the terms proportional to $\pazocal{Q}, \pazocal{R}$ and to $(S^{+(1)}S^{+(1)})\otimes(S^{-(1)}S^{-(1)})$, $(S^{-(1)}S^{+(1)})\otimes(S^{+(1)}S^{-(1)})$ partially cancel between the $j=0$ and $j=1$ contributions,
\begin{equation}
	O(\epsilon^2): \qquad \boxed{
	\begin{aligned}
			&-\tfrac38 (S^{+(1)} S^{+(1)} W_{\hat A})(S^{-(1)} S^{-(1)} W_{\hat B}) -\tfrac38 (S^{-(1)} S^{+(1)} W_{\hat A})(S^{+(1)}S^{-(1)} W_{\hat B}) \\
			&+  \tfrac18 (S^{+(1)}S^{-(1)} W_{\hat A})(S^{-(1)}S^{+(1)} W_{\hat B}) + \tfrac18 (S^{-(1)}S^{-(1)} W_{\hat A})(S^{+(1)}S^{+(1)} W_{\hat B})\\
			&+ \tfrac38 (\pazocal{Q} W_{\hat A})(S^{-(1)} W_{\hat B}) + \tfrac38 (S^{+(1)} W_{\hat A})(\pazocal{R} W_{\hat B}) \\
			&- \tfrac18 (\pazocal{R} W_{\hat A})(S^{+(1)} W_{\hat B}) - \tfrac18 (S^{-(1)} W_{\hat A})(\pazocal{Q} W_{\hat B})\\
			&+(S^{+(1)} W_{\hat A}) (S^{-(1)} W_{\hat B}) + \tfrac12 (S^{+(2)} W_{\hat A})(S^{-(2)} W_{\hat B})
	\end{aligned}}
\end{equation}

\subsection{Explicit form and simplification}

Collecting the results above, the star product for two spin symbols is, to second order in $\epsilon$,
\begin{align}
	\star = \mathbb{I} \otimes \mathbb{I}
	&+ \frac{\epsilon}{2} \left[ S^{-(1)} \otimes S^{+(1)} - S^{+(1)} \otimes S^{-(1)} \right] \notag\\
	&+ \epsilon^2 \Big[S^{+(1)} \otimes S^{-(1)} + \tfrac12 S^{+(2)} \otimes S^{-(2)} \notag\\
	&\qquad + \tfrac18 S^{+(1)} S^{-(1)} \otimes S^{-(1)}S^{+(1)} + \tfrac18 S^{-(1)} S^{-(1)} \otimes S^{+(1)}S^{+(1)} \notag\\
	&\qquad - \tfrac38 S^{-(1)}S^{+(1)} \otimes S^{+(1)} S^{-(1)} - \tfrac38 S^{+(1)}S^{+(1)} \otimes S^{-(1)}S^{-(1)} \notag\\
	&\qquad + \tfrac38 \pazocal{F} \otimes S^{-(1)} + \tfrac38 S^{+(1)} \otimes \pazocal{G} - \tfrac18 \pazocal{G} \otimes S^{+(1)} - \tfrac18 S^{-(1)} \otimes \pazocal{F}\Big] + O(\epsilon^3),
	\label{eq:star_full}
\end{align}
where
\begin{gather}
	S^{+(1)} = -\partial_\theta - i \csc\theta\, \partial_\phi, \qquad
	S^{-(1)} = -\partial_\theta + i \csc\theta\, \partial_\phi,\\
	S^{+(2)} = \partial_\theta^2 - \cot\theta\, \partial_\theta - \csc^2\theta\, \partial_\phi^2 + 2i \csc\theta\, \partial_\theta\partial_\phi - 2i \cot\theta \csc\theta\, \partial_\phi,\\
	S^{-(2)} = \partial_\theta^2 - \cot\theta\, \partial_\theta - \csc^2\theta\, \partial_\phi^2 - 2i \csc\theta\, \partial_\theta\partial_\phi + 2i \cot\theta \csc\theta\, \partial_\phi,\\
	\pazocal{F} = \csc^2\theta \left(2 \cot\theta\, \partial_\phi^2 - 2i \cos\theta\, \partial_\theta\partial_\phi + i \csc\theta\, \partial_\phi + \partial_\theta \right),\\
	\pazocal{G} = \csc^2\theta \left(2 \cot\theta\, \partial_\phi^2 + 2i \cos\theta\, \partial_\theta\partial_\phi - i \csc\theta\, \partial_\phi + \partial_\theta \right).
\end{gather}
The first-order term is recognized as the Poisson bracket on the sphere:
\begin{equation}
	\frac{\epsilon}{2} \left[ S^{-(1)} \otimes S^{+(1)} - S^{+(1)} \otimes S^{-(1)} \right] = i \epsilon \left(\lpartial\!_{\phi}\rpartial\!_{\cos\theta} - \lpartial\!_{\cos\theta}\rpartial\!_{\phi}\right) \equiv i \epsilon\, \{\leftarrow, \rightarrow\},
\end{equation}
so that Eq.~\eqref{eq:star_full} takes the compact form quoted in the main text,
\begin{equation}
	\star = 1 + i \epsilon\, \{\leftarrow, \rightarrow\} + \epsilon^2\, \pazocal{B}(\leftarrow, \rightarrow) + O(\epsilon^3),
	\label{eq:star_compact}
\end{equation}
with all remaining $O(\epsilon^2)$ terms of Eq.~\eqref{eq:star_full} absorbed into the symmetric bilinear operator $\pazocal{B}$.
Explicitly, acting on two symbols $W_{\hat A}(\theta,\phi)$, $W_{\hat B}(\theta,\phi)$,
\begin{equation}
	\begin{split}
		\pazocal{B}(W_{\hat A}, W_{\hat B}) = &\ \tfrac12\, \partial_\theta W_{\hat A}\, \partial_\theta W_{\hat B} - \tfrac12 \cot\theta\, \partial_\theta^2 W_{\hat A}\, \partial_\theta W_{\hat B} - \tfrac12 \cot\theta\, \partial_\theta W_{\hat A}\, \partial_\theta^2 W_{\hat B}\\
		&+ \tfrac34 \csc^4\theta\, \partial_\phi W_{\hat A}\, \partial_\phi W_{\hat B} + \csc^2\theta\, \partial_\theta\partial_\phi W_{\hat A}\, \partial_\theta\partial_\phi W_{\hat B} \\
		&- \tfrac12 \csc^2\theta\, \partial_\theta^2 W_{\hat A}\, \partial_\phi^2 W_{\hat B} - \tfrac12 \csc^2\theta\, \partial_\phi^2 W_{\hat A}\, \partial_\theta^2 W_{\hat B}\\
		&+ \tfrac14 \cos(2\theta)\csc^4\theta\, \partial_\phi W_{\hat A}\, \partial_\phi W_{\hat B} - \cot\theta \csc^2\theta\, \partial_\theta\partial_\phi W_{\hat A}\, \partial_\phi W_{\hat B} \\
		&- \tfrac12 \sin(2\theta)\csc^4\theta\, \partial_\phi W_{\hat A}\, \partial_\theta\partial_\phi W_{\hat B},
	\end{split}
	\label{eq:B_explicit}
\end{equation}
with all functions on the right-hand side evaluated at $(\theta,\phi)$.
The terms of Eq.~\eqref{eq:B_explicit} collect pairwise into the coefficients $\pazocal{B}_{ab}(\theta)$ of Table~\ref{tab:bilinear} in the main text, which makes the symmetry of $\pazocal{B}$ under exchange of its two arguments manifest.

\subsection{Bopp operators}

Because the star product is associative, left (and right) star multiplication by a fixed symbol is a representation of the operator algebra by differential operators on phase space, known as the Bopp representation~\cite{Bopp1956}.
For a bosonic mode, Eq.~\eqref{eq:moyal_boson} gives the familiar Bopp shifts
\begin{equation}
	x \star \circ = \left(x + \tfrac{i\hbar}{2}\rpartial\!_{p}\right)\circ, \qquad
	p \star \circ = \left(p - \tfrac{i\hbar}{2}\rpartial\!_{x}\right)\circ,
	\label{eq:bopp_boson}
\end{equation}
with the opposite signs for right multiplication; the shifted arguments obey $[x + \tfrac{i\hbar}{2}\partial_p,\, p - \tfrac{i\hbar}{2}\partial_x] = i\hbar$, so that the differential operator representing any $\hat{A}$ is obtained by substituting the shifts into $W_{\hat{A}}$.

For SU(2) no such closed form is available, but the same construction can be carried out order by order using Eq.~\eqref{eq:star_compact}.
Inserting the spin symbols $W_{\hat{S}^\pm} = \pazocal{S} e^{\pm i\phi}\sin(\theta)$ and $W_{\hat{S}^z} = \pazocal{S}\cos(\theta)$, with $\pazocal{S} \equiv \sqrt{S(S+1)}$, into the left or right slot of the star product yields second-order differential operators acting on an arbitrary symbol $\circ$,
\begin{equation}
	W_{\hat{A}} \star \circ = \pazocal{P}_{\hat{A}} \sum_{a} c^{\hat{A}}_a(\theta;\epsilon)\, \partial^a \circ + O(\epsilon^3),
	\qquad
	\partial^a \in \{1, \partial_\theta, \partial_\phi, \partial_\theta^2, \partial_\phi^2\},
	\label{eq:bopp_def}
\end{equation}
whose coefficients $c^{\hat{A}}_a$ and common prefactors $\pazocal{P}_{\hat{A}}$ are collected in Table~\ref{tab:bopp}; no mixed derivative $\partial_\theta\partial_\phi$ occurs at this order.
Since the expansion \eqref{eq:star_compact} is symmetric at orders $\epsilon^0$ and $\epsilon^2$ and antisymmetric at order $\epsilon^1$, left and right star multiplication differ only through the sign of the first-order term,
\begin{equation}
	W_{\hat{A}} \star \circ = \left(\pazocal{A}_0 + \epsilon \pazocal{A}_1 + \epsilon^2 \pazocal{A}_2\right)\circ,
	\qquad
	\circ \star W_{\hat{A}} = \left(\pazocal{A}_0 - \epsilon \pazocal{A}_1 + \epsilon^2 \pazocal{A}_2\right)\circ,
	\label{eq:bopp_leftright}
\end{equation}
i.e.\ they are related by $\epsilon \to -\epsilon$, which is why the two cases are covered by a single sign convention in Table~\ref{tab:bopp}.
It follows that only odd powers of $\epsilon$ survive in a star commutator and only even powers in a star anticommutator.

\begin{table*}[t]
\renewcommand{\arraystretch}{1.8}
\setlength{\tabcolsep}{8pt}
\begin{tabular}{ |c|c|c|c|c|c|c| }
\hline
$\hat{A}$ & $\pazocal{P}_{\hat{A}}$ & $1$ & $\partial_\theta$ & $\partial_\phi$ & $\partial_\theta^2$ & $\partial_\phi^2$ \\
\hline
$\hat{S}^+$ & $\pazocal{S}\, e^{i\phi}$ & $\sin(\theta)$
  & $\pm\epsilon + \epsilon^2\cos(\theta)$
  & $\left(\pm i\epsilon\, \cos(\theta) + \tfrac{i}{2}\epsilon^2\right)\csc(\theta)$
  & $\tfrac{1}{2}\epsilon^2 \sin(\theta)$
  & $\tfrac{1}{2}\epsilon^2 \csc(\theta)$ \\
$\hat{S}^-$ & $\pazocal{S}\, e^{-i\phi}$ & $\sin(\theta)$
  & $\mp\epsilon + \epsilon^2\cos\theta$
  & $\left(\pm i\epsilon\, \cos(\theta) - \tfrac{i}{2}\epsilon^2\right)\csc(\theta)$
  & $\tfrac{1}{2}\epsilon^2\sin(\theta)$
  & $\tfrac{1}{2}\epsilon^2\csc(\theta)$ \\
$\hat{S}^z$ & $\pazocal{S}$ & $\cos(\theta)$
  & $\tfrac{1}{2}\epsilon^2\cos(2\theta)\csc(\theta)$
  & $\mp i\epsilon$
  & $\tfrac{1}{2}\epsilon^2\cos(\theta)$
  & $\tfrac{1}{2}\epsilon^2 \cos(\theta)\csc^2(\theta)$ \\
\hline
\end{tabular}
\caption{
    Bopp representation of the spin raising, lowering and $z$-projection operators to $O(\epsilon^2)$ in the canonical coordinates $\theta$ and $\phi$, defined through Eq.~\eqref{eq:bopp_def}.
    Columns label the derivative monomial $\partial^a$ acting on an arbitrary symbol $\circ$, and $\pazocal{P}_{\hat{A}}$ is a common prefactor of the row, with $\pazocal{S} = \sqrt{S(S+1)}$.
    The upper sign applies to left star multiplication $W_{\hat{A}} \star \circ$ and the lower sign to right star multiplication $\circ \star W_{\hat{A}}$, the two being related by $\epsilon \to -\epsilon$ as in Eq.~\eqref{eq:bopp_leftright}.}
\label{tab:bopp}
\end{table*}

\subsection{Tensor products of star products} \label{tensorstarproduct}

Combining independent subsystems requires the tensor product of the individual star products.
Since each is a bidifferential operator built from products of partial derivatives, the tensor product simply multiplies pairs of such operators while tracking which side (via the left ($\leftarrow$) or right ($\rightarrow$) symbols) and variable of the (infix) star product each factor acts on.
For instance, $\lpartial\!_{x_1}\rpartial\!_{p_1} \otimes \lpartial\!_{p_2}\rpartial\!_{x_2} = \lpartial\!_{x_1 p_2}\rpartial\!_{p_1 x_2}$.
As an illustration, a single bosonic mode has star product $\star^b = e^{i\lambda \{\leftarrow,\rightarrow\}}$, with $\{\cdot,\cdot\}$ the canonical Poisson bracket; to $O(\hbar^2)$,
\begin{equation}
    \star^b = 1 + i\hbar \{\leftarrow,\rightarrow\} - \frac{\hbar^2}{2} \underbrace{\{\leftarrow,\rightarrow\}^2}_{\mathclap{\lpartial\!_{xx}\rpartial\!_{pp} - 2\lpartial\!_{xp}\rpartial\!_{px} + \lpartial\!_{pp}\rpartial\!_{xx}}} + O(\hbar^3).
\end{equation}
For two bosonic modes labelled $1,2$, with individual Poisson brackets $\{\leftarrow,\rightarrow\}_i$,
\begin{equation}
	\begin{split}
		\star^b_1 \otimes \star^b_2 = &\ 1 + i \lambda \{\leftarrow, \rightarrow\}_1 \otimes 1 + 1\otimes i \lambda \{\leftarrow, \rightarrow\}_2\\
		&- \frac{\lambda^2}{2} \{\leftarrow, \rightarrow\}_1^2 \otimes 1 - 1 \otimes \frac{\lambda^2}{2} \{\leftarrow, \rightarrow\}_2^2\\
		&- \lambda^2 \underbrace{\{\leftarrow, \rightarrow\}_1\otimes\{\leftarrow, \rightarrow\}_2}_{\mathclap{\lpartial\!_{x_1x_2} \rpartial\!_{p_1p_2} -  \lpartial\!_{x_1p_2} \rpartial\!_{p_1x_2} - \lpartial\!_{p_1x_2} \rpartial\!_{x_1p_2}+ \lpartial\!_{p_1p_2}\rpartial\!_{x_1x_2}}} + O(\lambda^3)\\
		= &\ 1 + i \lambda \{\leftarrow, \rightarrow\} - \frac{\lambda^2}{2} \{\leftarrow, \rightarrow\}^2 + O(\lambda^3), \qquad \{\leftarrow, \rightarrow\} \equiv \sum_i \{\leftarrow, \rightarrow\}_i,
	\end{split}
\end{equation}
in agreement with the exact many-body result $\star^b_{\rm mb} = e^{i\lambda \{\leftarrow,\rightarrow\}}$.
The same bookkeeping applies to Eq.~\eqref{eq:star_compact}.
Denoting the single-site Poisson bracket and $\pazocal{B}$ operator by $\{\leftarrow,\rightarrow\}_j$ and $\pazocal{B}_j$, the star product for a collection of spins indexed by $j \in J$ is
\begin{equation}
	\star = \bigotimes_{j \in J} \star_j = 1 + i \epsilon \{\leftarrow, \rightarrow\} - \epsilon^2 \sum_{\substack{j,k \in J\\j < k}} \{\leftarrow, \rightarrow\}_j \otimes \{\leftarrow, \rightarrow\}_k + \epsilon^2 \sum_{j \in J} \pazocal{B}_j(\leftarrow, \rightarrow) + O(\epsilon^3),
	\label{eq:star_multispin}
\end{equation}
with overall bracket $\{\leftarrow,\rightarrow\} = \sum_{j\in J} \{\leftarrow,\rightarrow\}_j$; this is the result quoted in the main text.

Finally, we can also write the expression for the star product for combined spin--boson systems.
Indexing bosonic modes by $i \in I$ and spins by $j \in J$, and writing
\begin{equation}
	p_i(\leftarrow, \rightarrow) =
	\begin{cases}
		\hbar\, \{\leftarrow, \rightarrow\}_i, & i \in I,\\[2pt]
		\epsilon\, \{\leftarrow, \rightarrow\}_i, & i \in J,
	\end{cases}
\end{equation}
the star product becomes
\begin{equation}
	\begin{split}
		\star^b \otimes \star = &\ 1 + i \hbar \{\leftarrow, \rightarrow\} + i \epsilon \{\leftarrow, \rightarrow\}
		-\frac{1}{2}\sum_{\substack{i,k \in I \cup J \\ i \neq k}} p_i(\leftarrow, \rightarrow) \otimes p_k(\leftarrow, \rightarrow)\\
		&-\frac{\hbar^2}{2} \sum_{i\in I} \{\leftarrow, \rightarrow\}^2_i + \epsilon^2 \sum_{j \in J} \pazocal{B}_j(\leftarrow, \rightarrow) + O(3),
	\end{split}
	\label{eq:star_combined}
\end{equation}
where $O(3)$ denotes terms of combined degree greater than two in $\hbar,\epsilon$.
Since both parameters are physically of order $\hbar$, Eq.~\eqref{eq:star_combined} is the star product to second order in $\hbar$ for an arbitrary collection of spins and bosonic modes.

\end{widetext}

\section{Derivation of the Fokker--Planck equation for the LMG model}
\label{app:LMG}

Here we derive Eqs.~\eqref{eq:LMG_drift}--\eqref{eq:LMG_diff} from the Lindblad QME \eqref{eq:LMG_QME_FULL}, restated here in terms of the measure-weighted Wigner distribution:
\begin{equation}
\begin{split}
    \frac{\dd}{\dd \tau}\overline{W}_{\hat{\rho}} &= -i\sin(\theta)\left[-\tilde{J}W_{\hat{S}^x} \star W_{\hat{S}^x} + \tilde{\Delta}W_{\hat{S}^z}, \frac{\overline{W}_{\hat{\rho}}}{\sin(\theta)}\right]_{\star} \\ &\qquad + \sin(\theta)\,
    \left(W_{\hat{S}^-} \star \frac{\overline{W}_{\hat{\rho}}}{\sin(\theta)} \star W_{\hat{S}^+}\right)\\
    &\qquad - \frac{\sin(\theta)}{2}\left\{W_{\hat{S}^+} \star W_{\hat{S}^-}, \frac{\overline{W}_{\hat{\rho}}}{\sin(\theta)}\right\}_{\star}.
\end{split}
\end{equation}
This will demonstrate the general structure of the approach, as well as why expansion of the star product to $O(\epsilon^2)$ rather than simply $O(\epsilon)$ is essential for the driven-dissipative setting.
We will need the primitive Wigner symbols
\begin{gather}
    W_{\hat{S}^{\pm}}
    = \pazocal{S}\,\sin(\theta)\, e^{\pm i\phi},\\
     W_{\hat{S}^{x}}
    = \pazocal{S}\,\sin(\theta)\cos\phi, \\
    W_{\hat{S}^{z}}
    = \pazocal{S}\,\cos(\theta),
    \label{eq:LMG_symbols}
\end{gather}
and find it helpful to introduce the selection functional
\begin{equation}
\pazocal{T}_C\left[\sum_{j} \epsilon^j \, c_j(\theta, \phi)\right] = \sum_{j\in C} \epsilon^j \, c_j(\theta, \phi)
\end{equation}
that picks out a specific set of powers of $\epsilon$ from an expansion.
With these, we may proceed through the terms of Eq.~\eqref{eq:wigner-lindblad}, building up the resulting Fokker--Planck equation as we do.

\subsection{Star commutator}

We begin with the coherent contribution,
\begin{equation}
    -i\sin(\theta)\left[-\tilde{J}W_{\hat{S}^x} \star W_{\hat{S}^x} + \tilde{\Delta}W_{\hat{S}^z}, \frac{\overline{W}_{\hat{\rho}}}{\sin(\theta)}\right]_{\star}.
\end{equation}
From Appendix~\ref{app:starproduct}, we know that only the $O(\epsilon)$ term of the star product expansion is antisymmetric to $O(\epsilon^2)$, and is thus the only one to contribute to the star commutator.
This allows us to rewrite the above simply in terms of the Poisson bracket:
\begin{equation}
    2\epsilon\sin(\theta)\left\{-\tilde{J}W_{\hat{S}^x} \star W_{\hat{S}^x} + \tilde{\Delta}W_{\hat{S}^z}, \frac{\overline{W}_{\hat{\rho}}}{\sin(\theta)}\right\}.
\end{equation}
Moreover, if we further expand the term $W_{\hat{S}^x} \star W_{\hat{S}^x}$ \footnote{This symbol is not difficult to calculate exactly, and is given in Ref.~\cite{VARILLY1989107}, but this approach demonstrates a useful simplification technique for more complicated symbols.}, it is clear that
\begin{equation}
\begin{split}
    &\pazocal{T}_{\{0,1,2\}}\left[2\epsilon\sin(\theta)\left\{-\tilde{J}W_{\hat{S}^x} \star W_{\hat{S}^x} + \tilde{\Delta}W_{\hat{S}^z}, \frac{\overline{W}_{\hat{\rho}}}{\sin(\theta)}\right\}\right]\\ &= 2\epsilon\sin(\theta)\left\{-\tilde{J}\pazocal{T}_{\{0,1\}}\left[W_{\hat{S}^x} \star W_{\hat{S}^x}\right] + \tilde{\Delta}W_{\hat{S}^z}, \frac{\overline{W}_{\hat{\rho}}}{\sin(\theta)}\right\}.
\end{split}
\end{equation}
As the Poisson bracket is antisymmetric, $\pazocal{T}_{\{0,1\}}\left[W_{\hat{S}^x} \star W_{\hat{S}^x}\right] = W_{\hat{S}^x}  W_{\hat{S}^x}$, so finally the coherent contribution simplifies to
\begin{equation}
    2\epsilon\sin(\theta)\left\{-\tilde{J}W_{\hat{S}^x} W_{\hat{S}^x} + \tilde{\Delta}W_{\hat{S}^z}, \frac{\overline{W}_{\hat{\rho}}}{\sin(\theta)}\right\}
\end{equation}
at $O(\epsilon^2)$.
Expanding the Poisson bracket and regrouping terms, we arrive at
\begin{equation}
    -\partial_\theta \left[\mu_{\text{sc}, \theta} \overline{W}_{\hat{\rho}}\right] -\partial_\phi \left[\mu_{\text{sc}, \phi} \overline{W}_{\hat{\rho}}\right],
\end{equation}
with $\lambda = 2 \epsilon \pazocal{S}$ and
\begin{equation}
    \mu_\text{sc} = \begin{pmatrix} \lambda\tilde{J}\pazocal{S}\sin(\theta) \sin(2\phi)\\[2pt]
        \lambda\tilde{\Delta} + 2\lambda\tilde{J}\pazocal{S}\,\cos(\theta)\cos^2(\phi)
    \end{pmatrix}.
\end{equation}
We observe that this is part, but not all, of Eq.~\eqref{eq:LMG_drift}: two terms, including the $\cot(\theta)$ term responsible for regularizing the final stochastic process at the poles, are absent and will be seen to originate from the dissipator.
We see that, for purely coherent evolution, the expansion of the star product to $O(\epsilon)$, as in Ref.~\cite{Klimov_2002}, is sufficient and yields a consistent Liouville evolution equation.

\subsection{Recycling term}

Proceeding to the recycling term
\begin{equation}
    \sin(\theta)\,
    \left(W_{\hat{S}^-} \star \frac{\overline{W}_{\hat{\rho}}}{\sin(\theta)} \star W_{\hat{S}^+}\right),
\end{equation}
we observe immediately that at $O(\epsilon^0)$ the star products are simply commutative scalar multiplication, and that all such terms will cancel exactly against those coming from the star anticommutator term considered in the next section.
We are thus immediately justified in restricting our attention to
\begin{equation}
\sin(\theta)\,\pazocal{T}_{\{1,2\}}\left[
    W_{\hat{S}^-} \star \frac{\overline{W}_{\hat{\rho}}}{\sin(\theta)} \star W_{\hat{S}^+}\right].
\end{equation}
Using the star product expansion of Appendix~\ref{app:starproduct}, we can see that the truncation is equivalent to
\begin{equation}
\begin{aligned}
&\begin{aligned}[t]
    \pazocal{T}_{\{1,2\}}&\left[W_{\hat{S}^-} \star \frac{\overline{W}_{\hat{\rho}}}{\sin(\theta)}\right] W_{\hat{S}^+}\\ &+ i\epsilon \left\{\pazocal{T}_{\{0,1\}}\left[W_{\hat{S}^-} \star \frac{\overline{W}_{\hat{\rho}}}{\sin(\theta)}\right], W_{\hat{S}^+}\right\}\\
    &+ \epsilon^2 \pazocal{B}\left( \pazocal{T}_{\{0\}}\left[W_{\hat{S}^-} \star \frac{\overline{W}_{\hat{\rho}}}{\sin(\theta)}\right], W_{\hat{S}^+}\right)
\end{aligned}\\
&=\begin{aligned}[t]
    &i \epsilon \left\{W_{\hat{S}^-},  \frac{\overline{W}_{\hat{\rho}}}{\sin(\theta)}\right\} W_{\hat{S}^+} +\epsilon^2 \pazocal{B}\left(W_{\hat{S}^-},  \frac{\overline{W}_{\hat{\rho}}}{\sin(\theta)}\right) W_{\hat{S}^+}
    \\& + i \epsilon \left\{W_{\hat{S}^-}  \frac{\overline{W}_{\hat{\rho}}}{\sin(\theta)}, W_{\hat{S}^+}\right\} \\ &-\epsilon^2 \left\{\left\{W_{\hat{S}^-}, \frac{\overline{W}_{\hat{\rho}}}{\sin(\theta)}\right\}, W_{\hat{S}^+}\right\} \\
    &+ \epsilon^2 \pazocal{B} \left( W_{\hat{S}^-} \frac{\overline{W}_{\hat{\rho}}}{\sin(\theta)}, W_{\hat{S}^+}\right).
\end{aligned}
\end{aligned} \label{rtexp}
\end{equation}
Taking into account the leading $\sin(\theta)$, the first and third of these terms together may be rewritten as
\begin{equation}
    2 \pazocal{S}^2 \epsilon \cos(\theta)\overline{W}_{\hat{\rho}} -\partial_j \left[\mu_{\text{rt}, j} \overline{W}_{\hat{\rho}}\right], \quad \mu_\text{\text{rt}} = \begin{pmatrix} \lambda\pazocal{S}\sin(\theta) \\[2pt]
        0
    \end{pmatrix}.
    \label{eq:murt1}
\end{equation}
This is the last of the $O(\epsilon)$ contributions to the drift Eq.~\eqref{eq:LMG_drift}, and we will see that the leading non-derivative term cancels with one in the next section.
As expected, the dissipator contributes a drift driving the trajectory towards the lower pole, near which the spin-down state has more probability mass.

Turning to the fourth term, it is worth noting that it is the only term of $O(\epsilon^2)$ that can be generated in this case by an expansion of the star product to only $O(\epsilon)$, and the only such term to yield second-order derivatives of $\overline{W}_{\hat{\rho}}$.
Were it possible to obtain a valid Fokker--Planck equation with non-zero diffusion by such a first-order expansion, this term would yield the diffusion component.
Upon expanding the nested Poisson brackets, however, one finds the following contribution:
\begin{equation}
    i \pazocal{S}^2 \epsilon^2 \csc(\theta) \partial_\phi \overline{W}_{\hat{\rho}}.
\end{equation}
This is purely imaginary, so it cannot be a valid component of a Fokker--Planck equation, and it is not cancelled by any other term obtainable by an $O(\epsilon)$ expansion of the star product (we will see it is cancelled by a combination of $\pazocal{B}$ terms).
We see, therefore, that one cannot obtain a consistent Fokker--Planck equation from a Lindbladian QME without expanding the star product to $O(\epsilon^2)$ as we have done.
This serves as an interesting contrast with the results of Ref.~\cite{zueco2007bopp}, where Bopp operators calculated only to $O(\epsilon)$ appeared sufficient to obtain a valid Fokker--Planck equation from a generalized Caldeira--Leggett master equation.
The Caldeira--Leggett equation is not a valid Lindbladian QME~\cite{breuer2002theory}, so the results are not contradictory.

Taken together, the second and fifth terms (arising from the $O(\epsilon^2)$ expansion of the star product) cancel off the imaginary contribution from the fourth.
Nevertheless, the remaining terms cannot be brought into a valid Fokker--Planck form on their own, so we proceed to the final term of the QME.

\subsection{Star anticommutator}

Finally, we consider the anticommutator contribution
\begin{equation}
    - \frac{\sin(\theta)}{2}\left\{W_{\hat{S}^+} \star W_{\hat{S}^-}, \frac{\overline{W}_{\hat{\rho}}}{\sin(\theta)}\right\}_{\star}.
\end{equation}
As already mentioned, the $O(\epsilon^0)$ terms cancel trivially against those in the recycling term, and we are left with
\begin{equation}
\begin{aligned}
    &\pazocal{T}_{\{1,2\}}\left[- \frac{\sin(\theta)}{2}\left\{W_{\hat{S}^+} \star W_{\hat{S}^-}, \frac{\overline{W}_{\hat{\rho}}}{\sin(\theta)}\right\}_{\star}\right]\\
    &=
    \begin{aligned}[t]
    &-i\epsilon\sin(\theta)  \left\{W_{\hat{S}^+},W_{\hat{S}^-} \right\}\frac{\overline{W}_{\hat{\rho}}}{\sin(\theta)}\\
    &-\epsilon^2\sin(\theta)  \pazocal{B}\left(W_{\hat{S}^+},W_{\hat{S}^-} \right)\frac{\overline{W}_{\hat{\rho}}}{\sin(\theta)}\\
    &-\epsilon^2\sin(\theta)\pazocal{B}  \left(W_{\hat{S}^+}W_{\hat{S}^-},  \frac{\overline{W}_{\hat{\rho}}}{\sin(\theta)} \right).
    \end{aligned}
\end{aligned}
\end{equation}
The first of these terms, at $O(\epsilon)$, simplifies to
\begin{equation}
    -2\pazocal{S}^2\epsilon\cos(\theta)\overline{W}_{\hat{\rho}},
\end{equation}
which is immediately seen to simply cancel the first term in Eq.~\eqref{eq:murt1}.

The final two terms, on their own, again cannot be brought to a valid Fokker--Planck form.
Combining them with the as-yet unincorporated second, fourth, and fifth terms of Eq.~\eqref{rtexp}, however, yields the following contribution:
\begin{equation}
    -\partial_j\left[\mu_{\text{rtac}, j} \overline{W}_{\hat{\rho}}\right] - \frac12 \partial_j\partial_k \left[ D_{\text{rtac},jk} \overline{W}_{\hat{\rho}}\right],
\end{equation}
with
\begin{equation}
    \mu_\text{\text{rtac}} = \begin{pmatrix} \frac12 \lambda^2 \cot(\theta) \\[2pt]
        0
    \end{pmatrix}, \quad D_{\text{rtac}} = \begin{pmatrix}
\lambda^2 & 0 \\
0 & \lambda^2 \cot^2(\theta)
    \end{pmatrix}.
\end{equation}
We see that $\mu_\text{sc} + \mu_\text{rt} + \mu_\text{rtac}$ now exactly matches Eq.~\eqref{eq:LMG_drift}, while $D_\text{rtac}$ is exactly Eq.~\eqref{eq:LMG_diff}, so the derivation is complete.
Note that both the recycling term and the anticommutator were required to obtain the correct diffusion matrix and drift regularizing that diffusion.

\bibliography{references.bib}

\begin{thebibliography}{120}%
\makeatletter
\providecommand \@ifxundefined [1]{%
 \@ifx{#1\undefined}
}%
\providecommand \@ifnum [1]{%
 \ifnum #1\expandafter \@firstoftwo
 \else \expandafter \@secondoftwo
 \fi
}%
\providecommand \@ifx [1]{%
 \ifx #1\expandafter \@firstoftwo
 \else \expandafter \@secondoftwo
 \fi
}%
\providecommand \natexlab [1]{#1}%
\providecommand \enquote  [1]{``#1''}%
\providecommand \bibnamefont  [1]{#1}%
\providecommand \bibfnamefont [1]{#1}%
\providecommand \citenamefont [1]{#1}%
\providecommand \href@noop [0]{\@secondoftwo}%
\providecommand \href [0]{\begingroup \@sanitize@url \@href}%
\providecommand \@href[1]{\@@startlink{#1}\@@href}%
\providecommand \@@href[1]{\endgroup#1\@@endlink}%
\providecommand \@sanitize@url [0]{\catcode `\\12\catcode `\$12\catcode
  `\&12\catcode `\#12\catcode `\^12\catcode `\_12\catcode `\%12\relax}%
\providecommand \@@startlink[1]{}%
\providecommand \@@endlink[0]{}%
\providecommand \url  [0]{\begingroup\@sanitize@url \@url }%
\providecommand \@url [1]{\endgroup\@href {#1}{\urlprefix }}%
\providecommand \urlprefix  [0]{URL }%
\providecommand \Eprint [0]{\href }%
\providecommand \doibase [0]{https://doi.org/}%
\providecommand \selectlanguage [0]{\@gobble}%
\providecommand \bibinfo  [0]{\@secondoftwo}%
\providecommand \bibfield  [0]{\@secondoftwo}%
\providecommand \translation [1]{[#1]}%
\providecommand \BibitemOpen [0]{}%
\providecommand \bibitemStop [0]{}%
\providecommand \bibitemNoStop [0]{.\EOS\space}%
\providecommand \EOS [0]{\spacefactor3000\relax}%
\providecommand \BibitemShut  [1]{\csname bibitem#1\endcsname}%
\let\auto@bib@innerbib\@empty
\bibitem [{\citenamefont {Carusotto}\ and\ \citenamefont
  {Ciuti}(2013)}]{Carusotto2013}%
  \BibitemOpen
  \bibfield  {author} {\bibinfo {author} {\bibfnamefont {I.}~\bibnamefont
  {Carusotto}}\ and\ \bibinfo {author} {\bibfnamefont {C.}~\bibnamefont
  {Ciuti}},\ }\bibfield  {title} {\bibinfo {title} {Quantum fluids of light},\
  }\href {https://doi.org/10.1103/RevModPhys.85.299} {\bibfield  {journal}
  {\bibinfo  {journal} {Rev. Mod. Phys.}\ }\textbf {\bibinfo {volume} {85}},\
  \bibinfo {pages} {299} (\bibinfo {year} {2013})}\BibitemShut {NoStop}%
\bibitem [{\citenamefont {Sieberer}\ \emph {et~al.}(2016)\citenamefont
  {Sieberer}, \citenamefont {Buchhold},\ and\ \citenamefont
  {Diehl}}]{Sieberer2016}%
  \BibitemOpen
  \bibfield  {author} {\bibinfo {author} {\bibfnamefont {L.~M.}\ \bibnamefont
  {Sieberer}}, \bibinfo {author} {\bibfnamefont {M.}~\bibnamefont {Buchhold}},\
  and\ \bibinfo {author} {\bibfnamefont {S.}~\bibnamefont {Diehl}},\ }\bibfield
   {title} {\bibinfo {title} {Keldysh field theory for driven open quantum
  systems},\ }\href {https://doi.org/10.1088/0034-4885/79/9/096001} {\bibfield
  {journal} {\bibinfo  {journal} {Rep. Prog. Phys.}\ }\textbf {\bibinfo
  {volume} {79}},\ \bibinfo {pages} {096001} (\bibinfo {year}
  {2016})}\BibitemShut {NoStop}%
\bibitem [{\citenamefont {Fazio}\ \emph {et~al.}(2025)\citenamefont {Fazio},
  \citenamefont {Keeling}, \citenamefont {Mazza},\ and\ \citenamefont
  {Schir\`o}}]{Fazio2025}%
  \BibitemOpen
  \bibfield  {author} {\bibinfo {author} {\bibfnamefont {R.}~\bibnamefont
  {Fazio}}, \bibinfo {author} {\bibfnamefont {J.}~\bibnamefont {Keeling}},
  \bibinfo {author} {\bibfnamefont {L.}~\bibnamefont {Mazza}},\ and\ \bibinfo
  {author} {\bibfnamefont {M.}~\bibnamefont {Schir\`o}},\ }\bibfield  {title}
  {\bibinfo {title} {Many-body open quantum systems},\ }\href
  {https://doi.org/10.21468/SciPostPhysLectNotes.99} {\bibfield  {journal}
  {\bibinfo  {journal} {SciPost Phys. Lect. Notes}\ }\textbf {\bibinfo {volume}
  {99}} (\bibinfo {year} {2025})}\BibitemShut {NoStop}%
\bibitem [{\citenamefont {Sieberer}\ \emph {et~al.}(2025)\citenamefont
  {Sieberer}, \citenamefont {Buchhold}, \citenamefont {Marino},\ and\
  \citenamefont {Diehl}}]{Sieberer2025}%
  \BibitemOpen
  \bibfield  {author} {\bibinfo {author} {\bibfnamefont {L.~M.}\ \bibnamefont
  {Sieberer}}, \bibinfo {author} {\bibfnamefont {M.}~\bibnamefont {Buchhold}},
  \bibinfo {author} {\bibfnamefont {J.}~\bibnamefont {Marino}},\ and\ \bibinfo
  {author} {\bibfnamefont {S.}~\bibnamefont {Diehl}},\ }\bibfield  {title}
  {\bibinfo {title} {Universality in driven open quantum matter},\ }\href
  {https://doi.org/10.1103/RevModPhys.97.025004} {\bibfield  {journal}
  {\bibinfo  {journal} {Rev. Mod. Phys.}\ }\textbf {\bibinfo {volume} {97}},\
  \bibinfo {pages} {025004} (\bibinfo {year} {2025})}\BibitemShut {NoStop}%
\bibitem [{\citenamefont {Baumann}\ \emph {et~al.}(2010)\citenamefont
  {Baumann}, \citenamefont {Guerlin}, \citenamefont {Brennecke},\ and\
  \citenamefont {Esslinger}}]{Baumann2010}%
  \BibitemOpen
  \bibfield  {author} {\bibinfo {author} {\bibfnamefont {K.}~\bibnamefont
  {Baumann}}, \bibinfo {author} {\bibfnamefont {C.}~\bibnamefont {Guerlin}},
  \bibinfo {author} {\bibfnamefont {F.}~\bibnamefont {Brennecke}},\ and\
  \bibinfo {author} {\bibfnamefont {T.}~\bibnamefont {Esslinger}},\ }\bibfield
  {title} {\bibinfo {title} {{D}icke quantum phase transition with a superfluid
  gas in an optical cavity},\ }\href {https://doi.org/10.1038/nature09009}
  {\bibfield  {journal} {\bibinfo  {journal} {Nature}\ }\textbf {\bibinfo
  {volume} {464}},\ \bibinfo {pages} {1301} (\bibinfo {year}
  {2010})}\BibitemShut {NoStop}%
\bibitem [{\citenamefont {Kirton}\ and\ \citenamefont
  {Keeling}(2017)}]{Kirton2017}%
  \BibitemOpen
  \bibfield  {author} {\bibinfo {author} {\bibfnamefont {P.}~\bibnamefont
  {Kirton}}\ and\ \bibinfo {author} {\bibfnamefont {J.}~\bibnamefont
  {Keeling}},\ }\bibfield  {title} {\bibinfo {title} {Suppressing and restoring
  the {D}icke superradiance transition by dephasing and decay},\ }\href
  {https://doi.org/10.1103/PhysRevLett.118.123602} {\bibfield  {journal}
  {\bibinfo  {journal} {Phys. Rev. Lett.}\ }\textbf {\bibinfo {volume} {118}},\
  \bibinfo {pages} {123602} (\bibinfo {year} {2017})}\BibitemShut {NoStop}%
\bibitem [{\citenamefont {Diehl}\ \emph {et~al.}(2008)\citenamefont {Diehl},
  \citenamefont {Micheli}, \citenamefont {Kantian}, \citenamefont {Kraus},
  \citenamefont {B\"uchler},\ and\ \citenamefont {Zoller}}]{Diehl2008}%
  \BibitemOpen
  \bibfield  {author} {\bibinfo {author} {\bibfnamefont {S.}~\bibnamefont
  {Diehl}}, \bibinfo {author} {\bibfnamefont {A.}~\bibnamefont {Micheli}},
  \bibinfo {author} {\bibfnamefont {A.}~\bibnamefont {Kantian}}, \bibinfo
  {author} {\bibfnamefont {B.}~\bibnamefont {Kraus}}, \bibinfo {author}
  {\bibfnamefont {H.~P.}\ \bibnamefont {B\"uchler}},\ and\ \bibinfo {author}
  {\bibfnamefont {P.}~\bibnamefont {Zoller}},\ }\bibfield  {title} {\bibinfo
  {title} {Quantum states and phases in driven open quantum systems with cold
  atoms},\ }\href {https://doi.org/10.1038/nphys1073} {\bibfield  {journal}
  {\bibinfo  {journal} {Nat. Phys.}\ }\textbf {\bibinfo {volume} {4}},\
  \bibinfo {pages} {878} (\bibinfo {year} {2008})}\BibitemShut {NoStop}%
\bibitem [{\citenamefont {Carr}\ \emph {et~al.}(2013)\citenamefont {Carr},
  \citenamefont {Ritter}, \citenamefont {Wade}, \citenamefont {Adams},\ and\
  \citenamefont {Weatherill}}]{Carr2013}%
  \BibitemOpen
  \bibfield  {author} {\bibinfo {author} {\bibfnamefont {C.}~\bibnamefont
  {Carr}}, \bibinfo {author} {\bibfnamefont {R.}~\bibnamefont {Ritter}},
  \bibinfo {author} {\bibfnamefont {C.~G.}\ \bibnamefont {Wade}}, \bibinfo
  {author} {\bibfnamefont {C.~S.}\ \bibnamefont {Adams}},\ and\ \bibinfo
  {author} {\bibfnamefont {K.~J.}\ \bibnamefont {Weatherill}},\ }\bibfield
  {title} {\bibinfo {title} {Nonequilibrium phase transition in a dilute
  {R}ydberg ensemble},\ }\href {https://doi.org/10.1103/PhysRevLett.111.113901}
  {\bibfield  {journal} {\bibinfo  {journal} {Phys. Rev. Lett.}\ }\textbf
  {\bibinfo {volume} {111}},\ \bibinfo {pages} {113901} (\bibinfo {year}
  {2013})}\BibitemShut {NoStop}%
\bibitem [{\citenamefont {Fitzpatrick}\ \emph {et~al.}(2017)\citenamefont
  {Fitzpatrick}, \citenamefont {Sundaresan}, \citenamefont {Li}, \citenamefont
  {Koch},\ and\ \citenamefont {Houck}}]{Fitzpatrick2017}%
  \BibitemOpen
  \bibfield  {author} {\bibinfo {author} {\bibfnamefont {M.}~\bibnamefont
  {Fitzpatrick}}, \bibinfo {author} {\bibfnamefont {N.~M.}\ \bibnamefont
  {Sundaresan}}, \bibinfo {author} {\bibfnamefont {A.~C.~Y.}\ \bibnamefont
  {Li}}, \bibinfo {author} {\bibfnamefont {J.}~\bibnamefont {Koch}},\ and\
  \bibinfo {author} {\bibfnamefont {A.~A.}\ \bibnamefont {Houck}},\ }\bibfield
  {title} {\bibinfo {title} {Observation of a dissipative phase transition in a
  one-dimensional circuit {QED} lattice},\ }\href
  {https://doi.org/10.1103/PhysRevX.7.011016} {\bibfield  {journal} {\bibinfo
  {journal} {Phys. Rev. X}\ }\textbf {\bibinfo {volume} {7}},\ \bibinfo {pages}
  {011016} (\bibinfo {year} {2017})}\BibitemShut {NoStop}%
\bibitem [{\citenamefont {Macieszczak}\ \emph {et~al.}(2016)\citenamefont
  {Macieszczak}, \citenamefont {Gu\c{t}\u{a}}, \citenamefont {Lesanovsky},\
  and\ \citenamefont {Garrahan}}]{Macieszczak2016}%
  \BibitemOpen
  \bibfield  {author} {\bibinfo {author} {\bibfnamefont {K.}~\bibnamefont
  {Macieszczak}}, \bibinfo {author} {\bibfnamefont {M.}~\bibnamefont
  {Gu\c{t}\u{a}}}, \bibinfo {author} {\bibfnamefont {I.}~\bibnamefont
  {Lesanovsky}},\ and\ \bibinfo {author} {\bibfnamefont {J.~P.}\ \bibnamefont
  {Garrahan}},\ }\bibfield  {title} {\bibinfo {title} {Towards a theory of
  metastability in open quantum dynamics},\ }\href
  {https://doi.org/10.1103/PhysRevLett.116.240404} {\bibfield  {journal}
  {\bibinfo  {journal} {Phys. Rev. Lett.}\ }\textbf {\bibinfo {volume} {116}},\
  \bibinfo {pages} {240404} (\bibinfo {year} {2016})}\BibitemShut {NoStop}%
\bibitem [{\citenamefont {Schau\ss{}}\ \emph {et~al.}(2012)\citenamefont
  {Schau\ss{}}, \citenamefont {Cheneau}, \citenamefont {Endres}, \citenamefont
  {Fukuhara}, \citenamefont {Hild}, \citenamefont {Omran}, \citenamefont
  {Pohl}, \citenamefont {Gross}, \citenamefont {Kuhr},\ and\ \citenamefont
  {Bloch}}]{Schauss2012}%
  \BibitemOpen
  \bibfield  {author} {\bibinfo {author} {\bibfnamefont {P.}~\bibnamefont
  {Schau\ss{}}}, \bibinfo {author} {\bibfnamefont {M.}~\bibnamefont {Cheneau}},
  \bibinfo {author} {\bibfnamefont {M.}~\bibnamefont {Endres}}, \bibinfo
  {author} {\bibfnamefont {T.}~\bibnamefont {Fukuhara}}, \bibinfo {author}
  {\bibfnamefont {S.}~\bibnamefont {Hild}}, \bibinfo {author} {\bibfnamefont
  {A.}~\bibnamefont {Omran}}, \bibinfo {author} {\bibfnamefont
  {T.}~\bibnamefont {Pohl}}, \bibinfo {author} {\bibfnamefont {C.}~\bibnamefont
  {Gross}}, \bibinfo {author} {\bibfnamefont {S.}~\bibnamefont {Kuhr}},\ and\
  \bibinfo {author} {\bibfnamefont {I.}~\bibnamefont {Bloch}},\ }\bibfield
  {title} {\bibinfo {title} {Observation of spatially ordered structures in a
  two-dimensional {R}ydberg gas},\ }\href {https://doi.org/10.1038/nature11596}
  {\bibfield  {journal} {\bibinfo  {journal} {Nature}\ }\textbf {\bibinfo
  {volume} {491}},\ \bibinfo {pages} {87} (\bibinfo {year} {2012})}\BibitemShut
  {NoStop}%
\bibitem [{\citenamefont {Lee}\ \emph {et~al.}(2011)\citenamefont {Lee},
  \citenamefont {H\"affner},\ and\ \citenamefont {Cross}}]{Lee2011}%
  \BibitemOpen
  \bibfield  {author} {\bibinfo {author} {\bibfnamefont {T.~E.}\ \bibnamefont
  {Lee}}, \bibinfo {author} {\bibfnamefont {H.}~\bibnamefont {H\"affner}},\
  and\ \bibinfo {author} {\bibfnamefont {M.~C.}\ \bibnamefont {Cross}},\
  }\bibfield  {title} {\bibinfo {title} {Antiferromagnetic phase transition in
  a nonequilibrium lattice of {R}ydberg atoms},\ }\href
  {https://doi.org/10.1103/PhysRevA.84.031402} {\bibfield  {journal} {\bibinfo
  {journal} {Phys. Rev. A}\ }\textbf {\bibinfo {volume} {84}},\ \bibinfo
  {pages} {031402(R)} (\bibinfo {year} {2011})}\BibitemShut {NoStop}%
\bibitem [{\citenamefont {Lee}\ \emph {et~al.}(2013)\citenamefont {Lee},
  \citenamefont {Gopalakrishnan},\ and\ \citenamefont
  {Lukin}}]{PhysRevLett.110.257204}%
  \BibitemOpen
  \bibfield  {author} {\bibinfo {author} {\bibfnamefont {T.~E.}\ \bibnamefont
  {Lee}}, \bibinfo {author} {\bibfnamefont {S.}~\bibnamefont
  {Gopalakrishnan}},\ and\ \bibinfo {author} {\bibfnamefont {M.~D.}\
  \bibnamefont {Lukin}},\ }\bibfield  {title} {\bibinfo {title} {Unconventional
  magnetism via optical pumping of interacting spin systems},\ }\href
  {https://doi.org/10.1103/PhysRevLett.110.257204} {\bibfield  {journal}
  {\bibinfo  {journal} {Phys. Rev. Lett.}\ }\textbf {\bibinfo {volume} {110}},\
  \bibinfo {pages} {257204} (\bibinfo {year} {2013})}\BibitemShut {NoStop}%
\bibitem [{\citenamefont {Fontaine}\ \emph {et~al.}(2022)\citenamefont
  {Fontaine}, \citenamefont {Squizzato}, \citenamefont {Baboux}, \citenamefont
  {Amelio}, \citenamefont {Lema{\^\i}tre}, \citenamefont {Morassi},
  \citenamefont {Sagnes}, \citenamefont {Le~Gratiet}, \citenamefont {Harouri},
  \citenamefont {Wouters}, \citenamefont {Carusotto}, \citenamefont {Amo},
  \citenamefont {Richard}, \citenamefont {Minguzzi}, \citenamefont {Canet},
  \citenamefont {Ravets},\ and\ \citenamefont {Bloch}}]{Fontaine2022}%
  \BibitemOpen
  \bibfield  {author} {\bibinfo {author} {\bibfnamefont {Q.}~\bibnamefont
  {Fontaine}}, \bibinfo {author} {\bibfnamefont {D.}~\bibnamefont {Squizzato}},
  \bibinfo {author} {\bibfnamefont {F.}~\bibnamefont {Baboux}}, \bibinfo
  {author} {\bibfnamefont {I.}~\bibnamefont {Amelio}}, \bibinfo {author}
  {\bibfnamefont {A.}~\bibnamefont {Lema{\^\i}tre}}, \bibinfo {author}
  {\bibfnamefont {M.}~\bibnamefont {Morassi}}, \bibinfo {author} {\bibfnamefont
  {I.}~\bibnamefont {Sagnes}}, \bibinfo {author} {\bibfnamefont
  {L.}~\bibnamefont {Le~Gratiet}}, \bibinfo {author} {\bibfnamefont
  {A.}~\bibnamefont {Harouri}}, \bibinfo {author} {\bibfnamefont
  {M.}~\bibnamefont {Wouters}}, \bibinfo {author} {\bibfnamefont
  {I.}~\bibnamefont {Carusotto}}, \bibinfo {author} {\bibfnamefont
  {A.}~\bibnamefont {Amo}}, \bibinfo {author} {\bibfnamefont {M.}~\bibnamefont
  {Richard}}, \bibinfo {author} {\bibfnamefont {A.}~\bibnamefont {Minguzzi}},
  \bibinfo {author} {\bibfnamefont {L.}~\bibnamefont {Canet}}, \bibinfo
  {author} {\bibfnamefont {S.}~\bibnamefont {Ravets}},\ and\ \bibinfo {author}
  {\bibfnamefont {J.}~\bibnamefont {Bloch}},\ }\bibfield  {title} {\bibinfo
  {title} {{K}ardar--{P}arisi--{Z}hang universality in a one-dimensional
  polariton condensate},\ }\href {https://doi.org/10.1038/s41586-022-05001-8}
  {\bibfield  {journal} {\bibinfo  {journal} {Nature}\ }\textbf {\bibinfo
  {volume} {608}},\ \bibinfo {pages} {687} (\bibinfo {year}
  {2022})}\BibitemShut {NoStop}%
\bibitem [{\citenamefont {Marcuzzi}\ \emph {et~al.}(2016)\citenamefont
  {Marcuzzi}, \citenamefont {Buchhold}, \citenamefont {Diehl},\ and\
  \citenamefont {Lesanovsky}}]{Marcuzzi2016}%
  \BibitemOpen
  \bibfield  {author} {\bibinfo {author} {\bibfnamefont {M.}~\bibnamefont
  {Marcuzzi}}, \bibinfo {author} {\bibfnamefont {M.}~\bibnamefont {Buchhold}},
  \bibinfo {author} {\bibfnamefont {S.}~\bibnamefont {Diehl}},\ and\ \bibinfo
  {author} {\bibfnamefont {I.}~\bibnamefont {Lesanovsky}},\ }\bibfield  {title}
  {\bibinfo {title} {Absorbing state phase transition with competing quantum
  and classical fluctuations},\ }\href
  {https://doi.org/10.1103/PhysRevLett.116.245701} {\bibfield  {journal}
  {\bibinfo  {journal} {Phys. Rev. Lett.}\ }\textbf {\bibinfo {volume} {116}},\
  \bibinfo {pages} {245701} (\bibinfo {year} {2016})}\BibitemShut {NoStop}%
\bibitem [{\citenamefont {Richerme}\ \emph {et~al.}(2014)\citenamefont
  {Richerme}, \citenamefont {Gong}, \citenamefont {Lee}, \citenamefont {Senko},
  \citenamefont {Smith}, \citenamefont {Foss-Feig}, \citenamefont {Michalakis},
  \citenamefont {Gorshkov},\ and\ \citenamefont {Monroe}}]{Richerme2014}%
  \BibitemOpen
  \bibfield  {author} {\bibinfo {author} {\bibfnamefont {P.}~\bibnamefont
  {Richerme}}, \bibinfo {author} {\bibfnamefont {Z.-X.}\ \bibnamefont {Gong}},
  \bibinfo {author} {\bibfnamefont {A.}~\bibnamefont {Lee}}, \bibinfo {author}
  {\bibfnamefont {C.}~\bibnamefont {Senko}}, \bibinfo {author} {\bibfnamefont
  {J.}~\bibnamefont {Smith}}, \bibinfo {author} {\bibfnamefont
  {M.}~\bibnamefont {Foss-Feig}}, \bibinfo {author} {\bibfnamefont
  {S.}~\bibnamefont {Michalakis}}, \bibinfo {author} {\bibfnamefont {A.~V.}\
  \bibnamefont {Gorshkov}},\ and\ \bibinfo {author} {\bibfnamefont
  {C.}~\bibnamefont {Monroe}},\ }\bibfield  {title} {\bibinfo {title}
  {Non-local propagation of correlations in quantum systems with long-range
  interactions},\ }\href {https://doi.org/10.1038/nature13450} {\bibfield
  {journal} {\bibinfo  {journal} {Nature}\ }\textbf {\bibinfo {volume} {511}},\
  \bibinfo {pages} {198} (\bibinfo {year} {2014})}\BibitemShut {NoStop}%
\bibitem [{\citenamefont {Foss-Feig}\ \emph {et~al.}(2015)\citenamefont
  {Foss-Feig}, \citenamefont {Gong}, \citenamefont {Clark},\ and\ \citenamefont
  {Gorshkov}}]{FossFeig2015}%
  \BibitemOpen
  \bibfield  {author} {\bibinfo {author} {\bibfnamefont {M.}~\bibnamefont
  {Foss-Feig}}, \bibinfo {author} {\bibfnamefont {Z.-X.}\ \bibnamefont {Gong}},
  \bibinfo {author} {\bibfnamefont {C.~W.}\ \bibnamefont {Clark}},\ and\
  \bibinfo {author} {\bibfnamefont {A.~V.}\ \bibnamefont {Gorshkov}},\
  }\bibfield  {title} {\bibinfo {title} {Nearly linear light cones in
  long-range interacting quantum systems},\ }\href
  {https://doi.org/10.1103/PhysRevLett.114.157201} {\bibfield  {journal}
  {\bibinfo  {journal} {Phys. Rev. Lett.}\ }\textbf {\bibinfo {volume} {114}},\
  \bibinfo {pages} {157201} (\bibinfo {year} {2015})}\BibitemShut {NoStop}%
\bibitem [{\citenamefont {Defenu}\ \emph {et~al.}(2023)\citenamefont {Defenu},
  \citenamefont {Donner}, \citenamefont {Macr{\`\i}}, \citenamefont {Pagano},
  \citenamefont {Ruffo},\ and\ \citenamefont {Trombettoni}}]{Defenu2023}%
  \BibitemOpen
  \bibfield  {author} {\bibinfo {author} {\bibfnamefont {N.}~\bibnamefont
  {Defenu}}, \bibinfo {author} {\bibfnamefont {T.}~\bibnamefont {Donner}},
  \bibinfo {author} {\bibfnamefont {T.}~\bibnamefont {Macr{\`\i}}}, \bibinfo
  {author} {\bibfnamefont {G.}~\bibnamefont {Pagano}}, \bibinfo {author}
  {\bibfnamefont {S.}~\bibnamefont {Ruffo}},\ and\ \bibinfo {author}
  {\bibfnamefont {A.}~\bibnamefont {Trombettoni}},\ }\bibfield  {title}
  {\bibinfo {title} {Long-range interacting quantum systems},\ }\href
  {https://doi.org/10.1103/RevModPhys.95.035002} {\bibfield  {journal}
  {\bibinfo  {journal} {Rev. Mod. Phys.}\ }\textbf {\bibinfo {volume} {95}},\
  \bibinfo {pages} {035002} (\bibinfo {year} {2023})}\BibitemShut {NoStop}%
\bibitem [{\citenamefont {Defenu}\ \emph {et~al.}(2024)\citenamefont {Defenu},
  \citenamefont {Lerose},\ and\ \citenamefont {Pappalardi}}]{Defenu2024}%
  \BibitemOpen
  \bibfield  {author} {\bibinfo {author} {\bibfnamefont {N.}~\bibnamefont
  {Defenu}}, \bibinfo {author} {\bibfnamefont {A.}~\bibnamefont {Lerose}},\
  and\ \bibinfo {author} {\bibfnamefont {S.}~\bibnamefont {Pappalardi}},\
  }\bibfield  {title} {\bibinfo {title} {Out-of-equilibrium dynamics of quantum
  many-body systems with long-range interactions},\ }\href
  {https://doi.org/10.1016/j.physrep.2024.04.005} {\bibfield  {journal}
  {\bibinfo  {journal} {Phys. Rep.}\ }\textbf {\bibinfo {volume} {1074}},\
  \bibinfo {pages} {1} (\bibinfo {year} {2024})}\BibitemShut {NoStop}%
\bibitem [{\citenamefont {Bohnet}\ \emph {et~al.}(2016)\citenamefont {Bohnet},
  \citenamefont {Sawyer}, \citenamefont {Britton}, \citenamefont {Wall},
  \citenamefont {Rey}, \citenamefont {Foss-Feig},\ and\ \citenamefont
  {Bollinger}}]{Bohnet2016}%
  \BibitemOpen
  \bibfield  {author} {\bibinfo {author} {\bibfnamefont {J.~G.}\ \bibnamefont
  {Bohnet}}, \bibinfo {author} {\bibfnamefont {B.~C.}\ \bibnamefont {Sawyer}},
  \bibinfo {author} {\bibfnamefont {J.~W.}\ \bibnamefont {Britton}}, \bibinfo
  {author} {\bibfnamefont {M.~L.}\ \bibnamefont {Wall}}, \bibinfo {author}
  {\bibfnamefont {A.~M.}\ \bibnamefont {Rey}}, \bibinfo {author} {\bibfnamefont
  {M.}~\bibnamefont {Foss-Feig}},\ and\ \bibinfo {author} {\bibfnamefont
  {J.~J.}\ \bibnamefont {Bollinger}},\ }\bibfield  {title} {\bibinfo {title}
  {Quantum spin dynamics and entanglement generation with hundreds of trapped
  ions},\ }\href {https://doi.org/10.1126/science.aad9958} {\bibfield
  {journal} {\bibinfo  {journal} {Science}\ }\textbf {\bibinfo {volume}
  {352}},\ \bibinfo {pages} {1297} (\bibinfo {year} {2016})}\BibitemShut
  {NoStop}%
\bibitem [{\citenamefont {Browaeys}\ and\ \citenamefont
  {Lahaye}(2020)}]{BrowaeysLahaye2020}%
  \BibitemOpen
  \bibfield  {author} {\bibinfo {author} {\bibfnamefont {A.}~\bibnamefont
  {Browaeys}}\ and\ \bibinfo {author} {\bibfnamefont {T.}~\bibnamefont
  {Lahaye}},\ }\bibfield  {title} {\bibinfo {title} {Many-body physics with
  individually controlled {R}ydberg atoms},\ }\href
  {https://doi.org/10.1038/s41567-019-0733-z} {\bibfield  {journal} {\bibinfo
  {journal} {Nat. Phys.}\ }\textbf {\bibinfo {volume} {16}},\ \bibinfo {pages}
  {132} (\bibinfo {year} {2020})}\BibitemShut {NoStop}%
\bibitem [{\citenamefont {Barreiro}\ \emph {et~al.}(2011)\citenamefont
  {Barreiro}, \citenamefont {M\"uller}, \citenamefont {Schindler},
  \citenamefont {Nigg}, \citenamefont {Monz}, \citenamefont {Chwalla},
  \citenamefont {Hennrich}, \citenamefont {Roos}, \citenamefont {Zoller},\ and\
  \citenamefont {Blatt}}]{Barreiro2011}%
  \BibitemOpen
  \bibfield  {author} {\bibinfo {author} {\bibfnamefont {J.~T.}\ \bibnamefont
  {Barreiro}}, \bibinfo {author} {\bibfnamefont {M.}~\bibnamefont {M\"uller}},
  \bibinfo {author} {\bibfnamefont {P.}~\bibnamefont {Schindler}}, \bibinfo
  {author} {\bibfnamefont {D.}~\bibnamefont {Nigg}}, \bibinfo {author}
  {\bibfnamefont {T.}~\bibnamefont {Monz}}, \bibinfo {author} {\bibfnamefont
  {M.}~\bibnamefont {Chwalla}}, \bibinfo {author} {\bibfnamefont
  {M.}~\bibnamefont {Hennrich}}, \bibinfo {author} {\bibfnamefont {C.~F.}\
  \bibnamefont {Roos}}, \bibinfo {author} {\bibfnamefont {P.}~\bibnamefont
  {Zoller}},\ and\ \bibinfo {author} {\bibfnamefont {R.}~\bibnamefont
  {Blatt}},\ }\bibfield  {title} {\bibinfo {title} {An open-system quantum
  simulator with trapped ions},\ }\href {https://doi.org/10.1038/nature09801}
  {\bibfield  {journal} {\bibinfo  {journal} {Nature}\ }\textbf {\bibinfo
  {volume} {470}},\ \bibinfo {pages} {486} (\bibinfo {year}
  {2011})}\BibitemShut {NoStop}%
\bibitem [{\citenamefont {Blatt}\ and\ \citenamefont
  {Roos}(2012)}]{BlattRoos2012}%
  \BibitemOpen
  \bibfield  {author} {\bibinfo {author} {\bibfnamefont {R.}~\bibnamefont
  {Blatt}}\ and\ \bibinfo {author} {\bibfnamefont {C.~F.}\ \bibnamefont
  {Roos}},\ }\bibfield  {title} {\bibinfo {title} {Quantum simulations with
  trapped ions},\ }\href {https://doi.org/10.1038/nphys2252} {\bibfield
  {journal} {\bibinfo  {journal} {Nat. Phys.}\ }\textbf {\bibinfo {volume}
  {8}},\ \bibinfo {pages} {277} (\bibinfo {year} {2012})}\BibitemShut {NoStop}%
\bibitem [{\citenamefont {Pezz\`e}\ \emph {et~al.}(2018)\citenamefont
  {Pezz\`e}, \citenamefont {Smerzi}, \citenamefont {Oberthaler}, \citenamefont
  {Schmied},\ and\ \citenamefont {Treutlein}}]{Pezze2018}%
  \BibitemOpen
  \bibfield  {author} {\bibinfo {author} {\bibfnamefont {L.}~\bibnamefont
  {Pezz\`e}}, \bibinfo {author} {\bibfnamefont {A.}~\bibnamefont {Smerzi}},
  \bibinfo {author} {\bibfnamefont {M.~K.}\ \bibnamefont {Oberthaler}},
  \bibinfo {author} {\bibfnamefont {R.}~\bibnamefont {Schmied}},\ and\ \bibinfo
  {author} {\bibfnamefont {P.}~\bibnamefont {Treutlein}},\ }\bibfield  {title}
  {\bibinfo {title} {Quantum metrology with nonclassical states of atomic
  ensembles},\ }\href {https://doi.org/10.1103/RevModPhys.90.035005} {\bibfield
   {journal} {\bibinfo  {journal} {Rev. Mod. Phys.}\ }\textbf {\bibinfo
  {volume} {90}},\ \bibinfo {pages} {035005} (\bibinfo {year}
  {2018})}\BibitemShut {NoStop}%
\bibitem [{\citenamefont {Degen}\ \emph {et~al.}(2017)\citenamefont {Degen},
  \citenamefont {Reinhard},\ and\ \citenamefont {Cappellaro}}]{Degen2017}%
  \BibitemOpen
  \bibfield  {author} {\bibinfo {author} {\bibfnamefont {C.~L.}\ \bibnamefont
  {Degen}}, \bibinfo {author} {\bibfnamefont {F.}~\bibnamefont {Reinhard}},\
  and\ \bibinfo {author} {\bibfnamefont {P.}~\bibnamefont {Cappellaro}},\
  }\bibfield  {title} {\bibinfo {title} {Quantum sensing},\ }\href
  {https://doi.org/10.1103/RevModPhys.89.035002} {\bibfield  {journal}
  {\bibinfo  {journal} {Rev. Mod. Phys.}\ }\textbf {\bibinfo {volume} {89}},\
  \bibinfo {pages} {035002} (\bibinfo {year} {2017})}\BibitemShut {NoStop}%
\bibitem [{\citenamefont {Verstraete}\ \emph {et~al.}(2009)\citenamefont
  {Verstraete}, \citenamefont {Wolf},\ and\ \citenamefont
  {Cirac}}]{Verstraete2009}%
  \BibitemOpen
  \bibfield  {author} {\bibinfo {author} {\bibfnamefont {F.}~\bibnamefont
  {Verstraete}}, \bibinfo {author} {\bibfnamefont {M.~M.}\ \bibnamefont
  {Wolf}},\ and\ \bibinfo {author} {\bibfnamefont {J.~I.}\ \bibnamefont
  {Cirac}},\ }\bibfield  {title} {\bibinfo {title} {Quantum computation and
  quantum-state engineering driven by dissipation},\ }\href
  {https://doi.org/10.1038/nphys1342} {\bibfield  {journal} {\bibinfo
  {journal} {Nat. Phys.}\ }\textbf {\bibinfo {volume} {5}},\ \bibinfo {pages}
  {633} (\bibinfo {year} {2009})}\BibitemShut {NoStop}%
\bibitem [{\citenamefont {Weimer}\ \emph {et~al.}(2021)\citenamefont {Weimer},
  \citenamefont {Kshetrimayum},\ and\ \citenamefont {Or\'us}}]{Weimer2021}%
  \BibitemOpen
  \bibfield  {author} {\bibinfo {author} {\bibfnamefont {H.}~\bibnamefont
  {Weimer}}, \bibinfo {author} {\bibfnamefont {A.}~\bibnamefont
  {Kshetrimayum}},\ and\ \bibinfo {author} {\bibfnamefont {R.}~\bibnamefont
  {Or\'us}},\ }\bibfield  {title} {\bibinfo {title} {Simulation methods for
  open quantum many-body systems},\ }\href
  {https://doi.org/10.1103/RevModPhys.93.015008} {\bibfield  {journal}
  {\bibinfo  {journal} {Rev. Mod. Phys.}\ }\textbf {\bibinfo {volume} {93}},\
  \bibinfo {pages} {015008} (\bibinfo {year} {2021})}\BibitemShut {NoStop}%
\bibitem [{\citenamefont {Kshetrimayum}\ \emph {et~al.}(2017)\citenamefont
  {Kshetrimayum}, \citenamefont {Weimer},\ and\ \citenamefont
  {Or\'us}}]{Kshetrimayum2017}%
  \BibitemOpen
  \bibfield  {author} {\bibinfo {author} {\bibfnamefont {A.}~\bibnamefont
  {Kshetrimayum}}, \bibinfo {author} {\bibfnamefont {H.}~\bibnamefont
  {Weimer}},\ and\ \bibinfo {author} {\bibfnamefont {R.}~\bibnamefont
  {Or\'us}},\ }\bibfield  {title} {\bibinfo {title} {A simple tensor network
  algorithm for two-dimensional steady states},\ }\href
  {https://doi.org/10.1038/s41467-017-01511-6} {\bibfield  {journal} {\bibinfo
  {journal} {Nat. Commun.}\ }\textbf {\bibinfo {volume} {8}},\ \bibinfo {pages}
  {1291} (\bibinfo {year} {2017})}\BibitemShut {NoStop}%
\bibitem [{\citenamefont {Mc~Keever}\ and\ \citenamefont
  {Szyma\'nska}(2021)}]{McKeever2021}%
  \BibitemOpen
  \bibfield  {author} {\bibinfo {author} {\bibfnamefont {C.}~\bibnamefont
  {Mc~Keever}}\ and\ \bibinfo {author} {\bibfnamefont {M.~H.}\ \bibnamefont
  {Szyma\'nska}},\ }\bibfield  {title} {\bibinfo {title} {Stable {iPEPO}
  tensor-network algorithm for dynamics of two-dimensional open quantum lattice
  models},\ }\href {https://doi.org/10.1103/PhysRevX.11.021035} {\bibfield
  {journal} {\bibinfo  {journal} {Phys. Rev. X}\ }\textbf {\bibinfo {volume}
  {11}},\ \bibinfo {pages} {021035} (\bibinfo {year} {2021})}\BibitemShut
  {NoStop}%
\bibitem [{\citenamefont {Britton}\ \emph {et~al.}(2012)\citenamefont
  {Britton}, \citenamefont {Sawyer}, \citenamefont {Keith}, \citenamefont
  {Wang}, \citenamefont {Freericks}, \citenamefont {Uys}, \citenamefont
  {Biercuk},\ and\ \citenamefont {Bollinger}}]{Britton2012}%
  \BibitemOpen
  \bibfield  {author} {\bibinfo {author} {\bibfnamefont {J.~W.}\ \bibnamefont
  {Britton}}, \bibinfo {author} {\bibfnamefont {B.~C.}\ \bibnamefont {Sawyer}},
  \bibinfo {author} {\bibfnamefont {A.~C.}\ \bibnamefont {Keith}}, \bibinfo
  {author} {\bibfnamefont {C.-C.~J.}\ \bibnamefont {Wang}}, \bibinfo {author}
  {\bibfnamefont {J.~K.}\ \bibnamefont {Freericks}}, \bibinfo {author}
  {\bibfnamefont {H.}~\bibnamefont {Uys}}, \bibinfo {author} {\bibfnamefont
  {M.~J.}\ \bibnamefont {Biercuk}},\ and\ \bibinfo {author} {\bibfnamefont
  {J.~J.}\ \bibnamefont {Bollinger}},\ }\bibfield  {title} {\bibinfo {title}
  {Engineered two-dimensional {I}sing interactions in a trapped-ion quantum
  simulator with hundreds of spins},\ }\href
  {https://doi.org/10.1038/nature10981} {\bibfield  {journal} {\bibinfo
  {journal} {Nature}\ }\textbf {\bibinfo {volume} {484}},\ \bibinfo {pages}
  {489} (\bibinfo {year} {2012})}\BibitemShut {NoStop}%
\bibitem [{\citenamefont {Zhang}\ \emph {et~al.}(2017)\citenamefont {Zhang},
  \citenamefont {Pagano}, \citenamefont {Hess}, \citenamefont {Kyprianidis},
  \citenamefont {Becker}, \citenamefont {Kaplan}, \citenamefont {Gorshkov},
  \citenamefont {Gong},\ and\ \citenamefont {Monroe}}]{Zhang2017}%
  \BibitemOpen
  \bibfield  {author} {\bibinfo {author} {\bibfnamefont {J.}~\bibnamefont
  {Zhang}}, \bibinfo {author} {\bibfnamefont {G.}~\bibnamefont {Pagano}},
  \bibinfo {author} {\bibfnamefont {P.~W.}\ \bibnamefont {Hess}}, \bibinfo
  {author} {\bibfnamefont {A.}~\bibnamefont {Kyprianidis}}, \bibinfo {author}
  {\bibfnamefont {P.}~\bibnamefont {Becker}}, \bibinfo {author} {\bibfnamefont
  {H.}~\bibnamefont {Kaplan}}, \bibinfo {author} {\bibfnamefont {A.~V.}\
  \bibnamefont {Gorshkov}}, \bibinfo {author} {\bibfnamefont {Z.-X.}\
  \bibnamefont {Gong}},\ and\ \bibinfo {author} {\bibfnamefont
  {C.}~\bibnamefont {Monroe}},\ }\bibfield  {title} {\bibinfo {title}
  {Observation of a many-body dynamical phase transition with a 53-qubit
  quantum simulator},\ }\href {https://doi.org/10.1038/nature24654} {\bibfield
  {journal} {\bibinfo  {journal} {Nature}\ }\textbf {\bibinfo {volume} {551}},\
  \bibinfo {pages} {601} (\bibinfo {year} {2017})}\BibitemShut {NoStop}%
\bibitem [{\citenamefont {Labuhn}\ \emph {et~al.}(2016)\citenamefont {Labuhn},
  \citenamefont {Barredo}, \citenamefont {Ravets}, \citenamefont
  {de~L\'es\'eleuc}, \citenamefont {Macr{\`\i}}, \citenamefont {Lahaye},\ and\
  \citenamefont {Browaeys}}]{Labuhn2016}%
  \BibitemOpen
  \bibfield  {author} {\bibinfo {author} {\bibfnamefont {H.}~\bibnamefont
  {Labuhn}}, \bibinfo {author} {\bibfnamefont {D.}~\bibnamefont {Barredo}},
  \bibinfo {author} {\bibfnamefont {S.}~\bibnamefont {Ravets}}, \bibinfo
  {author} {\bibfnamefont {S.}~\bibnamefont {de~L\'es\'eleuc}}, \bibinfo
  {author} {\bibfnamefont {T.}~\bibnamefont {Macr{\`\i}}}, \bibinfo {author}
  {\bibfnamefont {T.}~\bibnamefont {Lahaye}},\ and\ \bibinfo {author}
  {\bibfnamefont {A.}~\bibnamefont {Browaeys}},\ }\bibfield  {title} {\bibinfo
  {title} {Tunable two-dimensional arrays of single {R}ydberg atoms for
  realizing quantum {I}sing models},\ }\href
  {https://doi.org/10.1038/nature18274} {\bibfield  {journal} {\bibinfo
  {journal} {Nature}\ }\textbf {\bibinfo {volume} {534}},\ \bibinfo {pages}
  {667} (\bibinfo {year} {2016})}\BibitemShut {NoStop}%
\bibitem [{\citenamefont {Bernien}\ \emph {et~al.}(2017)\citenamefont
  {Bernien}, \citenamefont {Schwartz}, \citenamefont {Keesling}, \citenamefont
  {Levine}, \citenamefont {Omran}, \citenamefont {Pichler}, \citenamefont
  {Choi}, \citenamefont {Zibrov}, \citenamefont {Endres}, \citenamefont
  {Greiner}, \citenamefont {Vuleti\'c},\ and\ \citenamefont
  {Lukin}}]{Bernien2017}%
  \BibitemOpen
  \bibfield  {author} {\bibinfo {author} {\bibfnamefont {H.}~\bibnamefont
  {Bernien}}, \bibinfo {author} {\bibfnamefont {S.}~\bibnamefont {Schwartz}},
  \bibinfo {author} {\bibfnamefont {A.}~\bibnamefont {Keesling}}, \bibinfo
  {author} {\bibfnamefont {H.}~\bibnamefont {Levine}}, \bibinfo {author}
  {\bibfnamefont {A.}~\bibnamefont {Omran}}, \bibinfo {author} {\bibfnamefont
  {H.}~\bibnamefont {Pichler}}, \bibinfo {author} {\bibfnamefont
  {S.}~\bibnamefont {Choi}}, \bibinfo {author} {\bibfnamefont {A.~S.}\
  \bibnamefont {Zibrov}}, \bibinfo {author} {\bibfnamefont {M.}~\bibnamefont
  {Endres}}, \bibinfo {author} {\bibfnamefont {M.}~\bibnamefont {Greiner}},
  \bibinfo {author} {\bibfnamefont {V.}~\bibnamefont {Vuleti\'c}},\ and\
  \bibinfo {author} {\bibfnamefont {M.~D.}\ \bibnamefont {Lukin}},\ }\bibfield
  {title} {\bibinfo {title} {Probing many-body dynamics on a 51-atom quantum
  simulator},\ }\href {https://doi.org/10.1038/nature24622} {\bibfield
  {journal} {\bibinfo  {journal} {Nature}\ }\textbf {\bibinfo {volume} {551}},\
  \bibinfo {pages} {579} (\bibinfo {year} {2017})}\BibitemShut {NoStop}%
\bibitem [{\citenamefont {Yan}\ \emph {et~al.}(2013)\citenamefont {Yan},
  \citenamefont {Moses}, \citenamefont {Gadway}, \citenamefont {Covey},
  \citenamefont {Hazzard}, \citenamefont {Rey}, \citenamefont {Jin},\ and\
  \citenamefont {Ye}}]{Yan2013}%
  \BibitemOpen
  \bibfield  {author} {\bibinfo {author} {\bibfnamefont {B.}~\bibnamefont
  {Yan}}, \bibinfo {author} {\bibfnamefont {S.~A.}\ \bibnamefont {Moses}},
  \bibinfo {author} {\bibfnamefont {B.}~\bibnamefont {Gadway}}, \bibinfo
  {author} {\bibfnamefont {J.~P.}\ \bibnamefont {Covey}}, \bibinfo {author}
  {\bibfnamefont {K.~R.~A.}\ \bibnamefont {Hazzard}}, \bibinfo {author}
  {\bibfnamefont {A.~M.}\ \bibnamefont {Rey}}, \bibinfo {author} {\bibfnamefont
  {D.~S.}\ \bibnamefont {Jin}},\ and\ \bibinfo {author} {\bibfnamefont
  {J.}~\bibnamefont {Ye}},\ }\bibfield  {title} {\bibinfo {title} {Observation
  of dipolar spin-exchange interactions with lattice-confined polar
  molecules},\ }\href {https://doi.org/10.1038/nature12483} {\bibfield
  {journal} {\bibinfo  {journal} {Nature}\ }\textbf {\bibinfo {volume} {501}},\
  \bibinfo {pages} {521} (\bibinfo {year} {2013})}\BibitemShut {NoStop}%
\bibitem [{\citenamefont {Lepoutre}\ \emph {et~al.}(2019)\citenamefont
  {Lepoutre}, \citenamefont {Schachenmayer}, \citenamefont {Gabardos},
  \citenamefont {Zhu}, \citenamefont {Naylor}, \citenamefont {Mar\'echal},
  \citenamefont {Gorceix}, \citenamefont {Rey}, \citenamefont {Vernac},\ and\
  \citenamefont {Laburthe-Tolra}}]{Lepoutre2019}%
  \BibitemOpen
  \bibfield  {author} {\bibinfo {author} {\bibfnamefont {S.}~\bibnamefont
  {Lepoutre}}, \bibinfo {author} {\bibfnamefont {J.}~\bibnamefont
  {Schachenmayer}}, \bibinfo {author} {\bibfnamefont {L.}~\bibnamefont
  {Gabardos}}, \bibinfo {author} {\bibfnamefont {B.}~\bibnamefont {Zhu}},
  \bibinfo {author} {\bibfnamefont {B.}~\bibnamefont {Naylor}}, \bibinfo
  {author} {\bibfnamefont {E.}~\bibnamefont {Mar\'echal}}, \bibinfo {author}
  {\bibfnamefont {O.}~\bibnamefont {Gorceix}}, \bibinfo {author} {\bibfnamefont
  {A.~M.}\ \bibnamefont {Rey}}, \bibinfo {author} {\bibfnamefont
  {L.}~\bibnamefont {Vernac}},\ and\ \bibinfo {author} {\bibfnamefont
  {B.}~\bibnamefont {Laburthe-Tolra}},\ }\bibfield  {title} {\bibinfo {title}
  {Out-of-equilibrium quantum magnetism and thermalization in a spin-3
  many-body dipolar lattice system},\ }\href
  {https://doi.org/10.1038/s41467-019-09699-5} {\bibfield  {journal} {\bibinfo
  {journal} {Nat. Commun.}\ }\textbf {\bibinfo {volume} {10}},\ \bibinfo
  {pages} {1714} (\bibinfo {year} {2019})}\BibitemShut {NoStop}%
\bibitem [{\citenamefont {Patscheider}\ \emph {et~al.}(2020)\citenamefont
  {Patscheider}, \citenamefont {Zhu}, \citenamefont {Chomaz}, \citenamefont
  {Petter}, \citenamefont {Baier}, \citenamefont {Rey}, \citenamefont
  {Ferlaino},\ and\ \citenamefont {Mark}}]{Patscheider2020}%
  \BibitemOpen
  \bibfield  {author} {\bibinfo {author} {\bibfnamefont {A.}~\bibnamefont
  {Patscheider}}, \bibinfo {author} {\bibfnamefont {B.}~\bibnamefont {Zhu}},
  \bibinfo {author} {\bibfnamefont {L.}~\bibnamefont {Chomaz}}, \bibinfo
  {author} {\bibfnamefont {D.}~\bibnamefont {Petter}}, \bibinfo {author}
  {\bibfnamefont {S.}~\bibnamefont {Baier}}, \bibinfo {author} {\bibfnamefont
  {A.-M.}\ \bibnamefont {Rey}}, \bibinfo {author} {\bibfnamefont
  {F.}~\bibnamefont {Ferlaino}},\ and\ \bibinfo {author} {\bibfnamefont
  {M.~J.}\ \bibnamefont {Mark}},\ }\bibfield  {title} {\bibinfo {title}
  {Controlling dipolar exchange interactions in a dense three-dimensional array
  of large-spin fermions},\ }\href
  {https://doi.org/10.1103/PhysRevResearch.2.023050} {\bibfield  {journal}
  {\bibinfo  {journal} {Phys. Rev. Research}\ }\textbf {\bibinfo {volume}
  {2}},\ \bibinfo {pages} {023050} (\bibinfo {year} {2020})}\BibitemShut
  {NoStop}%
\bibitem [{\citenamefont {Bornet}\ \emph {et~al.}(2023)\citenamefont {Bornet},
  \citenamefont {Emperauger}, \citenamefont {Chen}, \citenamefont {Ye},
  \citenamefont {Block}, \citenamefont {Bintz}, \citenamefont {Boyd},
  \citenamefont {Barredo}, \citenamefont {Comparin}, \citenamefont {Mezzacapo},
  \citenamefont {Roscilde}, \citenamefont {Lahaye}, \citenamefont {Yao},\ and\
  \citenamefont {Browaeys}}]{Bornet2023}%
  \BibitemOpen
  \bibfield  {author} {\bibinfo {author} {\bibfnamefont {G.}~\bibnamefont
  {Bornet}}, \bibinfo {author} {\bibfnamefont {G.}~\bibnamefont {Emperauger}},
  \bibinfo {author} {\bibfnamefont {C.}~\bibnamefont {Chen}}, \bibinfo {author}
  {\bibfnamefont {B.}~\bibnamefont {Ye}}, \bibinfo {author} {\bibfnamefont
  {M.}~\bibnamefont {Block}}, \bibinfo {author} {\bibfnamefont
  {M.}~\bibnamefont {Bintz}}, \bibinfo {author} {\bibfnamefont {J.~A.}\
  \bibnamefont {Boyd}}, \bibinfo {author} {\bibfnamefont {D.}~\bibnamefont
  {Barredo}}, \bibinfo {author} {\bibfnamefont {T.}~\bibnamefont {Comparin}},
  \bibinfo {author} {\bibfnamefont {F.}~\bibnamefont {Mezzacapo}}, \bibinfo
  {author} {\bibfnamefont {T.}~\bibnamefont {Roscilde}}, \bibinfo {author}
  {\bibfnamefont {T.}~\bibnamefont {Lahaye}}, \bibinfo {author} {\bibfnamefont
  {N.~Y.}\ \bibnamefont {Yao}},\ and\ \bibinfo {author} {\bibfnamefont
  {A.}~\bibnamefont {Browaeys}},\ }\bibfield  {title} {\bibinfo {title}
  {Scalable spin squeezing in a dipolar {R}ydberg atom array},\ }\href
  {https://doi.org/10.1038/s41586-023-06414-9} {\bibfield  {journal} {\bibinfo
  {journal} {Nature}\ }\textbf {\bibinfo {volume} {621}},\ \bibinfo {pages}
  {728} (\bibinfo {year} {2023})}\BibitemShut {NoStop}%
\bibitem [{\citenamefont {Eckner}\ \emph {et~al.}(2023)\citenamefont {Eckner},
  \citenamefont {Darkwah~Oppong}, \citenamefont {Cao}, \citenamefont {Young},
  \citenamefont {Milner}, \citenamefont {Robinson}, \citenamefont {Ye},\ and\
  \citenamefont {Kaufman}}]{Eckner2023}%
  \BibitemOpen
  \bibfield  {author} {\bibinfo {author} {\bibfnamefont {W.~J.}\ \bibnamefont
  {Eckner}}, \bibinfo {author} {\bibfnamefont {N.}~\bibnamefont
  {Darkwah~Oppong}}, \bibinfo {author} {\bibfnamefont {A.}~\bibnamefont {Cao}},
  \bibinfo {author} {\bibfnamefont {A.~W.}\ \bibnamefont {Young}}, \bibinfo
  {author} {\bibfnamefont {W.~R.}\ \bibnamefont {Milner}}, \bibinfo {author}
  {\bibfnamefont {J.~M.}\ \bibnamefont {Robinson}}, \bibinfo {author}
  {\bibfnamefont {J.}~\bibnamefont {Ye}},\ and\ \bibinfo {author}
  {\bibfnamefont {A.~M.}\ \bibnamefont {Kaufman}},\ }\bibfield  {title}
  {\bibinfo {title} {Realizing spin squeezing with {R}ydberg interactions in an
  optical clock},\ }\href {https://doi.org/10.1038/s41586-023-06360-6}
  {\bibfield  {journal} {\bibinfo  {journal} {Nature}\ }\textbf {\bibinfo
  {volume} {621}},\ \bibinfo {pages} {734} (\bibinfo {year}
  {2023})}\BibitemShut {NoStop}%
\bibitem [{\citenamefont {Franke}\ \emph {et~al.}(2023)\citenamefont {Franke},
  \citenamefont {Muleady}, \citenamefont {Kaubruegger}, \citenamefont {Kranzl},
  \citenamefont {Blatt}, \citenamefont {Rey}, \citenamefont {Joshi},\ and\
  \citenamefont {Roos}}]{Franke2023}%
  \BibitemOpen
  \bibfield  {author} {\bibinfo {author} {\bibfnamefont {J.}~\bibnamefont
  {Franke}}, \bibinfo {author} {\bibfnamefont {S.~R.}\ \bibnamefont {Muleady}},
  \bibinfo {author} {\bibfnamefont {R.}~\bibnamefont {Kaubruegger}}, \bibinfo
  {author} {\bibfnamefont {F.}~\bibnamefont {Kranzl}}, \bibinfo {author}
  {\bibfnamefont {R.}~\bibnamefont {Blatt}}, \bibinfo {author} {\bibfnamefont
  {A.~M.}\ \bibnamefont {Rey}}, \bibinfo {author} {\bibfnamefont {M.~K.}\
  \bibnamefont {Joshi}},\ and\ \bibinfo {author} {\bibfnamefont {C.~F.}\
  \bibnamefont {Roos}},\ }\bibfield  {title} {\bibinfo {title}
  {Quantum-enhanced sensing on optical transitions through finite-range
  interactions},\ }\href {https://doi.org/10.1038/s41586-023-06472-z}
  {\bibfield  {journal} {\bibinfo  {journal} {Nature}\ }\textbf {\bibinfo
  {volume} {621}},\ \bibinfo {pages} {740} (\bibinfo {year}
  {2023})}\BibitemShut {NoStop}%
\bibitem [{\citenamefont {Levine}\ \emph {et~al.}(2019)\citenamefont {Levine},
  \citenamefont {Keesling}, \citenamefont {Semeghini}, \citenamefont {Omran},
  \citenamefont {Wang}, \citenamefont {Ebadi}, \citenamefont {Bernien},
  \citenamefont {Greiner}, \citenamefont {Vuleti\'c}, \citenamefont {Pichler},\
  and\ \citenamefont {Lukin}}]{Levine2019}%
  \BibitemOpen
  \bibfield  {author} {\bibinfo {author} {\bibfnamefont {H.}~\bibnamefont
  {Levine}}, \bibinfo {author} {\bibfnamefont {A.}~\bibnamefont {Keesling}},
  \bibinfo {author} {\bibfnamefont {G.}~\bibnamefont {Semeghini}}, \bibinfo
  {author} {\bibfnamefont {A.}~\bibnamefont {Omran}}, \bibinfo {author}
  {\bibfnamefont {T.~T.}\ \bibnamefont {Wang}}, \bibinfo {author}
  {\bibfnamefont {S.}~\bibnamefont {Ebadi}}, \bibinfo {author} {\bibfnamefont
  {H.}~\bibnamefont {Bernien}}, \bibinfo {author} {\bibfnamefont
  {M.}~\bibnamefont {Greiner}}, \bibinfo {author} {\bibfnamefont
  {V.}~\bibnamefont {Vuleti\'c}}, \bibinfo {author} {\bibfnamefont
  {H.}~\bibnamefont {Pichler}},\ and\ \bibinfo {author} {\bibfnamefont {M.~D.}\
  \bibnamefont {Lukin}},\ }\bibfield  {title} {\bibinfo {title} {Parallel
  implementation of high-fidelity multiqubit gates with neutral atoms},\ }\href
  {https://doi.org/10.1103/PhysRevLett.123.170503} {\bibfield  {journal}
  {\bibinfo  {journal} {Phys. Rev. Lett.}\ }\textbf {\bibinfo {volume} {123}},\
  \bibinfo {pages} {170503} (\bibinfo {year} {2019})}\BibitemShut {NoStop}%
\bibitem [{\citenamefont {Sulz}\ \emph {et~al.}(2024)\citenamefont {Sulz},
  \citenamefont {Lubich}, \citenamefont {Ceruti}, \citenamefont {Lesanovsky},\
  and\ \citenamefont {Carollo}}]{Sulz2024}%
  \BibitemOpen
  \bibfield  {author} {\bibinfo {author} {\bibfnamefont {D.}~\bibnamefont
  {Sulz}}, \bibinfo {author} {\bibfnamefont {C.}~\bibnamefont {Lubich}},
  \bibinfo {author} {\bibfnamefont {G.}~\bibnamefont {Ceruti}}, \bibinfo
  {author} {\bibfnamefont {I.}~\bibnamefont {Lesanovsky}},\ and\ \bibinfo
  {author} {\bibfnamefont {F.}~\bibnamefont {Carollo}},\ }\bibfield  {title}
  {\bibinfo {title} {Numerical simulation of long-range open quantum many-body
  dynamics with tree tensor networks},\ }\href
  {https://doi.org/10.1103/PhysRevA.109.022420} {\bibfield  {journal} {\bibinfo
   {journal} {Phys. Rev. A}\ }\textbf {\bibinfo {volume} {109}},\ \bibinfo
  {pages} {022420} (\bibinfo {year} {2024})}\BibitemShut {NoStop}%
\bibitem [{\citenamefont {Hryniuk}\ and\ \citenamefont
  {Szyma\'nska}(2024)}]{Hryniuk2024}%
  \BibitemOpen
  \bibfield  {author} {\bibinfo {author} {\bibfnamefont {D.~A.}\ \bibnamefont
  {Hryniuk}}\ and\ \bibinfo {author} {\bibfnamefont {M.~H.}\ \bibnamefont
  {Szyma\'nska}},\ }\bibfield  {title} {\bibinfo {title} {Tensor-network-based
  variational {M}onte {C}arlo approach to the non-equilibrium steady state of
  open quantum systems},\ }\href {https://doi.org/10.22331/q-2024-09-17-1475}
  {\bibfield  {journal} {\bibinfo  {journal} {Quantum}\ }\textbf {\bibinfo
  {volume} {8}},\ \bibinfo {pages} {1475} (\bibinfo {year} {2024})}\BibitemShut
  {NoStop}%
\bibitem [{\citenamefont {Hryniuk}\ and\ \citenamefont
  {Szyma{\'{n}}ska}(2026)}]{Hryniuk2026}%
  \BibitemOpen
  \bibfield  {author} {\bibinfo {author} {\bibfnamefont {D.~A.}\ \bibnamefont
  {Hryniuk}}\ and\ \bibinfo {author} {\bibfnamefont {M.~H.}\ \bibnamefont
  {Szyma{\'{n}}ska}},\ }\bibfield  {title} {\bibinfo {title} {Variational
  approach to open quantum systems with long-range competing interactions},\
  }\href {https://doi.org/10.1038/s42005-025-02479-2} {\bibfield  {journal}
  {\bibinfo  {journal} {Commun. Phys.}\ }\textbf {\bibinfo {volume} {9}},\
  \bibinfo {pages} {45} (\bibinfo {year} {2026})}\BibitemShut {NoStop}%
\bibitem [{\citenamefont {Dunham}\ and\ \citenamefont
  {Szymańska}(2025)}]{dunham2025efficienttimeevolution2d}%
  \BibitemOpen
  \bibfield  {author} {\bibinfo {author} {\bibfnamefont {J.}~\bibnamefont
  {Dunham}}\ and\ \bibinfo {author} {\bibfnamefont {M.~H.}\ \bibnamefont
  {Szymańska}},\ }\href {https://doi.org/10.48550/arXiv.2512.01781} {\bibinfo
  {title} {Efficient time evolution of {2D} open-quantum lattice models with
  long-range interactions using tensor networks}} (\bibinfo {year} {2025}),\
  \Eprint {https://arxiv.org/abs/2512.01781} {arXiv:2512.01781 [quant-ph]}
  \BibitemShut {NoStop}%
\bibitem [{\citenamefont {Wigner}(1932)}]{Wigner1932}%
  \BibitemOpen
  \bibfield  {author} {\bibinfo {author} {\bibfnamefont {E.~P.}\ \bibnamefont
  {Wigner}},\ }\bibfield  {title} {\bibinfo {title} {On the quantum correction
  for thermodynamic equilibrium},\ }\href
  {https://doi.org/10.1103/PhysRev.40.749} {\bibfield  {journal} {\bibinfo
  {journal} {Phys. Rev.}\ }\textbf {\bibinfo {volume} {40}},\ \bibinfo {pages}
  {749} (\bibinfo {year} {1932})}\BibitemShut {NoStop}%
\bibitem [{\citenamefont {Hillery}\ \emph {et~al.}(1984)\citenamefont
  {Hillery}, \citenamefont {O'Connell}, \citenamefont {Scully},\ and\
  \citenamefont {Wigner}}]{Hillery1984}%
  \BibitemOpen
  \bibfield  {author} {\bibinfo {author} {\bibfnamefont {M.}~\bibnamefont
  {Hillery}}, \bibinfo {author} {\bibfnamefont {R.~F.}\ \bibnamefont
  {O'Connell}}, \bibinfo {author} {\bibfnamefont {M.~O.}\ \bibnamefont
  {Scully}},\ and\ \bibinfo {author} {\bibfnamefont {E.~P.}\ \bibnamefont
  {Wigner}},\ }\bibfield  {title} {\bibinfo {title} {Distribution functions in
  physics: Fundamentals},\ }\href
  {https://doi.org/10.1016/0370-1573(84)90160-1} {\bibfield  {journal}
  {\bibinfo  {journal} {Phys. Rep.}\ }\textbf {\bibinfo {volume} {106}},\
  \bibinfo {pages} {121} (\bibinfo {year} {1984})}\BibitemShut {NoStop}%
\bibitem [{\citenamefont {Steel}\ \emph {et~al.}(1998)\citenamefont {Steel},
  \citenamefont {Olsen}, \citenamefont {Plimak}, \citenamefont {Drummond},
  \citenamefont {Tan}, \citenamefont {Collett}, \citenamefont {Walls},\ and\
  \citenamefont {Graham}}]{Steel1998}%
  \BibitemOpen
  \bibfield  {author} {\bibinfo {author} {\bibfnamefont {M.~J.}\ \bibnamefont
  {Steel}}, \bibinfo {author} {\bibfnamefont {M.~K.}\ \bibnamefont {Olsen}},
  \bibinfo {author} {\bibfnamefont {L.~I.}\ \bibnamefont {Plimak}}, \bibinfo
  {author} {\bibfnamefont {P.~D.}\ \bibnamefont {Drummond}}, \bibinfo {author}
  {\bibfnamefont {S.~M.}\ \bibnamefont {Tan}}, \bibinfo {author} {\bibfnamefont
  {M.~J.}\ \bibnamefont {Collett}}, \bibinfo {author} {\bibfnamefont {D.~F.}\
  \bibnamefont {Walls}},\ and\ \bibinfo {author} {\bibfnamefont
  {R.}~\bibnamefont {Graham}},\ }\bibfield  {title} {\bibinfo {title}
  {Dynamical quantum noise in trapped {B}ose-{E}instein condensates},\ }\href
  {https://doi.org/10.1103/PhysRevA.58.4824} {\bibfield  {journal} {\bibinfo
  {journal} {Phys. Rev. A}\ }\textbf {\bibinfo {volume} {58}},\ \bibinfo
  {pages} {4824} (\bibinfo {year} {1998})}\BibitemShut {NoStop}%
\bibitem [{\citenamefont {Sinatra}\ \emph {et~al.}(2002)\citenamefont
  {Sinatra}, \citenamefont {Lobo},\ and\ \citenamefont {Castin}}]{Sinatra2002}%
  \BibitemOpen
  \bibfield  {author} {\bibinfo {author} {\bibfnamefont {A.}~\bibnamefont
  {Sinatra}}, \bibinfo {author} {\bibfnamefont {C.}~\bibnamefont {Lobo}},\ and\
  \bibinfo {author} {\bibfnamefont {Y.}~\bibnamefont {Castin}},\ }\bibfield
  {title} {\bibinfo {title} {The truncated {W}igner method for {B}ose-condensed
  gases: limits of validity and applications},\ }\href
  {https://doi.org/10.1088/0953-4075/35/17/301} {\bibfield  {journal} {\bibinfo
   {journal} {J. Phys. B: At. Mol. Opt. Phys.}\ }\textbf {\bibinfo {volume}
  {35}},\ \bibinfo {pages} {3599} (\bibinfo {year} {2002})}\BibitemShut
  {NoStop}%
\bibitem [{\citenamefont {Blakie}\ \emph {et~al.}(2008)\citenamefont {Blakie},
  \citenamefont {Bradley}, \citenamefont {Davis}, \citenamefont {Ballagh},\
  and\ \citenamefont {Gardiner}}]{Blakie2008}%
  \BibitemOpen
  \bibfield  {author} {\bibinfo {author} {\bibfnamefont {P.~B.}\ \bibnamefont
  {Blakie}}, \bibinfo {author} {\bibfnamefont {A.~S.}\ \bibnamefont {Bradley}},
  \bibinfo {author} {\bibfnamefont {M.~J.}\ \bibnamefont {Davis}}, \bibinfo
  {author} {\bibfnamefont {R.~J.}\ \bibnamefont {Ballagh}},\ and\ \bibinfo
  {author} {\bibfnamefont {C.~W.}\ \bibnamefont {Gardiner}},\ }\bibfield
  {title} {\bibinfo {title} {Dynamics and statistical mechanics of ultra-cold
  {B}ose gases using c-field techniques},\ }\href
  {https://doi.org/10.1080/00018730802564254} {\bibfield  {journal} {\bibinfo
  {journal} {Adv. Phys.}\ }\textbf {\bibinfo {volume} {57}},\ \bibinfo {pages}
  {363} (\bibinfo {year} {2008})}\BibitemShut {NoStop}%
\bibitem [{\citenamefont {Polkovnikov}(2003)}]{Polkovnikov2003}%
  \BibitemOpen
  \bibfield  {author} {\bibinfo {author} {\bibfnamefont {A.}~\bibnamefont
  {Polkovnikov}},\ }\bibfield  {title} {\bibinfo {title} {Quantum corrections
  to the dynamics of interacting bosons: Beyond the truncated {W}igner
  approximation},\ }\href {https://doi.org/10.1103/PhysRevA.68.053604}
  {\bibfield  {journal} {\bibinfo  {journal} {Phys. Rev. A}\ }\textbf {\bibinfo
  {volume} {68}},\ \bibinfo {pages} {053604} (\bibinfo {year}
  {2003})}\BibitemShut {NoStop}%
\bibitem [{\citenamefont {Polkovnikov}(2010)}]{Polkovnikov2010}%
  \BibitemOpen
  \bibfield  {author} {\bibinfo {author} {\bibfnamefont {A.}~\bibnamefont
  {Polkovnikov}},\ }\bibfield  {title} {\bibinfo {title} {Phase space
  representation of quantum dynamics},\ }\href
  {https://doi.org/10.1016/j.aop.2010.02.006} {\bibfield  {journal} {\bibinfo
  {journal} {Ann. Phys.}\ }\textbf {\bibinfo {volume} {325}},\ \bibinfo {pages}
  {1790} (\bibinfo {year} {2010})}\BibitemShut {NoStop}%
\bibitem [{\citenamefont {Tosca}\ \emph
  {et~al.}(2026{\natexlab{a}})\citenamefont {Tosca}, \citenamefont {Carnazza},
  \citenamefont {Giacomelli},\ and\ \citenamefont {Ciuti}}]{Tosca2025}%
  \BibitemOpen
  \bibfield  {author} {\bibinfo {author} {\bibfnamefont {J.}~\bibnamefont
  {Tosca}}, \bibinfo {author} {\bibfnamefont {F.}~\bibnamefont {Carnazza}},
  \bibinfo {author} {\bibfnamefont {L.}~\bibnamefont {Giacomelli}},\ and\
  \bibinfo {author} {\bibfnamefont {C.}~\bibnamefont {Ciuti}},\ }\bibfield
  {title} {\bibinfo {title} {Variational multi-{G}aussian phase-space bosonic
  dynamics via automatic differentiation},\ }\href
  {https://doi.org/10.1103/q3m5-q44b} {\bibfield  {journal} {\bibinfo
  {journal} {Phys. Rev. X}\ }\textbf {\bibinfo {volume} {16}},\ \bibinfo
  {pages} {031077} (\bibinfo {year} {2026}{\natexlab{a}})}\BibitemShut
  {NoStop}%
\bibitem [{\citenamefont {Glauber}(1963)}]{Glauber1963}%
  \BibitemOpen
  \bibfield  {author} {\bibinfo {author} {\bibfnamefont {R.~J.}\ \bibnamefont
  {Glauber}},\ }\bibfield  {title} {\bibinfo {title} {Coherent and incoherent
  states of the radiation field},\ }\href
  {https://doi.org/10.1103/PhysRev.131.2766} {\bibfield  {journal} {\bibinfo
  {journal} {Phys. Rev.}\ }\textbf {\bibinfo {volume} {131}},\ \bibinfo {pages}
  {2766} (\bibinfo {year} {1963})}\BibitemShut {NoStop}%
\bibitem [{\citenamefont {Sudarshan}(1963)}]{Sudarshan1963}%
  \BibitemOpen
  \bibfield  {author} {\bibinfo {author} {\bibfnamefont {E.~C.~G.}\
  \bibnamefont {Sudarshan}},\ }\bibfield  {title} {\bibinfo {title}
  {Equivalence of semiclassical and quantum mechanical descriptions of
  statistical light beams},\ }\href
  {https://doi.org/10.1103/PhysRevLett.10.277} {\bibfield  {journal} {\bibinfo
  {journal} {Phys. Rev. Lett.}\ }\textbf {\bibinfo {volume} {10}},\ \bibinfo
  {pages} {277} (\bibinfo {year} {1963})}\BibitemShut {NoStop}%
\bibitem [{\citenamefont {Drummond}\ and\ \citenamefont
  {Gardiner}(1980)}]{DrummondGardiner1980}%
  \BibitemOpen
  \bibfield  {author} {\bibinfo {author} {\bibfnamefont {P.~D.}\ \bibnamefont
  {Drummond}}\ and\ \bibinfo {author} {\bibfnamefont {C.~W.}\ \bibnamefont
  {Gardiner}},\ }\bibfield  {title} {\bibinfo {title} {Generalised
  {P}-representations in quantum optics},\ }\href
  {https://doi.org/10.1088/0305-4470/13/7/018} {\bibfield  {journal} {\bibinfo
  {journal} {J. Phys. A: Math. Gen.}\ }\textbf {\bibinfo {volume} {13}},\
  \bibinfo {pages} {2353} (\bibinfo {year} {1980})}\BibitemShut {NoStop}%
\bibitem [{\citenamefont {Deuar}\ \emph {et~al.}(2021)\citenamefont {Deuar},
  \citenamefont {Ferrier}, \citenamefont {Matuszewski}, \citenamefont {Orso},\
  and\ \citenamefont {Szyma\'nska}}]{Deuar2021}%
  \BibitemOpen
  \bibfield  {author} {\bibinfo {author} {\bibfnamefont {P.}~\bibnamefont
  {Deuar}}, \bibinfo {author} {\bibfnamefont {A.}~\bibnamefont {Ferrier}},
  \bibinfo {author} {\bibfnamefont {M.}~\bibnamefont {Matuszewski}}, \bibinfo
  {author} {\bibfnamefont {G.}~\bibnamefont {Orso}},\ and\ \bibinfo {author}
  {\bibfnamefont {M.~H.}\ \bibnamefont {Szyma\'nska}},\ }\bibfield  {title}
  {\bibinfo {title} {Fully quantum scalable description of driven-dissipative
  lattice models},\ }\href {https://doi.org/10.1103/PRXQuantum.2.010319}
  {\bibfield  {journal} {\bibinfo  {journal} {PRX Quantum}\ }\textbf {\bibinfo
  {volume} {2}},\ \bibinfo {pages} {010319} (\bibinfo {year}
  {2021})}\BibitemShut {NoStop}%
\bibitem [{\citenamefont {Ferrier}\ \emph {et~al.}(2026)\citenamefont
  {Ferrier}, \citenamefont {Matuszewski}, \citenamefont {Deuar},\ and\
  \citenamefont {Szyma\'nska}}]{Ferrier2025}%
  \BibitemOpen
  \bibfield  {author} {\bibinfo {author} {\bibfnamefont {A.}~\bibnamefont
  {Ferrier}}, \bibinfo {author} {\bibfnamefont {M.}~\bibnamefont
  {Matuszewski}}, \bibinfo {author} {\bibfnamefont {P.}~\bibnamefont {Deuar}},\
  and\ \bibinfo {author} {\bibfnamefont {M.~H.}\ \bibnamefont {Szyma\'nska}},\
  }\bibfield  {title} {\bibinfo {title} {Antibunching in locally driven
  dissipative {L}ieb lattices},\ }\href {https://doi.org/10.1103/g22d-99w7}
  {\bibfield  {journal} {\bibinfo  {journal} {Phys. Rev. B}\ }\textbf {\bibinfo
  {volume} {114}},\ \bibinfo {pages} {185306} (\bibinfo {year}
  {2026})}\BibitemShut {NoStop}%
\bibitem [{\citenamefont {Gardiner}\ and\ \citenamefont
  {Zoller}(2004)}]{GardinerZoller2004}%
  \BibitemOpen
  \bibfield  {author} {\bibinfo {author} {\bibfnamefont {C.~W.}\ \bibnamefont
  {Gardiner}}\ and\ \bibinfo {author} {\bibfnamefont {P.}~\bibnamefont
  {Zoller}},\ }\href@noop {} {\emph {\bibinfo {title} {Quantum Noise}}},\
  \bibinfo {edition} {3rd}\ ed.\ (\bibinfo  {publisher} {Springer},\ \bibinfo
  {address} {Berlin},\ \bibinfo {year} {2004})\BibitemShut {NoStop}%
\bibitem [{\citenamefont {Wouters}\ and\ \citenamefont
  {Savona}(2009)}]{WoutersSavona2009}%
  \BibitemOpen
  \bibfield  {author} {\bibinfo {author} {\bibfnamefont {M.}~\bibnamefont
  {Wouters}}\ and\ \bibinfo {author} {\bibfnamefont {V.}~\bibnamefont
  {Savona}},\ }\bibfield  {title} {\bibinfo {title} {Stochastic classical field
  model for polariton condensates},\ }\href
  {https://doi.org/10.1103/PhysRevB.79.165302} {\bibfield  {journal} {\bibinfo
  {journal} {Phys. Rev. B}\ }\textbf {\bibinfo {volume} {79}},\ \bibinfo
  {pages} {165302} (\bibinfo {year} {2009})}\BibitemShut {NoStop}%
\bibitem [{\citenamefont {L{\"u}ders}\ \emph {et~al.}(2021)\citenamefont
  {L{\"u}ders}, \citenamefont {Pukrop}, \citenamefont {Rozas}, \citenamefont
  {Schneider}, \citenamefont {H{\"o}fling}, \citenamefont {Sperling},
  \citenamefont {Schumacher},\ and\ \citenamefont {A{\ss}mann}}]{Luders2021}%
  \BibitemOpen
  \bibfield  {author} {\bibinfo {author} {\bibfnamefont {C.}~\bibnamefont
  {L{\"u}ders}}, \bibinfo {author} {\bibfnamefont {M.}~\bibnamefont {Pukrop}},
  \bibinfo {author} {\bibfnamefont {E.}~\bibnamefont {Rozas}}, \bibinfo
  {author} {\bibfnamefont {C.}~\bibnamefont {Schneider}}, \bibinfo {author}
  {\bibfnamefont {S.}~\bibnamefont {H{\"o}fling}}, \bibinfo {author}
  {\bibfnamefont {J.}~\bibnamefont {Sperling}}, \bibinfo {author}
  {\bibfnamefont {S.}~\bibnamefont {Schumacher}},\ and\ \bibinfo {author}
  {\bibfnamefont {M.}~\bibnamefont {A{\ss}mann}},\ }\bibfield  {title}
  {\bibinfo {title} {Quantifying quantum coherence in polariton condensates},\
  }\href {https://doi.org/10.1103/PRXQuantum.2.030320} {\bibfield  {journal}
  {\bibinfo  {journal} {PRX Quantum}\ }\textbf {\bibinfo {volume} {2}},\
  \bibinfo {pages} {030320} (\bibinfo {year} {2021})}\BibitemShut {NoStop}%
\bibitem [{\citenamefont {Alnatah}\ \emph {et~al.}(2024)\citenamefont
  {Alnatah}, \citenamefont {Comaron}, \citenamefont {Mukherjee}, \citenamefont
  {Beaumariage}, \citenamefont {Pfeiffer}, \citenamefont {West}, \citenamefont
  {Baldwin}, \citenamefont {Szyma{\'n}ska},\ and\ \citenamefont
  {Snoke}}]{Alnatah2024}%
  \BibitemOpen
  \bibfield  {author} {\bibinfo {author} {\bibfnamefont {H.}~\bibnamefont
  {Alnatah}}, \bibinfo {author} {\bibfnamefont {P.}~\bibnamefont {Comaron}},
  \bibinfo {author} {\bibfnamefont {S.}~\bibnamefont {Mukherjee}}, \bibinfo
  {author} {\bibfnamefont {J.}~\bibnamefont {Beaumariage}}, \bibinfo {author}
  {\bibfnamefont {L.~N.}\ \bibnamefont {Pfeiffer}}, \bibinfo {author}
  {\bibfnamefont {K.}~\bibnamefont {West}}, \bibinfo {author} {\bibfnamefont
  {K.}~\bibnamefont {Baldwin}}, \bibinfo {author} {\bibfnamefont {M.~H.}\
  \bibnamefont {Szyma{\'n}ska}},\ and\ \bibinfo {author} {\bibfnamefont
  {D.~W.}\ \bibnamefont {Snoke}},\ }\bibfield  {title} {\bibinfo {title}
  {Critical fluctuations in a confined driven-dissipative quantum condensate},\
  }\href {https://doi.org/10.1126/sciadv.adi6762} {\bibfield  {journal}
  {\bibinfo  {journal} {Sci. Adv.}\ }\textbf {\bibinfo {volume} {10}},\
  \bibinfo {pages} {eadi6762} (\bibinfo {year} {2024})}\BibitemShut {NoStop}%
\bibitem [{\citenamefont {Dagvadorj}\ \emph {et~al.}(2015)\citenamefont
  {Dagvadorj}, \citenamefont {Fellows}, \citenamefont {Matyja\'skiewicz},
  \citenamefont {Marchetti}, \citenamefont {Carusotto},\ and\ \citenamefont
  {Szyma\'nska}}]{Dagvadorj2015}%
  \BibitemOpen
  \bibfield  {author} {\bibinfo {author} {\bibfnamefont {G.}~\bibnamefont
  {Dagvadorj}}, \bibinfo {author} {\bibfnamefont {J.~M.}\ \bibnamefont
  {Fellows}}, \bibinfo {author} {\bibfnamefont {S.}~\bibnamefont
  {Matyja\'skiewicz}}, \bibinfo {author} {\bibfnamefont {F.~M.}\ \bibnamefont
  {Marchetti}}, \bibinfo {author} {\bibfnamefont {I.}~\bibnamefont
  {Carusotto}},\ and\ \bibinfo {author} {\bibfnamefont {M.~H.}\ \bibnamefont
  {Szyma\'nska}},\ }\bibfield  {title} {\bibinfo {title} {Nonequilibrium phase
  transition in a two-dimensional driven open quantum system},\ }\href
  {https://doi.org/10.1103/PhysRevX.5.041028} {\bibfield  {journal} {\bibinfo
  {journal} {Phys. Rev. X}\ }\textbf {\bibinfo {volume} {5}},\ \bibinfo {pages}
  {041028} (\bibinfo {year} {2015})}\BibitemShut {NoStop}%
\bibitem [{\citenamefont {Comaron}\ \emph {et~al.}(2021)\citenamefont
  {Comaron}, \citenamefont {Carusotto}, \citenamefont {Szyma{\'n}ska},\ and\
  \citenamefont {Proukakis}}]{Comaron2021}%
  \BibitemOpen
  \bibfield  {author} {\bibinfo {author} {\bibfnamefont {P.}~\bibnamefont
  {Comaron}}, \bibinfo {author} {\bibfnamefont {I.}~\bibnamefont {Carusotto}},
  \bibinfo {author} {\bibfnamefont {M.~H.}\ \bibnamefont {Szyma{\'n}ska}},\
  and\ \bibinfo {author} {\bibfnamefont {N.~P.}\ \bibnamefont {Proukakis}},\
  }\bibfield  {title} {\bibinfo {title} {Non-equilibrium
  {Berezinskii--Kosterlitz--Thouless} transition in driven-dissipative
  condensates},\ }\href {https://doi.org/10.1209/0295-5075/133/17002}
  {\bibfield  {journal} {\bibinfo  {journal} {EPL}\ }\textbf {\bibinfo {volume}
  {133}},\ \bibinfo {pages} {17002} (\bibinfo {year} {2021})}\BibitemShut
  {NoStop}%
\bibitem [{\citenamefont {Wootters}(1987)}]{Wootters1987}%
  \BibitemOpen
  \bibfield  {author} {\bibinfo {author} {\bibfnamefont {W.~K.}\ \bibnamefont
  {Wootters}},\ }\bibfield  {title} {\bibinfo {title} {A {W}igner-function
  formulation of finite-state quantum mechanics},\ }\href
  {https://doi.org/10.1016/0003-4916(87)90176-X} {\bibfield  {journal}
  {\bibinfo  {journal} {Ann. Phys.}\ }\textbf {\bibinfo {volume} {176}},\
  \bibinfo {pages} {1} (\bibinfo {year} {1987})}\BibitemShut {NoStop}%
\bibitem [{\citenamefont {Schachenmayer}\ \emph
  {et~al.}(2015{\natexlab{a}})\citenamefont {Schachenmayer}, \citenamefont
  {Pikovski},\ and\ \citenamefont {Rey}}]{Schachenmayer2015PRX}%
  \BibitemOpen
  \bibfield  {author} {\bibinfo {author} {\bibfnamefont {J.}~\bibnamefont
  {Schachenmayer}}, \bibinfo {author} {\bibfnamefont {A.}~\bibnamefont
  {Pikovski}},\ and\ \bibinfo {author} {\bibfnamefont {A.~M.}\ \bibnamefont
  {Rey}},\ }\bibfield  {title} {\bibinfo {title} {Many-body quantum spin
  dynamics with {M}onte {C}arlo trajectories on a discrete phase space},\
  }\href {https://doi.org/10.1103/PhysRevX.5.011022} {\bibfield  {journal}
  {\bibinfo  {journal} {Phys. Rev. X}\ }\textbf {\bibinfo {volume} {5}},\
  \bibinfo {pages} {011022} (\bibinfo {year} {2015}{\natexlab{a}})}\BibitemShut
  {NoStop}%
\bibitem [{\citenamefont {Schachenmayer}\ \emph
  {et~al.}(2015{\natexlab{b}})\citenamefont {Schachenmayer}, \citenamefont
  {Pikovski},\ and\ \citenamefont {Rey}}]{Schachenmayer2015NJP}%
  \BibitemOpen
  \bibfield  {author} {\bibinfo {author} {\bibfnamefont {J.}~\bibnamefont
  {Schachenmayer}}, \bibinfo {author} {\bibfnamefont {A.}~\bibnamefont
  {Pikovski}},\ and\ \bibinfo {author} {\bibfnamefont {A.~M.}\ \bibnamefont
  {Rey}},\ }\bibfield  {title} {\bibinfo {title} {Dynamics of correlations in
  two-dimensional quantum spin models with long-range interactions: a
  phase-space {M}onte-{C}arlo study},\ }\href
  {https://doi.org/10.1088/1367-2630/17/6/065009} {\bibfield  {journal}
  {\bibinfo  {journal} {New J. Phys.}\ }\textbf {\bibinfo {volume} {17}},\
  \bibinfo {pages} {065009} (\bibinfo {year} {2015}{\natexlab{b}})}\BibitemShut
  {NoStop}%
\bibitem [{\citenamefont {Pi{\~n}eiro~Orioli}\ \emph
  {et~al.}(2018)\citenamefont {Pi{\~n}eiro~Orioli}, \citenamefont {Signoles},
  \citenamefont {Wildhagen}, \citenamefont {G{\"u}nter}, \citenamefont
  {Berges}, \citenamefont {Whitlock},\ and\ \citenamefont
  {Weidem{\"u}ller}}]{Orioli2018}%
  \BibitemOpen
  \bibfield  {author} {\bibinfo {author} {\bibfnamefont {A.}~\bibnamefont
  {Pi{\~n}eiro~Orioli}}, \bibinfo {author} {\bibfnamefont {A.}~\bibnamefont
  {Signoles}}, \bibinfo {author} {\bibfnamefont {H.}~\bibnamefont {Wildhagen}},
  \bibinfo {author} {\bibfnamefont {G.}~\bibnamefont {G{\"u}nter}}, \bibinfo
  {author} {\bibfnamefont {J.}~\bibnamefont {Berges}}, \bibinfo {author}
  {\bibfnamefont {S.}~\bibnamefont {Whitlock}},\ and\ \bibinfo {author}
  {\bibfnamefont {M.}~\bibnamefont {Weidem{\"u}ller}},\ }\bibfield  {title}
  {\bibinfo {title} {Relaxation of an isolated dipolar-interacting {R}ydberg
  quantum spin system},\ }\href
  {https://doi.org/10.1103/PhysRevLett.120.063601} {\bibfield  {journal}
  {\bibinfo  {journal} {Phys. Rev. Lett.}\ }\textbf {\bibinfo {volume} {120}},\
  \bibinfo {pages} {063601} (\bibinfo {year} {2018})}\BibitemShut {NoStop}%
\bibitem [{\citenamefont {Davidson}\ and\ \citenamefont
  {Polkovnikov}(2015)}]{DavidsonPolkovnikov2015}%
  \BibitemOpen
  \bibfield  {author} {\bibinfo {author} {\bibfnamefont {S.~M.}\ \bibnamefont
  {Davidson}}\ and\ \bibinfo {author} {\bibfnamefont {A.}~\bibnamefont
  {Polkovnikov}},\ }\bibfield  {title} {\bibinfo {title} {{SU(3)} semiclassical
  representation of quantum dynamics of interacting spins},\ }\href
  {https://doi.org/10.1103/PhysRevLett.114.045701} {\bibfield  {journal}
  {\bibinfo  {journal} {Phys. Rev. Lett.}\ }\textbf {\bibinfo {volume} {114}},\
  \bibinfo {pages} {045701} (\bibinfo {year} {2015})}\BibitemShut {NoStop}%
\bibitem [{\citenamefont {Zhu}\ \emph {et~al.}(2019)\citenamefont {Zhu},
  \citenamefont {Rey},\ and\ \citenamefont {Schachenmayer}}]{Zhu2019}%
  \BibitemOpen
  \bibfield  {author} {\bibinfo {author} {\bibfnamefont {B.}~\bibnamefont
  {Zhu}}, \bibinfo {author} {\bibfnamefont {A.~M.}\ \bibnamefont {Rey}},\ and\
  \bibinfo {author} {\bibfnamefont {J.}~\bibnamefont {Schachenmayer}},\
  }\bibfield  {title} {\bibinfo {title} {A generalized phase space approach for
  solving quantum spin dynamics},\ }\href
  {https://doi.org/10.1088/1367-2630/ab354d} {\bibfield  {journal} {\bibinfo
  {journal} {New J. Phys.}\ }\textbf {\bibinfo {volume} {21}},\ \bibinfo
  {pages} {082001} (\bibinfo {year} {2019})}\BibitemShut {NoStop}%
\bibitem [{\citenamefont {Nagao}\ \emph {et~al.}(2021)\citenamefont {Nagao},
  \citenamefont {Takasu}, \citenamefont {Takahashi},\ and\ \citenamefont
  {Danshita}}]{Nagao2021}%
  \BibitemOpen
  \bibfield  {author} {\bibinfo {author} {\bibfnamefont {K.}~\bibnamefont
  {Nagao}}, \bibinfo {author} {\bibfnamefont {Y.}~\bibnamefont {Takasu}},
  \bibinfo {author} {\bibfnamefont {Y.}~\bibnamefont {Takahashi}},\ and\
  \bibinfo {author} {\bibfnamefont {I.}~\bibnamefont {Danshita}},\ }\bibfield
  {title} {\bibinfo {title} {{SU(3)} truncated {W}igner approximation for
  strongly interacting {B}ose gases},\ }\href
  {https://doi.org/10.1103/PhysRevResearch.3.043091} {\bibfield  {journal}
  {\bibinfo  {journal} {Phys. Rev. Research}\ }\textbf {\bibinfo {volume}
  {3}},\ \bibinfo {pages} {043091} (\bibinfo {year} {2021})}\BibitemShut
  {NoStop}%
\bibitem [{\citenamefont {Wurtz}\ \emph {et~al.}(2018)\citenamefont {Wurtz},
  \citenamefont {Polkovnikov},\ and\ \citenamefont {Sels}}]{Wurtz2018}%
  \BibitemOpen
  \bibfield  {author} {\bibinfo {author} {\bibfnamefont {J.}~\bibnamefont
  {Wurtz}}, \bibinfo {author} {\bibfnamefont {A.}~\bibnamefont {Polkovnikov}},\
  and\ \bibinfo {author} {\bibfnamefont {D.}~\bibnamefont {Sels}},\ }\bibfield
  {title} {\bibinfo {title} {Cluster truncated {W}igner approximation in
  strongly interacting systems},\ }\href
  {https://doi.org/10.1016/j.aop.2018.06.001} {\bibfield  {journal} {\bibinfo
  {journal} {Ann. Phys.}\ }\textbf {\bibinfo {volume} {395}},\ \bibinfo {pages}
  {341} (\bibinfo {year} {2018})}\BibitemShut {NoStop}%
\bibitem [{\citenamefont {Czischek}\ \emph {et~al.}(2019)\citenamefont
  {Czischek}, \citenamefont {G{\"a}rttner}, \citenamefont {Oberthaler},
  \citenamefont {Kastner},\ and\ \citenamefont {Gasenzer}}]{Czischek2019}%
  \BibitemOpen
  \bibfield  {author} {\bibinfo {author} {\bibfnamefont {S.}~\bibnamefont
  {Czischek}}, \bibinfo {author} {\bibfnamefont {M.}~\bibnamefont
  {G{\"a}rttner}}, \bibinfo {author} {\bibfnamefont {M.}~\bibnamefont
  {Oberthaler}}, \bibinfo {author} {\bibfnamefont {M.}~\bibnamefont
  {Kastner}},\ and\ \bibinfo {author} {\bibfnamefont {T.}~\bibnamefont
  {Gasenzer}},\ }\bibfield  {title} {\bibinfo {title} {Quenches near
  criticality of the quantum {I}sing chain---power and limitations of the
  discrete truncated {W}igner approximation},\ }\href
  {https://doi.org/10.1088/2058-9565/aae3f7} {\bibfield  {journal} {\bibinfo
  {journal} {Quantum Sci. Technol.}\ }\textbf {\bibinfo {volume} {4}},\
  \bibinfo {pages} {014006} (\bibinfo {year} {2019})}\BibitemShut {NoStop}%
\bibitem [{\citenamefont {Sundar}\ \emph {et~al.}(2019)\citenamefont {Sundar},
  \citenamefont {Wang},\ and\ \citenamefont {Hazzard}}]{Sundar2019}%
  \BibitemOpen
  \bibfield  {author} {\bibinfo {author} {\bibfnamefont {B.}~\bibnamefont
  {Sundar}}, \bibinfo {author} {\bibfnamefont {K.~C.}\ \bibnamefont {Wang}},\
  and\ \bibinfo {author} {\bibfnamefont {K.~R.~A.}\ \bibnamefont {Hazzard}},\
  }\bibfield  {title} {\bibinfo {title} {Analysis of continuous and discrete
  {W}igner approximations for spin dynamics},\ }\href
  {https://doi.org/10.1103/PhysRevA.99.043627} {\bibfield  {journal} {\bibinfo
  {journal} {Phys. Rev. A}\ }\textbf {\bibinfo {volume} {99}},\ \bibinfo
  {pages} {043627} (\bibinfo {year} {2019})}\BibitemShut {NoStop}%
\bibitem [{\citenamefont {Khasseh}\ \emph {et~al.}(2020)\citenamefont
  {Khasseh}, \citenamefont {Russomanno}, \citenamefont {Schmitt}, \citenamefont
  {Heyl},\ and\ \citenamefont {Fazio}}]{Khasseh2020}%
  \BibitemOpen
  \bibfield  {author} {\bibinfo {author} {\bibfnamefont {R.}~\bibnamefont
  {Khasseh}}, \bibinfo {author} {\bibfnamefont {A.}~\bibnamefont {Russomanno}},
  \bibinfo {author} {\bibfnamefont {M.}~\bibnamefont {Schmitt}}, \bibinfo
  {author} {\bibfnamefont {M.}~\bibnamefont {Heyl}},\ and\ \bibinfo {author}
  {\bibfnamefont {R.}~\bibnamefont {Fazio}},\ }\bibfield  {title} {\bibinfo
  {title} {Discrete truncated {W}igner approach to dynamical phase transitions
  in {I}sing models after a quantum quench},\ }\href
  {https://doi.org/10.1103/PhysRevB.102.014303} {\bibfield  {journal} {\bibinfo
   {journal} {Phys. Rev. B}\ }\textbf {\bibinfo {volume} {102}},\ \bibinfo
  {pages} {014303} (\bibinfo {year} {2020})}\BibitemShut {NoStop}%
\bibitem [{\citenamefont {Kunimi}\ \emph {et~al.}(2021)\citenamefont {Kunimi},
  \citenamefont {Nagao}, \citenamefont {Goto},\ and\ \citenamefont
  {Danshita}}]{Kunimi2021}%
  \BibitemOpen
  \bibfield  {author} {\bibinfo {author} {\bibfnamefont {M.}~\bibnamefont
  {Kunimi}}, \bibinfo {author} {\bibfnamefont {K.}~\bibnamefont {Nagao}},
  \bibinfo {author} {\bibfnamefont {S.}~\bibnamefont {Goto}},\ and\ \bibinfo
  {author} {\bibfnamefont {I.}~\bibnamefont {Danshita}},\ }\bibfield  {title}
  {\bibinfo {title} {Performance evaluation of the discrete truncated {W}igner
  approximation for quench dynamics of quantum spin systems with long-range
  interactions},\ }\href {https://doi.org/10.1103/PhysRevResearch.3.013060}
  {\bibfield  {journal} {\bibinfo  {journal} {Phys. Rev. Research}\ }\textbf
  {\bibinfo {volume} {3}},\ \bibinfo {pages} {013060} (\bibinfo {year}
  {2021})}\BibitemShut {NoStop}%
\bibitem [{\citenamefont {Huber}\ \emph {et~al.}(2022)\citenamefont {Huber},
  \citenamefont {Rey},\ and\ \citenamefont {Rabl}}]{Huber2022}%
  \BibitemOpen
  \bibfield  {author} {\bibinfo {author} {\bibfnamefont {J.}~\bibnamefont
  {Huber}}, \bibinfo {author} {\bibfnamefont {A.~M.}\ \bibnamefont {Rey}},\
  and\ \bibinfo {author} {\bibfnamefont {P.}~\bibnamefont {Rabl}},\ }\bibfield
  {title} {\bibinfo {title} {Realistic simulations of spin squeezing and
  cooperative coupling effects in large ensembles of interacting two-level
  systems},\ }\href {https://doi.org/10.1103/PhysRevA.105.013716} {\bibfield
  {journal} {\bibinfo  {journal} {Phys. Rev. A}\ }\textbf {\bibinfo {volume}
  {105}},\ \bibinfo {pages} {013716} (\bibinfo {year} {2022})}\BibitemShut
  {NoStop}%
\bibitem [{\citenamefont {Mink}\ \emph {et~al.}(2022)\citenamefont {Mink},
  \citenamefont {Petrosyan},\ and\ \citenamefont {Fleischhauer}}]{Mink2022}%
  \BibitemOpen
  \bibfield  {author} {\bibinfo {author} {\bibfnamefont {C.~D.}\ \bibnamefont
  {Mink}}, \bibinfo {author} {\bibfnamefont {D.}~\bibnamefont {Petrosyan}},\
  and\ \bibinfo {author} {\bibfnamefont {M.}~\bibnamefont {Fleischhauer}},\
  }\bibfield  {title} {\bibinfo {title} {Hybrid discrete-continuous truncated
  {W}igner approximation for driven, dissipative spin systems},\ }\href
  {https://doi.org/10.1103/PhysRevResearch.4.043136} {\bibfield  {journal}
  {\bibinfo  {journal} {Phys. Rev. Research}\ }\textbf {\bibinfo {volume}
  {4}},\ \bibinfo {pages} {043136} (\bibinfo {year} {2022})}\BibitemShut
  {NoStop}%
\bibitem [{\citenamefont {Mink}\ and\ \citenamefont
  {Fleischhauer}(2023)}]{MinkFleischhauer2023}%
  \BibitemOpen
  \bibfield  {author} {\bibinfo {author} {\bibfnamefont {C.~D.}\ \bibnamefont
  {Mink}}\ and\ \bibinfo {author} {\bibfnamefont {M.}~\bibnamefont
  {Fleischhauer}},\ }\bibfield  {title} {\bibinfo {title} {Collective radiative
  interactions in the discrete truncated {W}igner approximation},\ }\href
  {https://doi.org/10.21468/SciPostPhys.15.6.233} {\bibfield  {journal}
  {\bibinfo  {journal} {SciPost Phys.}\ }\textbf {\bibinfo {volume} {15}},\
  \bibinfo {pages} {233} (\bibinfo {year} {2023})}\BibitemShut {NoStop}%
\bibitem [{\citenamefont {Merkel}\ \emph {et~al.}(2021)\citenamefont {Merkel},
  \citenamefont {Link}, \citenamefont {Luoma},\ and\ \citenamefont
  {Strunz}}]{Merkel2021}%
  \BibitemOpen
  \bibfield  {author} {\bibinfo {author} {\bibfnamefont {K.}~\bibnamefont
  {Merkel}}, \bibinfo {author} {\bibfnamefont {V.}~\bibnamefont {Link}},
  \bibinfo {author} {\bibfnamefont {K.}~\bibnamefont {Luoma}},\ and\ \bibinfo
  {author} {\bibfnamefont {W.~T.}\ \bibnamefont {Strunz}},\ }\bibfield  {title}
  {\bibinfo {title} {Phase space theory for open quantum systems with local and
  collective dissipative processes},\ }\href
  {https://doi.org/10.1088/1751-8121/abd155} {\bibfield  {journal} {\bibinfo
  {journal} {J. Phys. A: Math. Theor.}\ }\textbf {\bibinfo {volume} {54}},\
  \bibinfo {pages} {035303} (\bibinfo {year} {2021})}\BibitemShut {NoStop}%
\bibitem [{\citenamefont {Hosseinabadi}\ \emph {et~al.}(2025)\citenamefont
  {Hosseinabadi}, \citenamefont {Chelpanova},\ and\ \citenamefont
  {Marino}}]{Hosseinabadi2025}%
  \BibitemOpen
  \bibfield  {author} {\bibinfo {author} {\bibfnamefont {H.}~\bibnamefont
  {Hosseinabadi}}, \bibinfo {author} {\bibfnamefont {O.}~\bibnamefont
  {Chelpanova}},\ and\ \bibinfo {author} {\bibfnamefont {J.}~\bibnamefont
  {Marino}},\ }\bibfield  {title} {\bibinfo {title} {User-friendly truncated
  {W}igner approximation for dissipative spin dynamics},\ }\href
  {https://doi.org/10.1103/1wwv-k7hg} {\bibfield  {journal} {\bibinfo
  {journal} {PRX Quantum}\ }\textbf {\bibinfo {volume} {6}},\ \bibinfo {pages}
  {030344} (\bibinfo {year} {2025})}\BibitemShut {NoStop}%
\bibitem [{\citenamefont {Noel}\ \emph {et~al.}(2026)\citenamefont {Noel},
  \citenamefont {Fleischhauer},\ and\ \citenamefont {Lesanovsky}}]{Noel2026}%
  \BibitemOpen
  \bibfield  {author} {\bibinfo {author} {\bibfnamefont {V.}~\bibnamefont
  {Noel}}, \bibinfo {author} {\bibfnamefont {M.}~\bibnamefont {Fleischhauer}},\
  and\ \bibinfo {author} {\bibfnamefont {I.}~\bibnamefont {Lesanovsky}},\
  }\href {https://doi.org/10.48550/arXiv.2607.26029} {\bibinfo {title} {Path
  integral approach to the truncated {W}igner approximation of
  driven-dissipative spins}} (\bibinfo {year} {2026}),\ \Eprint
  {https://arxiv.org/abs/2607.26029} {arXiv:2607.26029 [quant-ph]} \BibitemShut
  {NoStop}%
\bibitem [{\citenamefont {Hartmann}\ \emph {et~al.}(2026)\citenamefont
  {Hartmann}, \citenamefont {Schlegel}, \citenamefont {Noel}, \citenamefont
  {Mink},\ and\ \citenamefont {Fleischhauer}}]{Hartmann2026}%
  \BibitemOpen
  \bibfield  {author} {\bibinfo {author} {\bibfnamefont {J.}~\bibnamefont
  {Hartmann}}, \bibinfo {author} {\bibfnamefont {T.}~\bibnamefont {Schlegel}},
  \bibinfo {author} {\bibfnamefont {V.}~\bibnamefont {Noel}}, \bibinfo {author}
  {\bibfnamefont {C.~D.}\ \bibnamefont {Mink}},\ and\ \bibinfo {author}
  {\bibfnamefont {M.}~\bibnamefont {Fleischhauer}},\ }\href
  {https://doi.org/10.48550/arXiv.2607.17295} {\bibinfo {title} {Truncated
  {W}igner approximation for spins in continuous phase space}} (\bibinfo {year}
  {2026}),\ \Eprint {https://arxiv.org/abs/2607.17295} {arXiv:2607.17295
  [quant-ph]} \BibitemShut {NoStop}%
\bibitem [{\citenamefont {Tosca}\ \emph
  {et~al.}(2026{\natexlab{b}})\citenamefont {Tosca}, \citenamefont {Li},
  \citenamefont {Carnazza},\ and\ \citenamefont {Ciuti}}]{Tosca2026}%
  \BibitemOpen
  \bibfield  {author} {\bibinfo {author} {\bibfnamefont {J.}~\bibnamefont
  {Tosca}}, \bibinfo {author} {\bibfnamefont {Z.}~\bibnamefont {Li}}, \bibinfo
  {author} {\bibfnamefont {F.}~\bibnamefont {Carnazza}},\ and\ \bibinfo
  {author} {\bibfnamefont {C.}~\bibnamefont {Ciuti}},\ }\href
  {https://doi.org/10.48550/arXiv.2604.01165} {\bibinfo {title} {Variational
  dynamics of open quantum spin systems in phase space}} (\bibinfo {year}
  {2026}{\natexlab{b}}),\ \Eprint {https://arxiv.org/abs/2604.01165}
  {arXiv:2604.01165 [quant-ph]} \BibitemShut {NoStop}%
\bibitem [{\citenamefont {Huber}\ \emph {et~al.}(2020)\citenamefont {Huber},
  \citenamefont {Kirton},\ and\ \citenamefont {Rabl}}]{Huber2020}%
  \BibitemOpen
  \bibfield  {author} {\bibinfo {author} {\bibfnamefont {J.}~\bibnamefont
  {Huber}}, \bibinfo {author} {\bibfnamefont {P.}~\bibnamefont {Kirton}},\ and\
  \bibinfo {author} {\bibfnamefont {P.}~\bibnamefont {Rabl}},\ }\bibfield
  {title} {\bibinfo {title} {Nonequilibrium magnetic phases in spin lattices
  with gain and loss},\ }\href {https://doi.org/10.1103/PhysRevA.102.012219}
  {\bibfield  {journal} {\bibinfo  {journal} {Phys. Rev. A}\ }\textbf {\bibinfo
  {volume} {102}},\ \bibinfo {pages} {012219} (\bibinfo {year}
  {2020})}\BibitemShut {NoStop}%
\bibitem [{\citenamefont {Huber}\ \emph {et~al.}(2021)\citenamefont {Huber},
  \citenamefont {Kirton},\ and\ \citenamefont {Rabl}}]{Huber2021}%
  \BibitemOpen
  \bibfield  {author} {\bibinfo {author} {\bibfnamefont {J.}~\bibnamefont
  {Huber}}, \bibinfo {author} {\bibfnamefont {P.}~\bibnamefont {Kirton}},\ and\
  \bibinfo {author} {\bibfnamefont {P.}~\bibnamefont {Rabl}},\ }\bibfield
  {title} {\bibinfo {title} {Phase-space methods for simulating the dissipative
  many-body dynamics of collective spin systems},\ }\href
  {https://doi.org/10.21468/SciPostPhys.10.2.045} {\bibfield  {journal}
  {\bibinfo  {journal} {SciPost Phys.}\ }\textbf {\bibinfo {volume} {10}},\
  \bibinfo {pages} {045} (\bibinfo {year} {2021})}\BibitemShut {NoStop}%
\bibitem [{\citenamefont {Maghrebi}\ and\ \citenamefont
  {Gorshkov}(2016)}]{PhysRevB.93.014307}%
  \BibitemOpen
  \bibfield  {author} {\bibinfo {author} {\bibfnamefont {M.~F.}\ \bibnamefont
  {Maghrebi}}\ and\ \bibinfo {author} {\bibfnamefont {A.~V.}\ \bibnamefont
  {Gorshkov}},\ }\bibfield  {title} {\bibinfo {title} {Nonequilibrium many-body
  steady states via {K}eldysh formalism},\ }\href
  {https://doi.org/10.1103/PhysRevB.93.014307} {\bibfield  {journal} {\bibinfo
  {journal} {Phys. Rev. B}\ }\textbf {\bibinfo {volume} {93}},\ \bibinfo
  {pages} {014307} (\bibinfo {year} {2016})}\BibitemShut {NoStop}%
\bibitem [{\citenamefont {Lindblad}(1976)}]{Lindblad1976}%
  \BibitemOpen
  \bibfield  {author} {\bibinfo {author} {\bibfnamefont {G.}~\bibnamefont
  {Lindblad}},\ }\bibfield  {title} {\bibinfo {title} {On the generators of
  quantum dynamical semigroups},\ }\href {https://doi.org/10.1007/BF01608499}
  {\bibfield  {journal} {\bibinfo  {journal} {Commun. Math. Phys.}\ }\textbf
  {\bibinfo {volume} {48}},\ \bibinfo {pages} {119} (\bibinfo {year}
  {1976})}\BibitemShut {NoStop}%
\bibitem [{\citenamefont {Gorini}\ \emph {et~al.}(1976)\citenamefont {Gorini},
  \citenamefont {Kossakowski},\ and\ \citenamefont {Sudarshan}}]{GKS1976}%
  \BibitemOpen
  \bibfield  {author} {\bibinfo {author} {\bibfnamefont {V.}~\bibnamefont
  {Gorini}}, \bibinfo {author} {\bibfnamefont {A.}~\bibnamefont
  {Kossakowski}},\ and\ \bibinfo {author} {\bibfnamefont {E.~C.~G.}\
  \bibnamefont {Sudarshan}},\ }\bibfield  {title} {\bibinfo {title} {Completely
  positive dynamical semigroups of {N}-level systems},\ }\href
  {https://doi.org/10.1063/1.522979} {\bibfield  {journal} {\bibinfo  {journal}
  {J. Math. Phys.}\ }\textbf {\bibinfo {volume} {17}},\ \bibinfo {pages} {821}
  (\bibinfo {year} {1976})}\BibitemShut {NoStop}%
\bibitem [{\citenamefont {Stratonovich}(1957)}]{stratonovich1957distributions}%
  \BibitemOpen
  \bibfield  {author} {\bibinfo {author} {\bibfnamefont {R.~L.}\ \bibnamefont
  {Stratonovich}},\ }\bibfield  {title} {\bibinfo {title} {On distributions in
  representation space},\ }\href
  {http://jetp.ras.ru/cgi-bin/e/index/e/4/6/p891?a=list} {\bibfield  {journal}
  {\bibinfo  {journal} {Sov. Phys. JETP}\ }\textbf {\bibinfo {volume} {4}},\
  \bibinfo {pages} {891} (\bibinfo {year} {1957})}\BibitemShut {NoStop}%
\bibitem [{\citenamefont {Várilly}\ and\ \citenamefont
  {Gracia-Bondía}(1989)}]{VARILLY1989107}%
  \BibitemOpen
  \bibfield  {author} {\bibinfo {author} {\bibfnamefont {J.~C.}\ \bibnamefont
  {Várilly}}\ and\ \bibinfo {author} {\bibfnamefont {J.~M.}\ \bibnamefont
  {Gracia-Bondía}},\ }\bibfield  {title} {\bibinfo {title} {The {M}oyal
  representation for spin},\ }\href
  {https://doi.org/10.1016/0003-4916(89)90262-5} {\bibfield  {journal}
  {\bibinfo  {journal} {Ann. Phys.}\ }\textbf {\bibinfo {volume} {190}},\
  \bibinfo {pages} {107} (\bibinfo {year} {1989})}\BibitemShut {NoStop}%
\bibitem [{\citenamefont {Klimov}\ and\ \citenamefont
  {Espinoza}(2002)}]{Klimov_2002}%
  \BibitemOpen
  \bibfield  {author} {\bibinfo {author} {\bibfnamefont {A.~B.}\ \bibnamefont
  {Klimov}}\ and\ \bibinfo {author} {\bibfnamefont {P.}~\bibnamefont
  {Espinoza}},\ }\bibfield  {title} {\bibinfo {title} {Moyal-like form of the
  star product for generalized {SU(2)} {S}tratonovich-{W}eyl symbols},\ }\href
  {https://doi.org/10.1088/0305-4470/35/40/305} {\bibfield  {journal} {\bibinfo
   {journal} {J. Phys. A: Math. Gen.}\ }\textbf {\bibinfo {volume} {35}},\
  \bibinfo {pages} {8435} (\bibinfo {year} {2002})}\BibitemShut {NoStop}%
\bibitem [{\citenamefont {Grossmann}(1976)}]{Grossmann1976}%
  \BibitemOpen
  \bibfield  {author} {\bibinfo {author} {\bibfnamefont {A.}~\bibnamefont
  {Grossmann}},\ }\bibfield  {title} {\bibinfo {title} {Parity operator and
  quantization of $\delta$-functions},\ }\href
  {https://doi.org/10.1007/BF01617867} {\bibfield  {journal} {\bibinfo
  {journal} {Commun. Math. Phys.}\ }\textbf {\bibinfo {volume} {48}},\ \bibinfo
  {pages} {191} (\bibinfo {year} {1976})}\BibitemShut {NoStop}%
\bibitem [{\citenamefont {Royer}(1977)}]{Royer1977}%
  \BibitemOpen
  \bibfield  {author} {\bibinfo {author} {\bibfnamefont {A.}~\bibnamefont
  {Royer}},\ }\bibfield  {title} {\bibinfo {title} {Wigner function as the
  expectation value of a parity operator},\ }\href
  {https://doi.org/10.1103/PhysRevA.15.449} {\bibfield  {journal} {\bibinfo
  {journal} {Phys. Rev. A}\ }\textbf {\bibinfo {volume} {15}},\ \bibinfo
  {pages} {449} (\bibinfo {year} {1977})}\BibitemShut {NoStop}%
\bibitem [{\citenamefont {Klimov}\ and\ \citenamefont
  {Chumakov}(2009)}]{klimov2009group}%
  \BibitemOpen
  \bibfield  {author} {\bibinfo {author} {\bibfnamefont {A.~B.}\ \bibnamefont
  {Klimov}}\ and\ \bibinfo {author} {\bibfnamefont {S.~M.}\ \bibnamefont
  {Chumakov}},\ }\href {https://doi.org/10.1002/9783527624003} {\emph {\bibinfo
  {title} {A Group-Theoretical Approach to Quantum Optics: Models of Atom-Field
  Interactions}}}\ (\bibinfo  {publisher} {Wiley-VCH},\ \bibinfo {address}
  {Weinheim},\ \bibinfo {year} {2009})\BibitemShut {NoStop}%
\bibitem [{\citenamefont {Groenewold}(1946)}]{Groenewold1946}%
  \BibitemOpen
  \bibfield  {author} {\bibinfo {author} {\bibfnamefont {H.~J.}\ \bibnamefont
  {Groenewold}},\ }\bibfield  {title} {\bibinfo {title} {On the principles of
  elementary quantum mechanics},\ }\href
  {https://doi.org/10.1016/S0031-8914(46)80059-4} {\bibfield  {journal}
  {\bibinfo  {journal} {Physica}\ }\textbf {\bibinfo {volume} {12}},\ \bibinfo
  {pages} {405} (\bibinfo {year} {1946})}\BibitemShut {NoStop}%
\bibitem [{\citenamefont {Moyal}(1949)}]{Moyal1949}%
  \BibitemOpen
  \bibfield  {author} {\bibinfo {author} {\bibfnamefont {J.~E.}\ \bibnamefont
  {Moyal}},\ }\bibfield  {title} {\bibinfo {title} {Quantum mechanics as a
  statistical theory},\ }\href {https://doi.org/10.1017/S0305004100000487}
  {\bibfield  {journal} {\bibinfo  {journal} {Math. Proc. Cambridge Philos.
  Soc.}\ }\textbf {\bibinfo {volume} {45}},\ \bibinfo {pages} {99} (\bibinfo
  {year} {1949})}\BibitemShut {NoStop}%
\bibitem [{\citenamefont {Bopp}(1956)}]{Bopp1956}%
  \BibitemOpen
  \bibfield  {author} {\bibinfo {author} {\bibfnamefont {F.}~\bibnamefont
  {Bopp}},\ }\bibfield  {title} {\bibinfo {title} {La m\'ecanique quantique
  est-elle une m\'ecanique statistique classique particuli\`ere?},\ }\href
  {http://www.numdam.org/item/AIHP_1956__15_2_81_0/} {\bibfield  {journal}
  {\bibinfo  {journal} {Ann. Inst. Henri Poincar\'e}\ }\textbf {\bibinfo
  {volume} {15}},\ \bibinfo {pages} {81} (\bibinfo {year} {1956})}\BibitemShut
  {NoStop}%
\bibitem [{\citenamefont {Gardiner}(2009)}]{Gardiner2009}%
  \BibitemOpen
  \bibfield  {author} {\bibinfo {author} {\bibfnamefont {C.~W.}\ \bibnamefont
  {Gardiner}},\ }\href@noop {} {\emph {\bibinfo {title} {Stochastic Methods: A
  Handbook for the Natural and Social Sciences}}},\ \bibinfo {edition} {4th}\
  ed.,\ Springer Series in Synergetics\ (\bibinfo  {publisher} {Springer},\
  \bibinfo {address} {Berlin},\ \bibinfo {year} {2009})\BibitemShut {NoStop}%
\bibitem [{\citenamefont {Lipkin}\ \emph {et~al.}(1965)\citenamefont {Lipkin},
  \citenamefont {Meshkov},\ and\ \citenamefont {Glick}}]{LMG1965}%
  \BibitemOpen
  \bibfield  {author} {\bibinfo {author} {\bibfnamefont {H.~J.}\ \bibnamefont
  {Lipkin}}, \bibinfo {author} {\bibfnamefont {N.}~\bibnamefont {Meshkov}},\
  and\ \bibinfo {author} {\bibfnamefont {A.~J.}\ \bibnamefont {Glick}},\
  }\bibfield  {title} {\bibinfo {title} {Validity of many-body approximation
  methods for a solvable model: ({I}). {E}xact solutions and perturbation
  theory},\ }\href {https://doi.org/10.1016/0029-5582(65)90862-X} {\bibfield
  {journal} {\bibinfo  {journal} {Nucl. Phys.}\ }\textbf {\bibinfo {volume}
  {62}},\ \bibinfo {pages} {188} (\bibinfo {year} {1965})}\BibitemShut
  {NoStop}%
\bibitem [{\citenamefont {Lawler}(2019)}]{lawler2018notes}%
  \BibitemOpen
  \bibfield  {author} {\bibinfo {author} {\bibfnamefont {G.~F.}\ \bibnamefont
  {Lawler}},\ }\href {http://www.math.uchicago.edu/~lawler/bessel18new.pdf}
  {\bibinfo {title} {Notes on the {B}essel process}},\ \bibinfo {howpublished}
  {Unpublished lecture notes} (\bibinfo {year} {2019})\BibitemShut {NoStop}%
\bibitem [{\citenamefont {R{\"{o}}{\ss}ler}(2010)}]{rossler2010runge}%
  \BibitemOpen
  \bibfield  {author} {\bibinfo {author} {\bibfnamefont {A.}~\bibnamefont
  {R{\"{o}}{\ss}ler}},\ }\bibfield  {title} {\bibinfo {title} {{R}unge--{K}utta
  methods for the strong approximation of solutions of stochastic differential
  equations},\ }\href {https://doi.org/10.1137/09076636X} {\bibfield  {journal}
  {\bibinfo  {journal} {SIAM J. Numer. Anal.}\ }\textbf {\bibinfo {volume}
  {48}},\ \bibinfo {pages} {922} (\bibinfo {year} {2010})}\BibitemShut
  {NoStop}%
\bibitem [{\citenamefont {Kac}\ \emph {et~al.}(1963)\citenamefont {Kac},
  \citenamefont {Uhlenbeck},\ and\ \citenamefont {Hemmer}}]{kac1963van}%
  \BibitemOpen
  \bibfield  {author} {\bibinfo {author} {\bibfnamefont {M.}~\bibnamefont
  {Kac}}, \bibinfo {author} {\bibfnamefont {G.~E.}\ \bibnamefont {Uhlenbeck}},\
  and\ \bibinfo {author} {\bibfnamefont {P.~C.}\ \bibnamefont {Hemmer}},\
  }\bibfield  {title} {\bibinfo {title} {On the van der {W}aals theory of the
  vapor-liquid equilibrium. {I}. {D}iscussion of a one-dimensional model},\
  }\href {https://doi.org/10.1063/1.1703946} {\bibfield  {journal} {\bibinfo
  {journal} {J. Math. Phys.}\ }\textbf {\bibinfo {volume} {4}},\ \bibinfo
  {pages} {216} (\bibinfo {year} {1963})}\BibitemShut {NoStop}%
\bibitem [{\citenamefont {Lee}\ \emph {et~al.}(2014)\citenamefont {Lee},
  \citenamefont {Chan},\ and\ \citenamefont {Yelin}}]{lee2014dissipative}%
  \BibitemOpen
  \bibfield  {author} {\bibinfo {author} {\bibfnamefont {T.~E.}\ \bibnamefont
  {Lee}}, \bibinfo {author} {\bibfnamefont {C.-K.}\ \bibnamefont {Chan}},\ and\
  \bibinfo {author} {\bibfnamefont {S.~F.}\ \bibnamefont {Yelin}},\ }\bibfield
  {title} {\bibinfo {title} {Dissipative phase transitions: Independent versus
  collective decay and spin squeezing},\ }\href
  {https://doi.org/10.1103/PhysRevA.90.052109} {\bibfield  {journal} {\bibinfo
  {journal} {Phys. Rev. A}\ }\textbf {\bibinfo {volume} {90}},\ \bibinfo
  {pages} {052109} (\bibinfo {year} {2014})}\BibitemShut {NoStop}%
\bibitem [{\citenamefont {Mc~Keever}(2022)}]{mckeever2022tensor}%
  \BibitemOpen
  \bibfield  {author} {\bibinfo {author} {\bibfnamefont {C.~P.}\ \bibnamefont
  {Mc~Keever}},\ }\emph {\bibinfo {title} {Tensor Network Simulation Methods
  for Open Quantum Lattice Models}},\ \href
  {https://discovery.ucl.ac.uk/id/eprint/10146915/} {Ph.D. thesis},\ \bibinfo
  {school} {University College London}, \bibinfo {address} {London, United
  Kingdom} (\bibinfo {year} {2022})\BibitemShut {NoStop}%
\bibitem [{\citenamefont {Strocchi}(2019)}]{Strocchi2019}%
  \BibitemOpen
  \bibfield  {author} {\bibinfo {author} {\bibfnamefont {F.}~\bibnamefont
  {Strocchi}},\ }\href {https://doi.org/10.1007/978-88-7642-660-5} {\emph
  {\bibinfo {title} {Symmetry Breaking in the Standard Model: A
  Non-Perturbative Outlook}}},\ \bibinfo {series} {Publications of the Scuola
  Normale Superiore}, Vol.~\bibinfo {volume} {19}\ (\bibinfo  {publisher}
  {Edizioni della Normale},\ \bibinfo {address} {Pisa},\ \bibinfo {year}
  {2019})\BibitemShut {NoStop}%
\bibitem [{\citenamefont {Berezinskii}(1971)}]{Berezinskii1971}%
  \BibitemOpen
  \bibfield  {author} {\bibinfo {author} {\bibfnamefont {V.~L.}\ \bibnamefont
  {Berezinskii}},\ }\bibfield  {title} {\bibinfo {title} {Destruction of
  long-range order in one-dimensional and two-dimensional systems having a
  continuous symmetry group {I}. {C}lassical systems},\ }\href
  {http://jetp.ras.ru/cgi-bin/e/index/e/32/3/p493?a=list} {\bibfield  {journal}
  {\bibinfo  {journal} {Sov. Phys. JETP}\ }\textbf {\bibinfo {volume} {32}},\
  \bibinfo {pages} {493} (\bibinfo {year} {1971})}\BibitemShut {NoStop}%
\bibitem [{\citenamefont {Kosterlitz}\ and\ \citenamefont
  {Thouless}(1973)}]{KosterlitzThouless1973}%
  \BibitemOpen
  \bibfield  {author} {\bibinfo {author} {\bibfnamefont {J.~M.}\ \bibnamefont
  {Kosterlitz}}\ and\ \bibinfo {author} {\bibfnamefont {D.~J.}\ \bibnamefont
  {Thouless}},\ }\bibfield  {title} {\bibinfo {title} {Ordering, metastability
  and phase transitions in two-dimensional systems},\ }\href
  {https://doi.org/10.1088/0022-3719/6/7/010} {\bibfield  {journal} {\bibinfo
  {journal} {J. Phys. C: Solid State Phys.}\ }\textbf {\bibinfo {volume} {6}},\
  \bibinfo {pages} {1181} (\bibinfo {year} {1973})}\BibitemShut {NoStop}%
\bibitem [{\citenamefont {Rabl}(2023)}]{rabl2023stochastic}%
  \BibitemOpen
  \bibfield  {author} {\bibinfo {author} {\bibfnamefont {P.}~\bibnamefont
  {Rabl}},\ }\href
  {https://iqtn.phys.strath.ac.uk/numerical-methods-for-many-body-open-quantum-systems-workshop-assets/}
  {\bibinfo {title} {Stochastic simulations of dissipative spin systems}},\
  \bibinfo {howpublished} {Talk presented at the International Quantum Tensor
  Network (IQTN) workshop on Numerical Methods for Many-Body Open Quantum
  Systems, Hotel Malmaison, Dundee} (\bibinfo {year} {2023}),\ \bibinfo {note}
  {{January} 9th--10th, 2023}\BibitemShut {NoStop}%
\bibitem [{\citenamefont {Senko}\ \emph {et~al.}(2015)\citenamefont {Senko},
  \citenamefont {Richerme}, \citenamefont {Smith}, \citenamefont {Lee},
  \citenamefont {Cohen}, \citenamefont {Retzker},\ and\ \citenamefont
  {Monroe}}]{Senko2015}%
  \BibitemOpen
  \bibfield  {author} {\bibinfo {author} {\bibfnamefont {C.}~\bibnamefont
  {Senko}}, \bibinfo {author} {\bibfnamefont {P.}~\bibnamefont {Richerme}},
  \bibinfo {author} {\bibfnamefont {J.}~\bibnamefont {Smith}}, \bibinfo
  {author} {\bibfnamefont {A.}~\bibnamefont {Lee}}, \bibinfo {author}
  {\bibfnamefont {I.}~\bibnamefont {Cohen}}, \bibinfo {author} {\bibfnamefont
  {A.}~\bibnamefont {Retzker}},\ and\ \bibinfo {author} {\bibfnamefont
  {C.}~\bibnamefont {Monroe}},\ }\bibfield  {title} {\bibinfo {title}
  {Realization of a quantum integer-spin chain with controllable
  interactions},\ }\href {https://doi.org/10.1103/PhysRevX.5.021026} {\bibfield
   {journal} {\bibinfo  {journal} {Phys. Rev. X}\ }\textbf {\bibinfo {volume}
  {5}},\ \bibinfo {pages} {021026} (\bibinfo {year} {2015})}\BibitemShut
  {NoStop}%
\bibitem [{\citenamefont {Hepworth}\ \emph {et~al.}(2025)\citenamefont
  {Hepworth}, \citenamefont {Ruttley}, \citenamefont {von Gierke},
  \citenamefont {Gregory}, \citenamefont {Guttridge},\ and\ \citenamefont
  {Cornish}}]{Hepworth2025}%
  \BibitemOpen
  \bibfield  {author} {\bibinfo {author} {\bibfnamefont {T.~R.}\ \bibnamefont
  {Hepworth}}, \bibinfo {author} {\bibfnamefont {D.~K.}\ \bibnamefont
  {Ruttley}}, \bibinfo {author} {\bibfnamefont {F.}~\bibnamefont {von Gierke}},
  \bibinfo {author} {\bibfnamefont {P.~D.}\ \bibnamefont {Gregory}}, \bibinfo
  {author} {\bibfnamefont {A.}~\bibnamefont {Guttridge}},\ and\ \bibinfo
  {author} {\bibfnamefont {S.~L.}\ \bibnamefont {Cornish}},\ }\bibfield
  {title} {\bibinfo {title} {Long-lived multilevel coherences and spin-1
  dynamics encoded in the rotational states of ultracold molecules},\ }\href
  {https://doi.org/10.1038/s41467-025-62275-y} {\bibfield  {journal} {\bibinfo
  {journal} {Nat. Commun.}\ }\textbf {\bibinfo {volume} {16}},\ \bibinfo
  {pages} {7131} (\bibinfo {year} {2025})}\BibitemShut {NoStop}%
\bibitem [{\citenamefont {Rui}\ \emph {et~al.}(2020)\citenamefont {Rui},
  \citenamefont {Wei}, \citenamefont {Rubio-Abadal}, \citenamefont {Hollerith},
  \citenamefont {Zeiher}, \citenamefont {Stamper-Kurn}, \citenamefont {Gross},\
  and\ \citenamefont {Bloch}}]{Rui2020}%
  \BibitemOpen
  \bibfield  {author} {\bibinfo {author} {\bibfnamefont {J.}~\bibnamefont
  {Rui}}, \bibinfo {author} {\bibfnamefont {D.}~\bibnamefont {Wei}}, \bibinfo
  {author} {\bibfnamefont {A.}~\bibnamefont {Rubio-Abadal}}, \bibinfo {author}
  {\bibfnamefont {S.}~\bibnamefont {Hollerith}}, \bibinfo {author}
  {\bibfnamefont {J.}~\bibnamefont {Zeiher}}, \bibinfo {author} {\bibfnamefont
  {D.~M.}\ \bibnamefont {Stamper-Kurn}}, \bibinfo {author} {\bibfnamefont
  {C.}~\bibnamefont {Gross}},\ and\ \bibinfo {author} {\bibfnamefont
  {I.}~\bibnamefont {Bloch}},\ }\bibfield  {title} {\bibinfo {title} {A
  subradiant optical mirror formed by a single structured atomic layer},\
  }\href {https://doi.org/10.1038/s41586-020-2463-x} {\bibfield  {journal}
  {\bibinfo  {journal} {Nature}\ }\textbf {\bibinfo {volume} {583}},\ \bibinfo
  {pages} {369} (\bibinfo {year} {2020})}\BibitemShut {NoStop}%
\bibitem [{\citenamefont {Rubies-Bigorda}\ \emph {et~al.}(2023)\citenamefont
  {Rubies-Bigorda}, \citenamefont {Ostermann},\ and\ \citenamefont
  {Yelin}}]{RubiesBigorda2023}%
  \BibitemOpen
  \bibfield  {author} {\bibinfo {author} {\bibfnamefont {O.}~\bibnamefont
  {Rubies-Bigorda}}, \bibinfo {author} {\bibfnamefont {S.}~\bibnamefont
  {Ostermann}},\ and\ \bibinfo {author} {\bibfnamefont {S.~F.}\ \bibnamefont
  {Yelin}},\ }\bibfield  {title} {\bibinfo {title} {Characterizing superradiant
  dynamics in atomic arrays via a cumulant expansion approach},\ }\href
  {https://doi.org/10.1103/PhysRevResearch.5.013091} {\bibfield  {journal}
  {\bibinfo  {journal} {Phys. Rev. Research}\ }\textbf {\bibinfo {volume}
  {5}},\ \bibinfo {pages} {013091} (\bibinfo {year} {2023})}\BibitemShut
  {NoStop}%
\bibitem [{\citenamefont {Ritsch}\ \emph {et~al.}(2013)\citenamefont {Ritsch},
  \citenamefont {Domokos}, \citenamefont {Brennecke},\ and\ \citenamefont
  {Esslinger}}]{Ritsch2013}%
  \BibitemOpen
  \bibfield  {author} {\bibinfo {author} {\bibfnamefont {H.}~\bibnamefont
  {Ritsch}}, \bibinfo {author} {\bibfnamefont {P.}~\bibnamefont {Domokos}},
  \bibinfo {author} {\bibfnamefont {F.}~\bibnamefont {Brennecke}},\ and\
  \bibinfo {author} {\bibfnamefont {T.}~\bibnamefont {Esslinger}},\ }\bibfield
  {title} {\bibinfo {title} {Cold atoms in cavity-generated dynamical optical
  potentials},\ }\href {https://doi.org/10.1103/RevModPhys.85.553} {\bibfield
  {journal} {\bibinfo  {journal} {Rev. Mod. Phys.}\ }\textbf {\bibinfo {volume}
  {85}},\ \bibinfo {pages} {553} (\bibinfo {year} {2013})}\BibitemShut
  {NoStop}%
\bibitem [{\citenamefont {Sheremet}\ \emph {et~al.}(2023)\citenamefont
  {Sheremet}, \citenamefont {Petrov}, \citenamefont {Iorsh}, \citenamefont
  {Poshakinskiy},\ and\ \citenamefont {Poddubny}}]{Sheremet2023}%
  \BibitemOpen
  \bibfield  {author} {\bibinfo {author} {\bibfnamefont {A.~S.}\ \bibnamefont
  {Sheremet}}, \bibinfo {author} {\bibfnamefont {M.~I.}\ \bibnamefont
  {Petrov}}, \bibinfo {author} {\bibfnamefont {I.~V.}\ \bibnamefont {Iorsh}},
  \bibinfo {author} {\bibfnamefont {A.~V.}\ \bibnamefont {Poshakinskiy}},\ and\
  \bibinfo {author} {\bibfnamefont {A.~N.}\ \bibnamefont {Poddubny}},\
  }\bibfield  {title} {\bibinfo {title} {Waveguide quantum electrodynamics:
  Collective radiance and photon-photon correlations},\ }\href
  {https://doi.org/10.1103/RevModPhys.95.015002} {\bibfield  {journal}
  {\bibinfo  {journal} {Rev. Mod. Phys.}\ }\textbf {\bibinfo {volume} {95}},\
  \bibinfo {pages} {015002} (\bibinfo {year} {2023})}\BibitemShut {NoStop}%
\bibitem [{\citenamefont {Aspelmeyer}\ \emph {et~al.}(2014)\citenamefont
  {Aspelmeyer}, \citenamefont {Kippenberg},\ and\ \citenamefont
  {Marquardt}}]{Aspelmeyer2014}%
  \BibitemOpen
  \bibfield  {author} {\bibinfo {author} {\bibfnamefont {M.}~\bibnamefont
  {Aspelmeyer}}, \bibinfo {author} {\bibfnamefont {T.~J.}\ \bibnamefont
  {Kippenberg}},\ and\ \bibinfo {author} {\bibfnamefont {F.}~\bibnamefont
  {Marquardt}},\ }\bibfield  {title} {\bibinfo {title} {Cavity optomechanics},\
  }\href {https://doi.org/10.1103/RevModPhys.86.1391} {\bibfield  {journal}
  {\bibinfo  {journal} {Rev. Mod. Phys.}\ }\textbf {\bibinfo {volume} {86}},\
  \bibinfo {pages} {1391} (\bibinfo {year} {2014})}\BibitemShut {NoStop}%
\bibitem [{\citenamefont {Bezanson}\ \emph {et~al.}(2017)\citenamefont
  {Bezanson}, \citenamefont {Edelman}, \citenamefont {Karpinski},\ and\
  \citenamefont {Shah}}]{doi:10.1137/141000671}%
  \BibitemOpen
  \bibfield  {author} {\bibinfo {author} {\bibfnamefont {J.}~\bibnamefont
  {Bezanson}}, \bibinfo {author} {\bibfnamefont {A.}~\bibnamefont {Edelman}},
  \bibinfo {author} {\bibfnamefont {S.}~\bibnamefont {Karpinski}},\ and\
  \bibinfo {author} {\bibfnamefont {V.~B.}\ \bibnamefont {Shah}},\ }\bibfield
  {title} {\bibinfo {title} {Julia: A fresh approach to numerical computing},\
  }\href {https://doi.org/10.1137/141000671} {\bibfield  {journal} {\bibinfo
  {journal} {SIAM Rev.}\ }\textbf {\bibinfo {volume} {59}},\ \bibinfo {pages}
  {65} (\bibinfo {year} {2017})}\BibitemShut {NoStop}%
\bibitem [{\citenamefont {Timofeev}(2025)}]{timofeev2025superfluidity}%
  \BibitemOpen
  \bibfield  {author} {\bibinfo {author} {\bibfnamefont {I.}~\bibnamefont
  {Timofeev}},\ }\emph {\bibinfo {title} {Superfluidity, Metastability, and
  Collective Spin Phenomena in Open Quantum Systems}},\ \href
  {https://discovery.ucl.ac.uk/id/eprint/10209991/} {Ph.D. thesis},\ \bibinfo
  {school} {University College London}, \bibinfo {address} {London, United
  Kingdom} (\bibinfo {year} {2025})\BibitemShut {NoStop}%
\bibitem [{Note1()}]{Note1}%
  \BibitemOpen
  \bibinfo {note} {This symbol is not difficult to calculate exactly, and is
  given in Ref.~\cite {VARILLY1989107}, but this approach demonstrates a useful
  simplification technique for more complicated symbols.}\BibitemShut {Stop}%
\bibitem [{\citenamefont {Zueco}\ and\ \citenamefont
  {Calvo}(2007)}]{zueco2007bopp}%
  \BibitemOpen
  \bibfield  {author} {\bibinfo {author} {\bibfnamefont {D.}~\bibnamefont
  {Zueco}}\ and\ \bibinfo {author} {\bibfnamefont {I.}~\bibnamefont {Calvo}},\
  }\bibfield  {title} {\bibinfo {title} {Bopp operators and phase-space spin
  dynamics: application to rotational quantum {B}rownian motion},\ }\href
  {https://doi.org/10.1088/1751-8113/40/17/015} {\bibfield  {journal} {\bibinfo
   {journal} {J. Phys. A: Math. Theor.}\ }\textbf {\bibinfo {volume} {40}},\
  \bibinfo {pages} {4635} (\bibinfo {year} {2007})}\BibitemShut {NoStop}%
\bibitem [{\citenamefont {Breuer}\ and\ \citenamefont
  {Petruccione}(2002)}]{breuer2002theory}%
  \BibitemOpen
  \bibfield  {author} {\bibinfo {author} {\bibfnamefont {H.-P.}\ \bibnamefont
  {Breuer}}\ and\ \bibinfo {author} {\bibfnamefont {F.}~\bibnamefont
  {Petruccione}},\ }\href@noop {} {\emph {\bibinfo {title} {The Theory of Open
  Quantum Systems}}}\ (\bibinfo  {publisher} {Oxford University Press},\
  \bibinfo {address} {Oxford},\ \bibinfo {year} {2002})\BibitemShut {NoStop}%
\end{thebibliography}%

\end{document}